\documentclass[11pt]{article}
\makeatletter \@addtoreset{equation}{subsection} \makeatother

\newcommand{\noi}{\vspace{12pt}\noindent}
\newcommand{\beq}{\begin{equation}}
\newcommand{\eeq}{\end{equation}}
\newcommand{\bea}{\begin{eqnarray}}
\newcommand{\eea}{\end{eqnarray}}

\newcommand{\e}[1]{{(\ref{#1})}}
\newcommand{\eq}[1]{{eq.\ (\ref{#1})}}
\newcommand{\es}[2]{{(\ref{#1}) and (\ref{#2})}}
\newcommand{\eqs}[2]{{eqs.\ (\ref{#1}) and (\ref{#2})}}
\newcommand{\Ref}[1]{{Ref.~\cite{#1}}}
\newcommand{\mb}[1]{{\mbox{${#1}$}}}
\newcommand{\equi}[1]{\stackrel{{#1}}{=}}

\newcommand{\ie}{{${ i.e., \ }$}}
\newcommand{\eg}{{${ e.g., \ }$}}

\newcommand{\cf}{{cf.\ }}
\newcommand{\resp}{{resp.\ }}
\newcommand{\aka}{{also known as }}
\newcommand{\wrt}{{with respect to }}
\newcommand{\wrtt}{{with respect to the }}

\newcommand{\wthot}{{with the help of the }}
\newcommand{\lhs}{{left--hand side }}
\newcommand{\rhs}{{right--hand side }}

\renewcommand{\propto}{\sim}
\renewcommand{\~}{ \ }
\renewcommand{\=}{ \ = \ }
\newcommand{\eps}{\varepsilon^{}}
\newcommand{\wedgecomma}{\stackrel{\wedge}{,}}

\newcommand{\otimescomma}{\stackrel{\otimes}{,}}

\newcommand{\cA}{{\cal A}}
\newcommand{\cF}{{\cal F}}
\newcommand{\cO}{{\cal O}}
\newcommand{\cR}{{\cal R}}
\newcommand{\cS}{{\cal S}}
\newcommand{\cV}{{\cal V}}
\newcommand{\BB}{B}
\newcommand{\CC}{C}

\newcommand{\PP}{P}
\newcommand{\SSS}{S}
\newcommand{\YY}{Y}
\newcommand{\cc}{c}
\newcommand{\pp}{p}
\newcommand{\sss}{s}
\newcommand{\yy}{y}
\newcommand{\ww}{w}
\newcommand{\dww}{dw}
\newcommand{\zz}{z}
\newcommand{\dzz}{dz}
\newcommand{\om}{\omega}
\newcommand{\sig}{\sigma}

\newcommand{\Sig}{\Sigma}
\newcommand{\Deltaone}{\Delta_{1}^{}}
\newcommand{\DeltaE}{\Delta_{E}^{}}
\newcommand{\Deltag}{\Delta_{g}^{}}
\newcommand{\Deltarho}{\Delta_{\rho}^{}}
\newcommand{\Deltarhog}{\Delta_{\rho^{}_{g}}^{}}
\newcommand{\nurho}{\nu_{\rho}^{}}
\newcommand{\nurhog}{\nu_{\rho^{}_{g}}^{}}
\newcommand{\nurhozero}{\nu_{\rho}^{(0)}}
\newcommand{\nurhogzero}{\nu_{\rho^{}_{g}}^{(0)}}
\newcommand{\rhog}{\rho^{}_{g}}
\newcommand{\sstate}{||s\rangle\rangle}
\newcommand{\ketrho}{|z,t\rangle\!{}^{}_{\rho}}
\newcommand{\brarho}{{}^{}_{\rho}\!\langle z,t|}

\newcommand{\mod}{{\rm mod}}

\newcommand{\str}{{\rm str}}
\newcommand{\sdet}{{\rm sdet}}

\newcommand{\cl}{{\rm cl}}

\newcommand{\Hf}{{1 \over 2}}

\newcommand{\Ih}{{i \over \hbar}}
\newcommand{\Hi}{{\hbar \over i}}

\newcommand{\twobyone}[2]{\left[\begin{array}{c}{#1} \cr
                                {#2} \end{array} \right]}

\newcommand{\twostack}[2]{\begin{array}{c} \lower.8ex\hbox{${#1}$}
                     \cr \raise.8ex\hbox{${#2}$} \end{array}}
\newcommand{\deder}[1]{{ 
 {\stackrel{\raise.2ex\hbox{$\leftarrow$}}{\delta^{r}}   } 
\over {   \delta {#1}}  }}
\newcommand{\dedel}[1]{{ 
 {\stackrel{\lower.3ex \hbox{$\rightarrow$}}{\delta^{\ell}}   }
 \over {   \delta {#1}}  }}
\newcommand{\papar}[1]{{ 
 {\stackrel{\raise.2ex\hbox{$\leftarrow$}}{\partial^{r}}   } 
\over {   \partial {#1}}  }}
\newcommand{\papal}[1]{{ 
 {\stackrel{\lower.3ex \hbox{$\rightarrow$}}{\partial^{\ell}}   }
 \over {   \partial {#1}}  }}
\newcommand{\rpa}[1]{{ 
 \stackrel{\raise.2ex\hbox{$\leftarrow$}}{\partial^{r}_{#1}}   }}
\newcommand{\lpa}[1]{{ 
 \stackrel{\lower.3ex\hbox{$\rightarrow$}}{\partial^{\ell}_{#1}}  }}

\newcommand{\larrow}[1]{\stackrel{\rightarrow}{#1}}

\newcommand{\proofbox}{\begin{flushright}
${\,\lower0.9pt\vbox{\hrule \hbox{\vrule
height 0.2 cm \hskip 0.2 cm \vrule height 0.2 cm}\hrule}\,}$
\end{flushright}}

\newtheorem{theorem}{Theorem}[section]

\newtheorem{lemma}[theorem]{Lemma}

\newtheorem{proposition}[theorem]{Proposition}

\begin{document}
\thispagestyle{empty}
\title{\Large{\bf A Comparative Study of Laplacians and \\
Schr\"odinger--Lichnerowicz--Weitzenb\"ock Identities \\ 
in Riemannian and Antisymplectic Geometry}}
\author{{\sc Igor~A.~Batalin}$^{a,b}$ and {\sc Klaus~Bering}$^{a,c}$ \\~\\
$^{a}$The Niels Bohr Institute\\The Niels Bohr International Academy\\
Blegdamsvej 17\\DK--2100 Copenhagen\\Denmark \\~\\
$^{b}$I.E.~Tamm Theory Division\\
P.N.~Lebedev Physics Institute\\Russian Academy of Sciences\\
53 Leninsky Prospect\\Moscow 119991\\Russia\\~\\
$^{c}$Institute for Theoretical Physics \& Astrophysics\\
Masaryk University\\Kotl\'a\v{r}sk\'a 2\\CZ--611 37 Brno\\Czech Republic}
\maketitle
\vfill
\begin{abstract}
We introduce an antisymplectic Dirac operator and antisymplectic gamma
matrices. We explore similarities between, on one hand, the
Schr\"odinger--Lichnerowicz formula for spinor bundles in Riemannian spin
geometry, which contains a zeroth--order term proportional to the Levi--Civita
scalar curvature, and, on the other hand, the nilpotent, Grassmann--odd,
second--order $\Delta$ operator in antisymplectic geometry, which in general
has a zeroth--order term proportional to the odd scalar curvature of an
arbitrary antisymplectic and torsionfree connection that is compatible with
the measure density. {}Finally, we discuss the close relationship with the 
two--loop scalar curvature term in the quantum Hamiltonian for a particle
in a curved Riemannian space.
\end{abstract}
\vfill
\begin{quote}
MSC number(s): 15A66; 53A55; 53B20; 58A50; 58C50. 
\hrule width 5.cm \vskip 2.mm \noindent 
$^{b}${\small E--mail:~{\tt batalin@lpi.ru}} \hspace{10mm}
$^{c}${\small E--mail:~{\tt bering@physics.muni.cz}} \\
\end{quote}

\newpage
\tableofcontents

\section{Introduction}
\label{secintro}

\noi
What do Riemannian and antisymplectic geometry have in common? The short
answer is that out of the \mb{2\times2=4} classical classes of even and odd,
Riemannian and symplectic geometries, they are the only two possibilities
that possess non--trivial Laplacians, scalar curvatures and
Weitzenb\"ock--type identities, \cf Table~\ref{gradtable}.
Our present investigation is partly spurred by the following remarkable fact. 
On one hand, one has the nilpotent, Grassmann--odd \mb{\Delta} operator, which
plays a fundamental r\^ole in antisymplectic geometry, and which helps encode
the BRST symmetry in the field--antifield formalism \cite{bv81,bv83,bv84}. It
can be written as \cite{bb07}
\beq
2\Delta\=2\Deltarho-\frac{R}{4}
\~\~\~\~\~\~\~\~\~\~\~\~\~\~\~\~({\rm antisymplectic})\label{deltaop}
\eeq
where \mb{\Deltarho} is the odd Laplacian, and \mb{R} is the odd scalar
curvature of an arbitrary antisymplectic, torsionfree and
\mb{\rho}--compatible connection \mb{\nabla^{(\Gamma)}\!=\!d\!+\!\Gamma}. On
the other hand, on a Riemannian spin manifold, one has the
Schr\"odinger--Lichnerowicz formula \cite{schroed32,lich63}
\beq
D^{(\sig)}D^{(\sig)}\=\Delta^{(\sig)}_{\rhog}-\frac{R}{4}
\~\~\~\~\~\~\~\~\~\~\~\~\~\~\~\~({\rm Riemannian})\label{squared}
\eeq
where \mb{D^{(\sig)}} is the Dirac operator, \mb{\Delta^{(\sig)}_{\rhog}} is
the spinor Laplacian, and \mb{R} is the scalar Levi--Civita curvature. 
The formula \e{deltaop} has been multiplied with a factor of \mb{2} to ease
comparison with formula \e{squared} because of the standard practice to
normalize odd Laplacians with an internal factor \mb{1/2}. In both formulas
\es{deltaop}{squared}, the coefficient in front of the zeroth--order scalar
curvature term is exactly the same, namely minus a quarter! Of course, there
are crucial differences between \eqs{deltaop}{squared}. The second--order
operators in \eq{deltaop} act on scalar functions, while the Dirac operator
\mb{D^{(\sig)}} and the Laplacian \mb{\Delta^{(\sig)}_{\rhog}} in \eq{squared}
act on spinors, as the index ``\mb{\sig}'' is meant to indicate. (The
subscript \mb{\rhog\!\equiv\!\sqrt{g}} refers to the canonical Riemannian
density.)

\noi
Our investigation can roughly be divided in three parts. The first part (which
is mainly covered in Subsections~\ref{secriemmetric}--\ref{secevenkhuda},
\ref{secriemonshell} and \ref{secantisympmetric}--\ref{secoddkhuda}) is to
define a Grassmann--even Riemannian analogue of the odd \mb{\Delta} operator
\e{deltaop}, that takes scalars in scalars:
\beq
\Deltarhog-\frac{R}{4}
\~\~\~\~\~\~\~\~\~\~\~\~\~\~\~\~({\rm Riemannian})\~.
\label{evenonshelldeltaop}
\eeq
Here \mb{\Deltarhog} is the Laplace--Beltrami operator and \mb{R} is the
Levi--Civita scalar curvature. The zeroth--order term \mb{-R/4} in the even
operator \e{evenonshelldeltaop} is special in several ways (as compared to
other choices of the zeroth--order term). {}For instance, the even operator
\e{evenonshelldeltaop} with this particular zeroth--order term \mb{-R/4} is
closely related to the quantum Hamiltonian \mb{\hat{H}} for a particle moving
in the Riemannian manifold \cite{dewitt57,dewitt64,berezin71,mizrahi75,gj76,
dewitt92,pvn93,dbpsvn96,bsvn98,bastian05}, \cf Subsection~\ref{secparticle}. 
Central to our investigation is the fact that the zeroth--order term \mb{-R/4}
also possesses a special mathematical property. To see this property, one notes
that it is possible to uniquely identify how all zeroth--order terms depend on
the canonical Riemannian density \mb{\rhog}, due to a classification of scalar
invariants, see Proposition~\ref{propositioneven}. Therefore it is possible to
consistently replace all the appearances of \mb{\rhog} with an arbitrary
density \mb{\rho}. One may now show that the \mb{\rho}--lifted version of the
operator \e{evenonshelldeltaop} is the unique operator such that the
\mb{\sqrt{\rho}}--conjugated operator is independent of \mb{\rho}. That's the
special property. This has parallels to antisymplectic geometry, where the odd
\mb{\Delta} operator \e{deltaop} shares a similar characterization. In
antisymplectic geometry, the \mb{\sqrt{\rho}}--conjugated operator
\beq
\DeltaE\=\sqrt{\rho}\Delta\frac{1}{\sqrt{\rho}}
\~\~\~\~\~\~\~\~\~\~\~\~\~\~\~\~({\rm antisymplectic})
\label{deltae}
\eeq
is precisely Khudaverdian's \mb{\DeltaE} operator
\cite{k99,kv02,k02,k04,bbd06,b06,b07}. The \mb{\DeltaE} operator \e{deltae} is
distinguished by being nilpotent and independent of \mb{\rho}. In fact, when
one tracks the equations in detail, it is possible to see that the same
coefficient \mb{-1/4} in front of the odd and even scalar curvature terms in
\eqs{deltaop}{evenonshelldeltaop} is not a coincidence, but indeed follows from
the same underlying principle of \mb{\rho}--independence. Thus it establishes
a bridge between the odd and even operators
\es{deltaop}{evenonshelldeltaop}. 

\noi
We should also mention that the even operator \e{evenonshelldeltaop} is often
compared with the conformally covariant Laplacian 
\beq
\Deltarhog-\frac{(N-2)R}{(N-1)4}
\~\~\~\~\~\~\~\~\~\~\~\~\~\~\~\~({\rm Riemannian})
\eeq
where \mb{N\!=\!\dim(M)} is the dimension of the Riemannian manifold \mb{M}.
The zeroth--order term \mb{-R/4} corresponds to \mb{N\!=\!\infty}. 

\noi
The second part (which is covered in
Subsections~\ref{secriemgamma}--\ref{secriemslformula}) is to check within
Riemannian geometry, if there is a bridge between the even operator
\e{evenonshelldeltaop} that acts on scalar functions, and the square of the
Dirac operator \e{squared} that acts on the spinor bundle \mb{\cS}. There is a
well--defined group--theoretical procedure how to compare scalars and spinors.
{}Firstly, the Dirac operator is extended to a Dirac operator that acts on the
bispinor bundle \mb{\cS\otimes\cS^{T}}. The Clebsch--Gordan decomposition
\mb{\cS\otimes\cS^{T}=\underline{\bf 1}\oplus\ldots}, in turn, contains a
singlet representation, \ie a scalar invariant, which is denoted as
\mb{\sstate}. Thus one just has to project the square of the bispinor Dirac
operator to the singlet representation to obtain an operator that acts on
scalars. Somewhat surprisingly, the operator turns out to be just the bare
Laplace--Beltrami operator \mb{\Deltarhog} with {\em no} zeroth--order term at
all, \cf Theorem~\ref{theoremriem4}. Roughly speaking, after the projection to
the singlet state \mb{\sstate}, the \mb{-R/4} curvature term in the spinor
sector \mb{\cS} is canceled by an opposite amount \mb{+R/4} in the transposed
spinor sector \mb{\cS^{T}}. So we have to conclude for the second part, that
the above group--theoretical procedure yields {\em no} relation between the
even operator \e{evenonshelldeltaop} that acts on scalar functions, and the
square of the Dirac operator \e{squared}, despite the fact that they both
contain the same \mb{-R/4} term!

\noi
The third part develops the antisymplectic side. It is spurred by the
following questions.
\begin{enumerate}
\item
Do there exist antisymplectic Clifford algebras and spinors?
\item
Does there exists a natural spinor generalization \mb{\Delta^{(\sig)}} of the
odd \mb{\Delta} operator \e{deltaop}, which takes antisymplectic spinors to
antisymplectic spinors?
\item
Can the odd \mb{\Delta^{(\sig)}} operator from question \mb{2} be written as a 
square 
\beq
\Delta^{(\sig)}\~\equi{?}\~D^{(\sig)}\star D^{(\sig)}
\~\~\~\~\~\~\~\~\~\~\~\~\~\~\~\~({\rm antisymplectic})
\label{deltastar}
\eeq
of an antisymplectic Dirac operator
\mb{D^{(\sig)}\!=\!\gamma^{A}\nabla^{(\sig)}_{A}}, where ``\mb{\star}'' is
a Fermionic multiplication, \mb{\eps(\star)=1}, and \mb{\gamma^{A}} 
are antisymplectic \mb{\gamma} matrices?
\end{enumerate}

\noi
The answers, which will be derived in detail in Sections~\ref{secantisymp}
and \ref{secantisympspingeom}, are, by most standards, ``{\em no}'' to
question 3, and ``{\em yes, there exists a first--order formalism, but there
is no second--order formalism}'' to question 1 and 2. Here the first-- and
second--order formalism refer to the realizations of the Lie--algebras of
infinitesimal frame and coordinate changes in terms of first-- and
second--order differential operators, respectively. The obstacle in
\eq{deltastar} lies in the definition of the \mb{\star} multiplication.
We shall, however, introduce a Fermionic nilpotent parameter \mb{\theta} that 
can be though of as the inverse \mb{\star^{-1}}, but since such \mb{\theta} 
parameter by definition is not invertible, the \mb{\star} multiplication
itself becomes meaningless. The trick is therefore, roughly speaking, to
multiply both side of \eq{deltastar} with \mb{\theta\!\equiv\!\star^{-1}}, \cf
Theorem~\ref{theoremantisymp0} and Theorem~\ref{theoremantisymp1}.

\noi
At the coarsest level, the main text is organized into \mb{3 \times 2 = 6}
sections. The three Sections~\ref{secgeneral}--\ref{secantisymp} are devoted
to general (=not necessarily spin) manifolds, while the next three
Sections~\ref{secgeneralspin}--\ref{secantisympspingeom} deal exclusively with
spin manifolds. Sections~\ref{secriem} and \ref{secriemspingeom} consider the
Riemannian case, and Sections~\ref{secantisymp} and \ref{secantisympspingeom}
consider the antisymplectic case, while Sections~\ref{secgeneral} and
\ref{secgeneralspin} consider the general theory that is common for both
Riemannian and antisymplectic case. The general theory
Sections~\ref{secgeneral} and \ref{secgeneralspin} explain differential
geometry, such as, connections, torsion tensors, vielbeins, flat and curved
exterior forms, etc., in the context of supermanifolds, where sign factors are
important. The Riemannian curvature tensor, the Ricci tensor and the scalar
curvature are considered in
Subsections~\ref{secriemcurv}--\ref{secriccitensor},
\ref{secmetricriemcurv}--\ref{secriemscalar} and 
\ref{secantisympriemcurv}--\ref{secoddscalar}. {}Finally,
Section~\ref{secconcl} has our conclusions.

\subsection{General Remarks About Notation}
\label{secnotation}

\noi
Adjectives from supermathematics such as ``graded'', ``super'', etc., are
implicitly implied. The sign conventions are such that two exterior forms
\mb{\xi} and \mb{\eta}, of  Grassmann--parity \mb{\eps_{\xi}}, \mb{\eps_{\eta}}
and of form--degree \mb{p^{}_{\xi}}, \mb{p^{}_{\eta}}, commute in the following
graded sense:
\beq 
 \eta \wedge \xi \=
(-1)^{\eps_{\xi}\eps_{\eta}+p^{}_{\xi}p^{}_{\eta}}\xi\wedge\eta
\~\equiv\~(-1)^{\vec{\eps}^{}_{\xi}\cdot\vec{\eps}^{}_{\eta}}\xi\wedge\eta
\label{etawedgexi}
\eeq
inside the exterior algebra. The pair \mb{(\eps,p)} acts as a
\mb{2}--dimensional vector--valued Grassmann--parity
\beq
\vec{\eps}\~:=\~\twobyone{\eps}{p\~(\mod\~2)}\~,
\eeq
as indicated in the second equality of \eq{etawedgexi}. The first component 
carries ordinary Grassmann--parity \mb{\eps}, while the second component
carries form--parity, \ie form degree modulo two. The exterior wedge symbol
``\mb{\wedge}'' is often not written explicitly, as it is redundant
information that can be deduced from the Grassmann-- and form--parity. The
commutator \mb{[F,G]} and anticommutator \mb{\{F,G\}^{}_{+}} of two operators
\mb{F} and \mb{G} are 
\bea
[F,G]\~&:=&FG-(-1)^{\eps_{F}\eps_{G}+p^{}_{F}p^{}_{G}}GF
\~\equiv\~FG-(-1)^{\vec{\eps}^{}_{F}\cdot\vec{\eps}^{}_{G}}GF\~,\label{com} \\
\{F,G\}^{}_{+}&:=&FG+(-1)^{\eps_{F}\eps_{G}+p^{}_{F}p^{}_{G}}GF
\~\equiv\~FG+(-1)^{\vec{\eps}^{}_{F}\cdot\vec{\eps}^{}_{G}}GF\~.\label{anticom}
\eea
The commutator \e{com} fulfills the Jacobi identity
\beq
\sum_{{\rm cycl.}\~F,G,H}(-1)^{\vec{\eps}^{}_{F}\cdot\vec{\eps}^{}_{H}}
[F,[G,H]]\= 0\~. \label{jacid}
\eeq
The transposed of a product of operators is:
\beq
(FG)^{T}\=(-1)^{\eps_{F}\eps_{G}+p^{}_{F}p^{}_{G}}G^{T}F^{T}
\~\equiv\~(-1)^{\vec{\eps}^{}_{F}\cdot\vec{\eps}^{}_{G}}G^{T}F^{T}\~.
\eeq
Covariant and exterior derivatives will always be from the left, while partial
derivatives can be from either left or right. We shall sometimes use round
parenthesis ``\mb{()}'' to indicate how far derivatives act, see \eg eqs.\
\e{rhocomp}, \e{evennurho0}, \es{evennu1}{evennu2} below.

\begin{table}[t]
\caption{The \mb{2\times2=4} classical geometries and their symmetries
\cite{kv02}. Only even Riemannian and antisymplectic geometries have
non--trivial Laplacians, scalar curvatures and Weitzenb\"ock--type identities.}

\label{gradtable}
\begin{center}
\begin{tabular}{|c||c||c|}  \hline
&Even Geometry& Odd Geometry\\ \hline\hline
&$g=\YY^{A} g^{}_{AB} \vee \YY^{B}$
&$g=\YY^{A} g^{}_{AB} \vee \YY^{B}$ \\
Riemannian&$\eps(g^{}_{AB})=\eps_{A}+\eps_{B}$
&$\eps(g^{}_{AB})=\eps_{A}+\eps_{B}+1$ \\
Covariant&$g^{}_{BA}=-(-1)^{(\eps_{A}+1)(\eps_{B}+1)}g^{}_{AB}$
&$g^{}_{BA}=(-1)^{\eps_{A}\eps_{B}}g^{}_{AB}$ \\
Metric&Symmetric&Symmetric \\ 
&No Closeness Relation&No Closeness Relation \\ \hline
Inverse&$\eps(g_{}^{AB})=\eps_{A}+\eps_{B}$
&$\eps(g_{}^{AB})=\eps_{A}+\eps_{B}+1$ \\
Riemannian&$g_{}^{BA}=(-1)^{\eps_{A}\eps_{B}}g_{}^{AB}$
&$g_{}^{BA}=(-1)^{(\eps_{A}+1)(\eps_{B}+1)}g_{}^{AB}$ \\
Contravariant&Symmetric&Skewsymmetric \\
Metric&Even Laplacian&No Laplacian    \\ \hline \hline
&$\omega=\Hf\CC^{A} \omega^{}_{AB} \wedge \CC^{B}$
&$E=\Hf\CC^{A} E^{}_{AB} \wedge \CC^{B}$ \\
Symplectic&$\eps(\omega^{}_{AB})=\eps_{A}+\eps_{B}$
&$\eps(E^{}_{AB})=\eps_{A}+\eps_{B}+1$ \\
Covariant&$\omega^{}_{BA}=(-1)^{(\eps_{A}+1)(\eps_{B}+1)}\omega^{}_{AB}$
&$E^{}_{BA}=-(-1)^{\eps_{A}\eps_{B}}E^{}_{AB}$ \\
Two--Form&Skewsymmetric&Skewsymmetric \\ 
&Closeness Relation&Closeness Relation\\ \hline
Inverse&$\eps(\omega_{}^{AB})=\eps_{A}+\eps_{B}$
&$\eps(E_{}^{AB})=\eps_{A}+\eps_{B}+1$ \\
Symplectic&$\omega_{}^{BA}=-(-1)^{\eps_{A}\eps_{B}}\omega_{}^{AB}$
&$E_{}^{BA}=-(-1)^{(\eps_{A}+1)(\eps_{B}+1)}E_{}^{AB}$ \\
Contravariant&Skewsymmetric&Symmetric \\
Tensor&No Laplacian&Odd Laplacian  
\\ \hline
\end{tabular}
\end{center}
\end{table}

\section{General Theory}
\label{secgeneral}

\subsection{Connection \mb{\nabla^{(\Gamma)}=d+\Gamma}}
\label{secdiff}

\noi
Let there be given a manifold \mb{M} with local coordinates \mb{\zz^{A}} of
Grassmann--parity \mb{\eps(\zz^{A})=\eps_{A}} (and form--degree
\mb{p(\zz^{A})=0}). Assume that \mb{M} is endowed with a measure density
\mb{\rho}. Let \mb{\Gamma(TM)} denote the set of sections in the tangent
bundle \mb{TM}, \ie the set of vector fields on \mb{M}. Let \mb{M} be endowed
with a tangent bundle connection
\mb{\nabla^{(\Gamma)}=d+\Gamma=\dzz^{A}\otimes\nabla^{(\Gamma)}_{A}
\~:\~\Gamma(TM)\times\Gamma(TM)\to\Gamma(TM)} 
\beq
\nabla^{(\Gamma)}_{A}
\= \papal{\zz^{A}}+\partial^{r}_{B}\~\Gamma^{B}{}_{AC} \larrow{\dzz^{C}}\~.
\label{nablaagammadef}
\eeq
Here \mb{\partial^{r}_{A}\!\equiv\!(-1)^{\eps_{A}}\partial^{\ell}_{A}} are
not usual partial derivatives. In particular, they do not act on the
Christoffel symbols \mb{\Gamma^{B}{}_{AC}} in \eq{nablaagammadef}. 
Rather they are a dual basis to the one--forms \mb{\larrow{\dzz^{A}}}:
\beq
\larrow{\dzz^{A}}(\partial^{r}_{B})\=\delta^{A}_{B}\~,\~\~\~\~\~\~\~\~
\eps(\larrow{\dzz^{A}})\=\eps_{A}\=\eps(\partial^{r}_{A})\~.
\eeq
Phrased differently, the \mb{\partial^{r}_{A}} are merely bookkeeping devices,
that transform as right partial derivatives under general coordinate
transformations. (To be able to distinguish them from true partial
derivatives, the differentiation variable \mb{\zz^{A}} on a true partial
derivative \mb{\partial/\partial \zz^{A}} is written explicitly.) {}For fixed
index ``\mb{A}'' in \eq{nablaagammadef}, the Christoffel symbol
\mb{\Gamma^{B}{}_{AC}} is a matrix \wrt index ``\mb{B}'' and index ``\mb{C}'',
and \mb{\partial^{r}_{B}\~\Gamma^{B}{}_{AC} \larrow{\dzz^{C}}} is the
corresponding linear operator: \mb{TM \to TM}. (We shall often refer to a
linear operator by its matrix, and vice--versa.)

\noi
The form--parities \mb{p(\larrow{\dzz^{A}})\!=\!p(\partial^{r}_{A})} are
either all \mb{0} or all \mb{1}, depending on applications, whereas a
\mb{1}--form \mb{\dzz^{A}} with no arrow ``\mb{\rightarrow}'' always
carries odd form--parity \mb{p(\dzz^{A})\!=\!1} (and Grassmann--parity
\mb{\eps(\dzz^{A})\!=\!\eps_{A}}).

\subsection{Torsion}
\label{secgammatorsion}

\noi
The torsion tensor \mb{T^{(\Gamma)}:\~\Gamma(TM)\times\Gamma(TM)\to\Gamma(TM)}
is defined as
\bea
T^{(\Gamma)}&\equiv&\Hf \dzz^{A} \wedge \partial^{r}_{B}\~
T^{(\Gamma)B}{}_{AC}\~\dzz^{C}
\~:=\~ [\nabla^{(\Gamma)} \wedgecomma {\rm Id} ] \cr
&=& [\dzz^{A} \papal{\zz^{A}}
+\dzz^{A}\~\partial^{r}_{B}\~\Gamma^{B}{}_{AD}\larrow{\dzz^{D}}
\~\wedgecomma \partial^{r}_{C}\~\dzz^{C} ] 
\= \dzz^{A} \wedge \partial^{r}_{B}\~\Gamma^{B}{}_{AC}\~\dzz^{C}
\~.\label{torsiongammadef}
\eea
where it is implicitly understood that there are no contractions with base
manifold indices, in this case index ``\mb{A}'' and index ``\mb{C}''. 
As expected, the torsion tensor is just an antisymmetrization of the 
Christoffel symbol \mb{\Gamma^{B}{}_{AC}} \wrtt lower indices,
\beq
 T^{(\Gamma)A}{}_{BC}\~:=\~\Gamma^{A}{}_{BC}
+(-1)^{(\eps_{B}+1)(\eps_{C}+1)}(B \leftrightarrow C)\~.
\label{torsiongamma}
\eeq
In particular, the Christoffel symbol
\beq
\Gamma^{A}{}_{BC}\=-(-1)^{(\eps_{B}+1)(\eps_{C}+1)}(B \leftrightarrow C)
\label{torsionfree}
\eeq
is symmetric \wrtt lower indices when the connection is torsionfree.

\subsection{Divergence}
\label{secdivergence}

\noi
A connection \mb{\nabla^{(\Gamma)}} can be used to define a divergence of a
Bosonic vector field \mb{X^{A}} as 
\beq
\str(\nabla^{(\Gamma)} X)
\~\equiv\~(-1)^{\eps_{A}}(\nabla^{(\Gamma)}_{A}X)^{A}\=
((-1)^{\eps_{A}}\papal{\zz^{A}}+\Gamma^{B}{}_{BA})X^{A}
\~,\~\~\~\~\~\~\~\~\~\eps_{X}~=~0~.\label{divgamma}
\eeq
On the other hand, the divergence is defined in terms of \mb{\rho} as
\beq
{\rm div}_{\rho}^{}X\~:=\~\frac{(-1)^{\eps_{A}}}{\rho}
\papal{\zz^{A}}(\rho X^{A})\~. \label{divrho}
\eeq
See \Ref{yks02} for a mathematical exposition of divergence operators on 
supermanifolds. The \mb{\nabla^{(\Gamma)}} connection is called compatible
with the measure density \mb{\rho} if
\beq
\Gamma^{B}{}_{BA} \= (-1)^{\eps_{A}}(\papal{\zz^{A}}\ln\rho)\~.\label{rhocomp}
\eeq
In this case, the two definitions \es{divgamma}{divrho} of divergence agree,
\cf \Ref{bb07}.

\subsection{The Riemann Curvature}
\label{secriemcurv}

\noi
We discuss in this Subsection~\ref{secriemcurv} the Riemann curvature tensor
on a supermanifold \cite{b97}. See \Ref{dewitt92} and \Ref{lavrov04} for
related discussions. The Riemann curvature \mb{R^{(\Gamma)}} is defined as
(half) the commutator of the \mb{\nabla^{(\Gamma)}} connection
\e{nablaagammadef},
\bea
 R^{(\Gamma)}&=&\Hf [\nabla^{(\Gamma)} \wedgecomma \nabla^{(\Gamma)}]
\= -\Hf \dzz^{B} \wedge \dzz^{A}\~\otimes\~
[\nabla^{(\Gamma)}_{A},\nabla^{(\Gamma)}_{B}] \cr
&=&-\Hf \dzz^{B} \wedge \dzz^{A}\~\otimes\~\partial^{r}_{D}\~R^{D}{}_{ABC}\~
\larrow{\dzz^{C}}\~,\label{riemtensordef1}
\eea
where it is implicitly understood that there are no contractions with base
manifold indices, in this case index ``\mb{A}'' and index ``\mb{B}''. 
({}For a torsionfree connection such contractions vanish, and there is no
ambiguity.)
\bea
R^{D}{}_{ABC}&=&\larrow{\dzz^{D}}
\left([\nabla^{(\Gamma)}_{A},\nabla^{(\Gamma)}_{B}]\partial^{r}_{C}\right) \cr
&=&(-1)^{\eps_{D}\eps_{A}}(\papal{\zz^{A}}\Gamma^{D}{}_{BC})
+\Gamma^{D}{}_{AE}\~\Gamma^{E}{}_{BC}
-(-1)^{\eps_{A}\eps_{B}}(A\leftrightarrow B)\~,\label{riemtensor1} 
\eea
Note that
the order of indices in the Riemann curvature tensor \mb{R^{D}{}_{ABC}} is
non--standard. This is to minimize appearances of Grassmann sign factors.
Alternatively, the Riemann curvature tensor may be defined as
\beq
R(X,Y)Z \= \left([\nabla^{(\Gamma)}_{X},\nabla^{(\Gamma)}_{Y}]
-\nabla^{(\Gamma)}_{[X,Y]}\right)Z
\= Y^{B}X^{A}R_{AB}{}^{D}{}_{C}Z^{C}\~\partial^{\ell}_{D}\~,
\label{riemtensordef2}
\eeq
where \mb{X=X^{A}\partial^{\ell}_{A}}, \mb{Y=Y^{B}\partial^{\ell}_{B}} and
\mb{Z=Z^{C}\partial^{\ell}_{C}} are left vector field of even Grassmann-- and
form--parity. The Riemann curvature tensor \mb{R_{AB}{}^{D}{}_{C}} reads in
local coordinates
\beq
R_{AB}{}^{D}{}_{C} \= (-1)^{\eps_{D}(\eps_{A}+\eps_{B})}R^{D}{}_{ABC}
\= (\papal{\zz^{A}}\Gamma_{B}{}^{D}{}_{C})
+(-1)^{\eps_{B}\eps_{D}}\Gamma_{A}{}^{D}{}_{E}\~\Gamma^{E}{}_{BC}
-(-1)^{\eps_{A}\eps_{B}}(A\leftrightarrow B)\~. 
\label{riemtensor3}
\eeq 
Here we have introduced a reordered Christoffel symbol
\beq
\Gamma_{A}{}^{B}{}_{C} \~:=\~(-1)^{\eps_{A}\eps_{B}}\Gamma^{B}{}_{AC}~.
\label{reorderedgamma}
\eeq
It is sometimes useful to reorder the indices in the Riemann curvature tensors
as
\beq
R_{ABC}{}^{D} \= ([\nabla_{A},\nabla_{B}]\partial^{\ell}_{C})^{D}
\= (-1)^{\eps_{C}(\eps_{D}+1)}R_{AB}{}^{D}{}_{C}\~.
\label{riemtensor4}
\eeq
Note that all expressions \e{riemtensor1}, \es{riemtensor3}{riemtensor4}
of Riemann curvature tensor are antisymmetric under an 
\mb{(A \leftrightarrow B)} exchange of index ``\mb{A}'' and ``\mb{B}''.
The first Bianchi identity reads (in the torsionfree case):
\beq
0~=~\sum_{{\rm cycl.}\~A,B,C}(-1)^{\eps_{A}\eps_{C}}R_{ABC}{}^{D}~.
\label{1bianchiid}
\eeq
We have exceptionally used the convention \mb{p(\partial^{\ell}_{A})\!=\!0} in
\eqs{riemtensordef2}{riemtensor4}.

\subsection{The Ricci Tensor}
\label{secricci2form}

\noi
The Ricci tensor is defined as
\beq
R_{AB}\~:=\~R^{C}{}_{CAB}\~.\label{defriccitensor}
\eeq
The Ricci tensor becomes symmetric
\bea
R_{AB}&=& \frac{(-1)^{\eps_{C}}}{\rho}
\papal{\zz^{C}}(\rho \Gamma^{C}{}_{AB}) 
- (\papal{\zz^{A}}\ln\rho\papar{\zz^{B}})
- \Gamma_{A}{}^{D}{}_{C}\~\Gamma^{C}{}_{DB} \cr
&=& -(-1)^{(\eps_{A}+1)(\eps_{B}+1)}(A \leftrightarrow B)\~,
\label{riccitensorsym}
\eea
when the \mb{\nabla^{(\Gamma)}} connection is torsionfree
\mb{T^{(\Gamma)}\!=\!0} and \mb{\rho}--compatible \e{rhocomp}.

\subsection{The Ricci Two--Form}
\label{secriccitensor}

\noi
The Ricci two--form is defined as
\beq
\cR_{AB}\~:=\~R_{AB}{}^{C}{}_{C}(-1)^{\eps_{C}}\=
-(-1)^{\eps_{A}\eps_{B}}(A \leftrightarrow B)\~.\label{defricci2form}
\eeq
The Ricci two--form vanishes
\beq
\cR_{AB} \= 0~,\label{ricci2formvanish}
\eeq
when the \mb{\nabla^{(\Gamma)}} connection is torsionfree
\mb{T^{(\Gamma)}\!=\!0} and \mb{\rho}--compatible \e{rhocomp}.

\subsection{Covariant Tensors}
\label{seccurvedtensor}

\noi
Let 
\beq
\Omega_{mn}(M)\~:=\~\Gamma\left(\bigwedge{}^{m}(T^{*}M)
\otimes\bigvee{}^{n}(T^{*}M)\right)
\label{curvedanmalgebra}
\eeq
be the vector space of \mb{(0,m\!+\!n)}--tensors
\mb{\eta^{}_{A_{1}\cdots A_{m}B_{1}\cdots B_{n}}(\zz)} 
that are antisymmetric \wrtt first \mb{m} indices \mb{A_{1}\ldots A_{m}}, 
and symmetric \wrtt last \mb{n} indices \mb{B_{1}\ldots B_{n}}.
As usual, it is practical to introduce a coordinate--free notation
\beq
\eta(\zz;\CC;\YY)
\= \frac{1}{m!n!}\CC^{A_{m}}\wedge\cdots\wedge \CC^{A_{1}}\~
\eta^{}_{A_{1}\cdots A_{m}B_{1}\cdots B_{n}}(\zz)\otimes \YY^{B_{n}}
\vee\cdots\vee \YY^{B_{1}}\~.
\label{curvedcoordinatefree}
\eeq
Here the variables \mb{\YY^{A}} are symmetric counterparts to the
one--form basis \mb{\CC^{A}\equiv \dzz^{A}}. 
\beq
\begin{array}{rclrclrcl}
\CC^{A} \wedge \CC^{B}&=& -(-1)^{\eps_{A}\eps_{B}} \CC^{B} \wedge \CC^{A}
\~,\~\~\~\~& \eps(\CC^{A}) &=& \eps_{A}\~,\~\~\~\~& p(\CC^{A}) &=& 1\~, \\ 
\YY^{A} \vee \YY^{B}&=& (-1)^{\eps_{A}\eps_{B}} \YY^{B} \vee \YY^{A}
\~,& \eps(\YY^{A}) &=& \eps_{A}\~,& p(\YY^{A}) &=& 0\~.
\end{array}\label{curvedcom}
\eeq
The covariant derivative can be realized on covariant tensors 
\mb{\eta\in\Omega_{mn}(M)} by a linear differential operator
\beq
\nabla^{(T)}_{A}
\= \papal{\zz^{A}}-\Gamma_{A}{}^{B}{}_{C}\~T^{C}{}_{B}\~,
\label{nablaagammarealiz}
\eeq
where
\beq
T^{A}{}_{B}\~:=\~\CC^{A}\papal{\CC^{B}}+\YY^{A}\papal{\YY^{B}}
\label{tgenerator}
\eeq
are themselves linear differential operators. They are generators of the 
general linear (\mb{=gl}) Lie--algebra,
\beq
[T^{A}{}_{B},T^{C}{}_{D}] \= \delta_{B}^{C}\~ T^{A}{}_{D}
-(-1)^{(\eps_{A}+\eps_{B})(\eps_{C}+\eps_{D})}\delta_{D}^{A}\~ T^{C}{}_{B}\~.
\label{tliealg}
\eeq
It is important for the implementation \e{nablaagammarealiz} to make sense that
\mb{\eta} carries no explicit indices, \ie all indices should be paired as
indicated in \eq{curvedcoordinatefree}. The Lie--algebra \e{tliealg} reflects
infinitesimal coordinate transformation, \ie diffeomorphism invariance.

\subsection{Coordinate Transformations}
\label{seccoordtransf}

\noi
Consider for simplicity a one--form
\mb{\eta=\eta^{}_{A}(\zz)\CC^{A}\in\Omega_{10}(M)}. The covariant derivative 
reads 
\beq
(\nabla^{}_{A}\eta)^{}_{C}
\=(\papal{\zz^{A}}\eta^{}_{C})-\eta^{}_{B}\~\Gamma^{B}{}_{AC}\~.
\label{curvedetacovder}
\eeq
Under a coordinate transformation \mb{\zz^{A}\to\zz^{\prime A}} one has
\bea
\eta^{}_{A}&=&\eta^{\prime}_{B} (\zz^{\prime B}\papar{\zz^{A}})\~, \\
\CC^{\prime A}&=&(\zz^{\prime A}\papar{\zz^{B}})\CC^{B}
\=\CC^{B}(\papal{\zz^{B}}\zz^{\prime A})\~, \\
(-1)^{\eps_{A}\eps_{B}}(\zz^{\prime B}\papar{\zz^{D}})\Gamma^{D}{}_{AC}
&=&(\papal{\zz^{A}}\zz^{\prime B}\papar{\zz^{C}})
+(\papal{\zz^{A}}\zz^{\prime D})\Gamma^{\prime}_{D}{}^{B}{}_{E}^{}
(\zz^{\prime E}\papar{\zz^{C}})\~, \label{gammatrans}
\eea
so that the covariant derivative transforms covariantly,
\beq
(\nabla^{}_{A}\eta)^{}_{D}
\=(\papal{\zz^{A}}\zz^{\prime B})
(\nabla^{}_{\prime B}\eta^{\prime})^{}_{C}(\zz^{\prime C}\papar{\zz^{D}})\~.
\label{covderetacoordtrans}
\eeq

\section{Riemannian Geometry}
\label{secriem}

\subsection{Metric}
\label{secriemmetric}

\noi
Let there be given a (pseudo) Riemannian metric, \ie a covariant symmetric 
\mb{(0,2)} tensor field
\beq
g\=\YY^{A}\~ g^{}_{AB} \vee \YY^{B}\~\in\~\Omega_{02}(M)\~,
\label{defg}
\eeq
of Grassmann--parity
\beq
\eps(g^{}_{AB})\=\eps_{A}+\eps_{B}
\~,\~\~\~\~\~\~\~\~\~\eps(g)\=0\~,\~\~\~\~\~\~\~\~\~p(g^{}_{AB})\=0\~,
\label{epsg}
\eeq
and of symmetry 
\beq
g^{}_{BA}\=-(-1)^{(\eps_{A}+1)(\eps_{B}+1)}g^{}_{AB}\~. \label{symg}
\eeq
We shall not need nor discuss positivity/reality/Hermiticity--conditions 
in this paper (except for the application to a particle in a curved space,
\cf Subsection~\ref{secparticle}). The symmetry \e{symg}
becomes more transparent if one reorders the Riemannian metric as
\beq
g \= \YY^{B} \vee \YY^{A} \tilde{g}^{}_{AB}\~,\label{defgtilde1}
\eeq
where
\beq
\tilde{g}^{}_{AB}\~:=\~g^{}_{AB}(-1)^{\eps_{B}}\~.\label{defgtilde2}
\eeq
Then the symmetry \e{symg} simply reads
\beq
\tilde{g}^{}_{BA} \= (-1)^{\eps_{A}\eps_{B}}\tilde{g}^{}_{AB}\~.
\label{symgtilde}
\eeq
The Riemannian metric \mb{g^{}_{AB}} is assumed to be non--degenerate, \ie
there exists an inverse contravariant symmetric \mb{(2,0)} tensor field 
\mb{g^{AB}} such that 
\beq
g^{}_{AB}\~g_{}^{BC} \= \delta_{A}^{C}\~.\label{ginv}
\eeq
The inverse \mb{g^{AB}} has Grassmann--parity
\beq
\eps(g_{}^{AB}) \= \eps_{A}+\eps_{B}\~, \label{epsginv}
\eeq
and symmetry 
\beq
g_{}^{BA} \= (-1)^{\eps_{A}\eps_{B}}g_{}^{AB}\~. \label{symginv}
\eeq
The canonical density on a Riemannian manifold is
\beq
\rhog\~:=\~\sqrt{g}\~:=\~\sqrt{\sdet(g^{}_{AB})}\~. \label{riemrho}
\eeq
This should be compared with the antisymplectic case, where the density
\mb{\rho} is kept arbitrary, since there is no canonical choice \cite{b97}. To
ease comparison, we shall temporarily allow for arbitrary densities \mb{\rho}
in the Riemannian case as well.

\subsection{Laplacian \mb{\Deltarho}}
\label{seclapl}

\noi
A Laplacian \mb{\Deltarho}, which takes scalar functions to scalar functions, 
can be constructed from the inverse metric \mb{g_{}^{AB}} and a (not
necessarily canonical) density \mb{\rho},
\beq
\Deltarho\~:=\~
\frac{(-1)^{\eps_{A}}}{\rho}\papal{\zz^{A}}\rho g^{AB}\papal{\zz^{B}}
\~,\~\~\~\~\~\~\~\~\~\~\~\~\eps(\Deltarho)\=0
\~,\~\~\~\~\~\~\~\~\~\~\~\~p(\Deltarho)\=0\~.\label{evenlapl}
\eeq 
A metric bracket \mb{(f,g)} of two functions \mb{f\!=\!f(\zz)} and
\mb{g\!=\!g(\zz)} can be defined via a double commutator with the Laplacian,
acting on the constant unit function \mb{1},
\bea
(f,g)&:=&\Hf[[\stackrel{\rightarrow}{\Delta}^{}_{\rho},f],g]1
\~\equiv\~\Hf\Deltarho(fg) - \Hf(\Deltarho f)g
- \Hf f(\Deltarho g) + \Hf fg (\Deltarho 1) \cr
&=&(f\papar{\zz^{A}})g^{AB}(\papal{\zz^{B}}g)
\=(-1)^{\eps_{f}\eps_{g}}(g,f)\~.
\label{metricbracket}
\eea 
There are {\em no} closeness relations (\resp Jacobi identities) associated 
with the Riemannian \mb{g^{}_{AB}} metric \e{defgtilde1} (\resp metric 
\mb{(\cdot,\cdot)} bracket \e{metricbracket}) in contrast to symplectic 
situations. In fact, even if such closeness relations and Jacobi identities
were to be artificially enforced in one coordinate patch, they would not
transform covariantly under general coordinate transformations
\mb{\zz^{A}\to\zz^{\prime B}}. See also Subsection 3.1 in \Ref{kv04}.

\subsection{Two--cocycle \mb{\nu(\rho^{\prime};\rho,g)}}
\label{seceventwocycle}

\noi
It is possible to introduce a Riemannian analogue of the two--cocycle of
Khudaverdian and Voronov \cite{kv02,bbd06,bb07}. The two--cocycle 
\mb{\nu(\rho^{\prime};\rho,g)}  is a function of a measure density
\mb{\rho^{\prime}} \wrt a reference system \mb{(\rho,g)},
\beq
\nu(\rho^{\prime};\rho,g) \~:=\~\sqrt{{\rho \over \rho^{\prime}}}
(\Deltarho\sqrt{{\rho^{\prime} \over \rho}})
\=\nu^{(0)}_{\rho^{\prime}}-\nurhozero\~,
\label{relmass}
\eeq
where 
\beq
\nurhozero\~:=\~
\frac{1}{\sqrt{\rho}}(\Deltaone\sqrt{\rho}) \label{evennurho0}
\= -\sqrt{\rho}(\Deltarho\frac{1}{\sqrt{\rho}})
\= (\Deltaone\ln\sqrt{\rho}) + (\ln\sqrt{\rho},\ln\sqrt{\rho}) \~.
\eeq
Here \mb{\Deltaone} is the Laplacian \e{evenlapl} with \mb{\rho\!=\!1}.
The expression \e{relmass} acts as a scalar under general coordinate
transformations, and satisfies the following two--cocycle condition:
\beq
\nu(\rho_{1};\rho_{2},g)+\nu(\rho_{2};\rho_{3},g)
+\nu(\rho_{3};\rho_{1},g)\=0\~.
\label{cocyclecondition}
\eeq
In fact, it is a two--coboundary, because we shall prove in the next 
Subsection~\ref{secevennu}, that there exists a scalar \mb{\nurho}, such 
that
\beq
\nu(\rho^{\prime};\rho,g)\=\nu^{}_{\rho^{\prime}}-\nurho\~.
\label{absmass}
\eeq

\subsection{Scalar \mb{\nurho}}
\label{secevennu}

\noi
A Grassmann--even function \mb{\nurho} can be constructed from the metric
\mb{g} and a (not necessarily canonical) density \mb{\rho} as
\beq
\nurho\~:=\~\nurhozero+\frac{\nu^{(1)}}{4}
-\frac{\nu^{(2)}}{8}-\frac{\nu^{(3)}}{16}\~,
\label{evennurho}
\eeq
where \mb{\nurhozero} is given by \eq{evennurho0}, and
\bea
\nu^{(1)}&:=&(-1)^{\eps_{A}}(\papal{\zz^{A}}g^{AB}
\papar{\zz^{B}})(-1)^{\eps_{B}}\~,\label{evennu1} \\
\nu^{(2)}&:=&-(-1)^{\eps_{C}}(\zz^{C},(\zz^{B},\zz^{A}))
(\papal{\zz^{A}}g^{}_{BC}) \cr
&=&-(-1)^{(\eps_{A}+1)(\eps_{D}+1)}
(\papal{\zz^{D}}g_{}^{AB})g^{}_{BC}(g_{}^{CD}\papar{\zz^{A}})
\~,\label{evennu2} \\
\nu^{(3)}&:=&(-1)^{\eps_{A}}(g^{}_{AB},g_{}^{BA})
\~.\label{evennu3}
\eea
Here \mb{(\cdot,\cdot)} is the metric bracket \e{metricbracket}.

\begin{lemma}
The even quantity \mb{\nurho} is a scalar, \ie it does not depend on the 
coordinate system.
\label{lemmaeven}
\end{lemma}

\noi
{\sc Proof of Lemma~\ref{lemmaeven}:}\~
Under an arbitrary infinitesimal coordinate transformation
\mb{\delta\zz^{A}=X^{A}}, one calculates (by using methods similar to the 
antisymplectic case \cite{b06})
\bea
\delta\nurhozero
&=&-\Hf \Deltaone{\rm div}^{}_{1}X\~,\label{devennurho0} \\
\delta\nu^{(1)}&=& 2 \Deltaone{\rm div}^{}_{1}X 
+ (-1)^{\epsilon_{C}}(\papal{\zz^{C}}g_{}^{AB})
(\papal{\zz^{B}}\papal{\zz^{A}}X^{C})~, \label{devennu1} \\
\delta\nu^{(2)}&=&2(-1)^{\epsilon_{C}}(\papal{\zz^{C}}g_{}^{AB})
(\papal{\zz^{B}}\papal{\zz^{A}}X^{C})
+2(-1)^{\epsilon_{A}}g^{}_{AB}(g_{}^{BC},\papal{\zz^{C}}X^{A})
\~,\label{devennu2} \\
\delta\nu^{(3)}&=&-4(-1)^{\epsilon_{A}}g^{}_{AB}
(g_{}^{BC},\papal{\zz^{C}}X^{A})\~.\label{devennu3}
\eea
One easily sees that while the four constituents \mb{\nurhozero},
\mb{\nu^{(1)}}, \mb{\nu^{(2)}} and \mb{\nu^{(3)}} separately have non--trivial
transformation properties, the linear combination \mb{\nurho} in \eq{evennurho}
is indeed a scalar.
\proofbox

\noi
Spurred by what happens in the antisymplectic case \cite{bb07}, we would like
to classify which zeroth--order term \mb{\nu} one could add to the Laplacian
\e{evenlapl}. The following Proposition~\ref{propositioneven} is designed to
answer this question.

\begin{proposition}[Classification of \mb{2}--order differential invariants] 
If a function \mb{\nu\!=\!\nu(\zz)} has the following properties:
\begin{enumerate}
\item
The function \mb{\nu} is a scalar.
\item
\mb{\nu(\zz)} is a polynomial of the metric \mb{g^{}_{AB}(\zz)}, the density
\mb{\rho(\zz)}, their inverses, and \mb{\zz}--derivatives thereof in the point
\mb{\zz}. 
\item
\mb{\nu} is invariant under constant rescaling of the density 
\mb{\rho\to\lambda\rho}, where \mb{\lambda} is a
\mb{\zz}--independent parameter.
\item
\mb{\nu} scales as \mb{\nu\to \lambda\nu} under constant Weyl scaling
\mb{g_{}^{AB}\to\lambda g_{}^{AB}}, where \mb{\lambda} is a
\mb{\zz}--independent parameter.
\item
Each term in \mb{\nu} contains precisely two \mb{\zz}--derivatives.
\end{enumerate}
Then \mb{\nu} is of the form
\beq
\nu\=\alpha\~\nurho+\beta\~\nurhog
+\gamma\~(\ln\frac{\rho}{\rhog},\ln\frac{\rho}{\rhog})\~,
\label{threelittlepigs}
\eeq
where \mb{\alpha}, \mb{\beta} and \mb{\gamma} are three arbitrary
\mb{\zz}--independent parameters. 
\label{propositioneven}
\end{proposition}

\noi
{\em Remarks}: Conditions 1--5 are imposed, because the Laplacian \e{evenlapl} 
has these properties. Note that if one collects the \mb{\rho}--dependence into
a function of \mb{\ln\rho} and its \mb{\zz}--derivatives, the conditions 2 and
3 both exclude undifferentiated \mb{\ln\rho}--dependence (because \mb{\ln\rho}
is not a finite polynomial in \mb{\rho} and \mb{\rho^{-1}}, and because
\mb{\ln\rho\!\to\!\ln\rho+\ln\lambda} is not invariant, respectively).
So scalars like \mb{\nurho\!\ln(\rho/\rhog)} are excluded from our
considerations.

\noi
{\sc Sketched proof of Proposition~\ref{propositioneven}:}\~
The first idea of the proof is to replace condition 1 with a weaker condition
\begin{itemize}
\item[$1^{\prime}$.]{\em The function \mb{\nu} is invariant under affine 
coordinate transformations}
\mb{\zz^{A}\to \zz^{\prime B}=\Lambda^{B}{}_{A}\zz^{A}+\lambda^{B}}.
\end{itemize}
Secondly, recall that every polynomial is a finite linear combinations of 
monomials. One can argue that if \mb{\nu(\zz)} is a polynomial that satisfy
condition \mb{1^{\prime}} plus conditions 2--5 of
Proposition~\ref{propositioneven}, then each of its constituent monomials 
(that contributes nontrivially) must by themselves satisfy condition
\mb{1^{\prime}} plus conditions 2--5.
Thus one can limit the search (for a linear basis) to monomials. 
It follows from lengthy but straightforward combinatorial arguments that 
a basis for the polynomials \mb{\nu} that satisfy condition \mb{1^{\prime}}
plus conditions 2--5 is:
\beq
\nurhozero\~,\~\~
\nurhogzero\~,\~\~ \nu^{(1)}\~,\~\~ \nu^{(2)}\~,\~\~ \nu^{(3)}\~,\~\~ 
\nu^{(4)}\~,\~\~ \nu_{\rho}^{(5)}\~,\~\~\nu_{\rhog}^{(5)}\~,
\~\~\nu_{\rho}^{(6)}\~,\~\~ \nu_{\rhog}^{(6)}\~,\~\~ \nu_{\rho}^{(7)}\~,
\label{greateleven}
\eeq  
where \mb{\nurhozero}, \mb{\nu^{(1)}}, \mb{\nu^{(2)}}, \mb{\nu^{(3)}} were 
defined above, and
\bea
\nu^{(4)}&:=&(-1)^{\eps_{A}}
(\papal{\zz^{A}}g_{}^{AB})g^{}_{BC}(g_{}^{CD}\papar{\zz^{D}})(-1)^{\eps_{D}}
\~,\label{evennu4} \\
\nu_{\rho}^{(5)}&:=&(-1)^{\eps_{A}}(\papal{\zz^{A}}g_{}^{AB})
(\papal{\zz^{B}}\ln\rho)\~,\label{evennurho5} \\
\nu_{\rho}^{(6)}&:=&(\ln\rho,\ln\rho)\~,\label{evennurho6} \\
\nu_{\rho}^{(7)}&:=&(\ln\rho,\ln\rhog)\~.\label{evennurho7}
\eea
Thirdly, under an arbitrary infinitesimal coordinate transformation
\mb{\delta\zz^{A}=X^{A}}, one calculates 
\bea
\delta\nu^{(4)}&=& 2(-1)^{\eps_{A}}(\papal{\zz^{A}}g_{}^{AB})
(\papal{\zz^{B}} {\rm div}^{}_{1}X)\cr
&& +2g_{}^{AB}(\papal{\zz^{B}}\papal{\zz^{A}}X^{C})
g^{}_{CD}(g_{}^{DE}\papar{\zz^{E}})(-1)^{\eps_{E}}
~, \label{devennu4} \\
\delta\nu_{\rho}^{(5)}&=& (\ln\rho,{\rm div}^{}_{1}X)
-(-1)^{\eps_{A}}(\papal{\zz^{A}}g_{}^{AB})
(\papal{\zz^{B}} {\rm div}^{}_{1}X) \cr
&& +g_{}^{AB}(\papal{\zz^{B}}\papal{\zz^{A}}X^{C})(\papal{\zz^{C}}\ln\rho)
~, \label{devennurho5} \\
\delta\nu_{\rho}^{(6)}&=& -2(\ln\rho,{\rm div}^{}_{1}X)
~, \label{devennurho6} \\
\delta\nu_{\rho}^{(7)}&=& -(\ln(\rhog\rho),{\rm div}^{}_{1}X)
~. \label{devennurho7} 
\eea
It is easy to check that the only linear combinations of the basis elements
\e{greateleven} that satisfy condition 1, are given by formula 
\e{threelittlepigs}.
\proofbox

\subsection{\mb{\Delta} And \mb{\Deltag}}
\label{secevenkhuda}

\noi
The Riemannian analogue \mb{\Deltag} of Khudaverdian's \mb{\DeltaE} operator
\cite{k99,kv02,k02,k04,bbd06,b06,b07} is defined as
\beq
\Deltag\~:=\~\Deltaone+\frac{\nu^{(1)}}{4}
-\frac{\nu^{(2)}}{8}-\frac{\nu^{(3)}}{16}\~.
\label{deltagdef}
\eeq
We will prove below that the \mb{\Deltag} operator \e{deltagdef} takes
semidensities to semidensities. It is obviously manifestly independent of
\mb{\rho}. Next, we define a Riemannian analogue of the Grassmann--odd
nilpotent \mb{\Delta} operator in antisymplectic geometry \cite{bb07}. The
even \mb{\Delta} operator, which takes scalar functions to scalar functions,
is defined for arbitrary \mb{\rho} as
\beq
\Delta\~:=\~\Deltarho+\nurho\~.\label{evendelta}
\eeq
This \mb{\Delta} operator \e{evendelta} is well--defined, because of 
Lemma~\ref{lemmaeven}. One may prove (by using methods similar to the
antisymplectic case \cite{b06,bb07}), that the two operators \mb{\Delta} and 
\mb{\Deltag} are related via a similarity--transformation with
\mb{\sqrt{\rho}},
\beq
\Deltag\=\sqrt{\rho}\Delta\frac{1}{\sqrt{\rho}}\~.
\label{deltagdelta}
\eeq
{\sc Proof of \eq{deltagdelta}}:~~Let \mb{\sigma} denote an arbitrary argument
for the \mb{\Deltag} operator. (The argument \mb{\sigma} is a semidensity, but
we shall not use this fact.) Then, it follows from the explicit \mb{\nurho}
formula \e{evennurho} that
\bea
(\Deltag\sigma)
&=&(\Deltaone\sigma)+(\frac{\nu^{(1)}}{4}
-\frac{\nu^{(2)}}{8}-\frac{\nu^{(3)}}{16})\sigma
\=(\Deltaone\sigma)-(\Deltaone\sqrt{\rho})\frac{\sigma}{\sqrt{\rho}}
+\nurho\sigma \cr
&=&\sqrt{\rho}(\Deltaone\frac{\sigma}{\sqrt{\rho}}) 
+2(\sqrt{\rho},\frac{\sigma}{\sqrt{\rho}})+\nurho\sigma
\=\sqrt{\rho}(\Deltarho\frac{\sigma}{\sqrt{\rho}})+\nurho\sigma 
\=\sqrt{\rho}(\Delta\frac{\sigma}{\sqrt{\rho}})~.
\label{deltagdeletaproof}
\eea
\proofbox
Eq.\ \e{deltagdelta} shows that the \mb{\Deltag} operator \e{deltagdef} takes
semidensities to semidensities. The \mb{\Delta} operator \e{evendelta} has, in
turn, the remarkable property that the \mb{\sqrt{\rho}}--conjugated
operator \mb{\sqrt{\rho}\Delta\frac{1}{\sqrt{\rho}}} is independent of
\mb{\rho}. This is strikingly similar to what happens in the antisymplectic
case, \cf Subsection~\ref{secoddkhuda}. It is interesting to investigate how
unique this property is? Consider a primed operator
\beq
\Delta^{\prime}\~:=\~\Delta+\nu=\Deltarho+\nurho+\nu\~,\label{evendeltaprime}
\eeq 
where \mb{\nu} is a most general zeroth--order term. (We will in this paper
not consider the possibility of changing second-- and first--order parts of
Laplace operators, \ie we will only consider changes to the zeroth--order term
for simplicity.) It is easy to see from \eqs{deltagdelta}{evendeltaprime} that
the corresponding \mb{\sqrt{\rho}}--conjugated operator
\mb{\sqrt{\rho}\Delta^{\prime}\frac{1}{\sqrt{\rho}}} is independent of
\mb{\rho} if and only if the shift term \mb{\nu} is \mb{\rho}--independent.
On the other hand, by invoking Proposition~\ref{propositioneven}, one sees
that \mb{\nu} is \mb{\rho}--independent if and only if \mb{\nu=\beta\nurhog}
is proportional to \mb{\nurhog}. So an operator of the form
\mb{\Delta^{\prime}=\Delta+\beta\nurhog}, for arbitrary coefficient \mb{\beta},
is the most general operator with this property. This is the minimal answer
one could possibly have hoped for, since a \mb{\rho}--independence argument
will never be able to detect the presence of a \mb{\rho}--independent shift
term like \mb{\beta\nurhog}.

\subsection{Levi--Civita Connection}
\label{seclcconn}

\noi
A connection \mb{\nabla^{(\Gamma)}} is called {\em metric}, if it preserves the
metric \mb{g},
\beq
0 \= (\nabla^{(\Gamma)}_{A}\tilde{g})^{}_{BC}
\= (\papal{\zz^{A}}\tilde{g}^{}_{BC})
-\left((-1)^{\eps_{A}\eps_{B}}\Gamma^{}_{BAC}
+(-1)^{\eps_{B}\eps_{C}}(B \leftrightarrow C)\right)\~. \label{connmetric}
\eeq
Here we have lowered the Christoffel symbol with the metric
\beq
\Gamma_{ABC}^{}\~:=\~g_{AD}\Gamma^{D}{}_{BC}(-1)^{\eps_{C}}\~.
\label{lowerriemconn}
\eeq
The metric condition \e{connmetric} reads in terms of the contravariant 
inverse metric 
\beq
0\=(\nabla^{(\Gamma)}_{A}g)^{BC}
\~\equiv\~(\papal{\zz^{A}}g^{BC})+\left(\Gamma_{A}{}^{B}{}_{D}g^{DC}
+(-1)^{\eps_{B}\eps_{C}}(B\leftrightarrow C)\right)~.
\label{upperriemconn}
\eeq
The Levi--Civita connection is the unique connection \mb{\nabla^{(\Gamma)}} 
that is both torsionfree \mb{T^{(\Gamma)}\!=\!0} and metric \e{connmetric}.
The Levi--Civita formula for the lowered Christoffel symbol in terms of
derivatives of the metric reads
\beq
2 \Gamma^{}_{CAB} \= (-1)^{\eps_{A}\eps_{C}}(\papal{\zz^{A}}\tilde{g}_{CB})
+(-1)^{(\eps_{A}+\eps_{C})\eps_{B}}(\papal{\zz^{B}}\tilde{g}_{CA})
-(\papal{\zz^{C}}\tilde{g}_{AB})\~. \label{lcformula}
\eeq
A density \mb{\rho} is compatible \e{rhocomp} with the Levi--Civita
Christoffel symbol \e{lcformula} if and only if \mb{\rho} is proportional to
the canonical density \e{riemrho}.

\subsection{The Riemann Curvature}
\label{secmetricriemcurv}

\noi
{}For a metric connection \mb{\nabla^{(\Gamma)}}, we prefer to work with a
\mb{(0,4)} Riemann tensor (as opposed to a \mb{(1,3)} tensor) by lowering the
upper index with the metric \e{defg}. In terms of Christoffel symbols it is
easiest to work with expression \e{riemtensor1}:
\bea
R_{D,ABC}&:=&g^{}_{DE}R^{E}{}_{ABC}(-1)^{\eps_{C}}\cr
&=&(-1)^{\eps_{A}\eps_{D}}\left(\papal{\zz^{A}}\Gamma_{DBC}
+(-1)^{\eps_{E}(\eps_{A}+\eps_{D}+1)+\eps_{C}}
\Gamma_{EAD}\Gamma^{E}{}_{BC}\right) \cr
&&-(-1)^{\eps_{A}\eps_{B}}(A\leftrightarrow B)\~.\label{riemtensor1low}
\eea
In the second equality of \eq{riemtensor1low} is used the metric condition
\e{connmetric}. If the metric condition \e{connmetric} is used one more time
on the first term in \eq{riemtensor1low}, one derives the following
skewsymmetry
\beq
R_{D,ABC} \= -(-1)^{(\eps_{A}+\eps_{B})(\eps_{C}+\eps_{D})
+\eps_{C}\eps_{D}}(C\leftrightarrow D)\~.
\label{riemtensor1lowsym}
\eeq
This skewsymmetry becomes clearer if one instead starts from expression 
\e{riemtensor4} and define
\beq
R_{AB,CD}\~:=\~R_{ABC}{}^{E}\tilde{g}_{ED}
\= (-1)^{\eps_{D}(\eps_{A}+\eps_{B}+\eps_{C})}R_{D,ABC}\~.
\label{riemtensor4low}
\eeq
Then the skewsymmetry \e{riemtensor1lowsym} simply translates into a
skewsymmetry between the third and fourth index:
\beq
R_{AB,CD} \= -(-1)^{\eps_{C}\eps_{D}}(C\leftrightarrow D)\~.
\label{riemtensor4lowsym}
\eeq
We note that the torsionfree condition has not been used so far in this 
Section~\ref{secmetricriemcurv}.
The first Bianchi identity \e{1bianchiid} reads (in the torsionfree case):
\beq
0~=~\sum_{{\rm cycl.}\~A,B,C}(-1)^{\eps_{A}\eps_{C}}R_{AB,CD}~.
\label{riem1bianchiid}
\eeq
The \mb{(A \leftrightarrow B)} antisymmetry, the \mb{(C \leftrightarrow D)} 
antisymmetry \e{riemtensor4lowsym} and the first Bianchi identity
\e{riem1bianchiid} imply that Riemann curvature tensor \mb{R_{AB,CD}} is 
symmetric \wrt an \mb{(AB\leftrightarrow CD)} exchange of two pairs of indices:
\beq
R_{AB,CD}\=(-1)^{(\eps_{A}+\eps_{B})(\eps_{C}+\eps_{D})}(AB\leftrightarrow CD)
\~.
\label{riemtwopairs}
\eeq
This, in turn, implies that there is a version of the first Bianchi identity 
\e{riem1bianchiid}, where one sums cyclically over the three last indices:
\beq
0~=~\sum_{{\rm cycl.}\~B,C,D}(-1)^{\eps_{B}\eps_{D}}R_{AB,CD}~.
\label{riem1bianchiidprime}
\eeq
It is interesting to compare Riemann tensors in the Riemannian case with
the antisymplectic case. In both cases, the \mb{(A \leftrightarrow B)}
antisymmetry and the Bianchi identity \e{riem1bianchiid} hold, but the
\mb{(C \leftrightarrow D)} antisymmetry \e{riemtensor4lowsym} turns in the
antisymplectic case into an \mb{(C \leftrightarrow D)} symmetry
\e{antisymptensor4lowsym}, and there is {\em no} antisymplectic analogue of
the \mb{(AB\leftrightarrow CD)} exchange symmetry \e{riemtwopairs}, \cf
Subsection~\ref{secantisympriemcurv}.

\subsection{Scalar Curvature}
\label{secriemscalar}

\noi
The scalar curvature is defined as
\beq
R\~:=\~(-1)^{\eps_{B}}g^{BA}R^{}_{AB}\=(-1)^{\eps_{A}}R^{}_{AB}g^{BA}\~.
\label{evenscalarcurv}
\eeq

\begin{proposition}
The Levi--Civita scalar curvature \mb{R} is proportional to the scalar
\mb{\nurhog},
\beq
R\= -4\nurhog\~.\label{evenrnurho}
\eeq
\label{propositionevenrnu}
\end{proposition}

\noi
{\sc Sketched proof of Proposition~\ref{propositionevenrnu}:}\~Straightforward
calculations shows that
\beq
R\=-4\nurhogzero-\nu^{(1)}
+(-1)^{\eps_{A}}g^{AB}\~\Gamma_{B}{}^{D}{}_{C}\~\Gamma^{C}{}_{DA}\~,
\label{earlyr}
\eeq
where
\beq
2(-1)^{\eps_{A}}g^{AB}\~\Gamma_{B}{}^{D}{}_{C}\~\Gamma^{C}{}_{DA}
\=-(-1)^{\eps_{A}+\eps_{B}}\Gamma^{A}{}_{BC}(g^{CB}\papar{\zz^{A}})
\=\nu^{(2)}+\frac{\nu^{(3)}}{2}\~.\label{gammagamma}
\eeq
\proofbox

\noi
As a corollary of Proposition~\ref{propositionevenrnu} one gets that the
\mb{\nurho} scalar \e{evennurho} for arbitrary \mb{\rho} is given by the
formula
\beq
\nurho\=\nu(\rho;\rhog,g)+\nu^{}_{\rhog}\=\sqrt{{\rhog \over \rho}}
(\Delta^{}_{\rhog}\sqrt{{\rho \over \rhog}})-\frac{R}{4}\~.
\eeq

\subsection{The \mb{\Delta} Operator At \mb{\rho=\rhog}}
\label{secriemonshell}

\noi
When one restricts to \mb{\rho=\rhog}, the \mb{\Delta} operator \e{evendelta}
reduces to the Laplace--Beltrami operator minus a quarter of the Levi--Civita
scalar curvature:
\beq
\Delta|^{}_{\rho=\rhog}\=\Deltarhog+\nurhog
\=\Deltarhog-\frac{R}{4}\~. \label{evenanaloguedelta}
\eeq
This is the even operator \e{evenonshelldeltaop} already mentioned in the
Introduction. But the important question is: Does the zeroth--order term
\mb{\nurhog\!=\!-R/4} in the operator \e{evenanaloguedelta} have a property
that distinguish it from all the other zeroth--order terms? Yes, in the
following sense:
\begin{enumerate}
\item
{}Firstly, consider the most general \mb{\rho}--independent operator of the
form 
\beq 
\Deltarhog+\nu\~, \label{beltramiansatz1}
\eeq
where \mb{\Deltarhog} is the Laplace--Beltrami operator and \mb{\nu} is a 
general zeroth--order term. (Here it is important that we only allow
\mb{\rho}--independent \mb{\nu}'s from the very beginning.)
\item
Secondly, apply Proposition~\ref{propositioneven} to classify the possible
zeroth--order terms \mb{\nu}. In detail, one sees that \mb{\nu=\beta\nurhog}
is proportional to \mb{\nurhog} for some proportionality factor \mb{\beta}.
Hence the operator \e{beltramiansatz1} is actually
\beq 
\Deltarhog+\beta\nurhog\~. \label{beltramiansatz2}
\eeq
\item
Thirdly, replace the canonical density \mb{\rhog\to\rho} by an arbitrary
density \mb{\rho}. In other words, replace the \mb{\rho}--independent operator
\e{beltramiansatz2} with the corresponding \mb{\rho}--dependent operator
\beq
\Delta^{\prime}:=\Deltarho+\beta\nurho~. \label{liftedbeltramiansatz}
\eeq
More rigorously, one should consider an algebra homomorphism 
\mb{s:\cA^{}_{g}\to\cA^{}_{\rho,g}} from the algebra \mb{\cA^{}_{g}}
of differential operators, that only depend on the metric \mb{g}, to the 
algebra \mb{\cA^{}_{\rho,g}} of differential operators, that depend on both
the density \mb{\rho} and the metric \mb{g}. The \mb{s} homomorphism should
satisfy \mb{\pi\circ s={\rm Id}_{\cA^{}_{g}}}, where
\mb{\pi:\cA^{}_{\rho,g}\to\cA^{}_{g}} denotes the restriction map
\mb{|^{}_{\rho=\rhog}} and ``\mb{\circ}'' denotes composition. Clearly such a 
procedure is in general highly ambiguous, but in the present situation, where 
we are only interested in the \mb{\rho}--extension of just two operators,
namely the second--order operator \mb{\Deltarhog} and the zeroth--order
operator \mb{\nurhog}, there is a preferred candidate for the \mb{s}
homomorphism in this sector, \ie \mb{\Deltarhog\stackrel{s}{\mapsto}\Deltarho}
and \mb{\nurhog\stackrel{s}{\mapsto}\nurho}, respectively.
\item
{}Fourthly, apply the \mb{\sqrt{\rho}}--independence argument of
Subsection~\ref{secevenkhuda}. It follows 
that the \mb{\sqrt{\rho}}--conjugated \mb{\Delta^{\prime}} 
operator \mb{\sqrt{\rho}\Delta^{\prime}\frac{1}{\sqrt{\rho}}} becomes 
independent of \mb{\rho} if and only if \mb{\beta\!=\!1}. (In the 
antisymplectic case \mb{\Delta^{\prime}} is also nilpotent if and only if
\mb{\beta\!=\!1}.) Thus we conclude that the coefficient \mb{\beta\!=\!1}, 
and hence the even \mb{\Delta} operator \e{evendelta} is singled out.
\item
{}Fifthly, restrict to \mb{\rho=\rhog}. Hence one arrives at the preferred
operator \e{evenanaloguedelta}. 
\end{enumerate}

\noi
Needless to say, that the above argument depends crucially on the order of the
above five steps. In particular, if step 3 is performed before step 1 and 2,
\ie if one considers the most general \mb{\rho}--dependent zeroth--order term 
\mb{\nu} from the very beginning, the \mb{\beta} coefficient in front of the 
zeroth--order term \mb{\nurhog} would remain arbitrary.

\subsection{Particle In Curved Space}
\label{secparticle}

\noi
In this Subsection~\ref{secparticle} we indicate how the \mb{\Delta} operator
\e{evendelta} is related to quantization of a particle in a curved Riemannian
target space \cite{dewitt57,dewitt64,berezin71,mizrahi75,gj76,dewitt92,pvn93,
dbpsvn96,bsvn98,bastian05} with a measure density \mb{\rho} not necessarily
equal to the canonical density \e{riemrho}. The classical Hamiltonian action
\mb{S^{}_{\cl}} is
\beq
S^{}_{\cl}\=\int\!dt\left(\pp^{}_{A}\dot{\zz}^{A}-H^{}_{\cl}\right)
\~,\~\~\~\~\~\~\~\~
H^{}_{\cl}\=\Hf\pp^{}_{A}\pp^{}_{B}g^{BA}+V\~,\~\~\~\~\~\~\~\~
\mb{\{\zz^{A},\pp^{}_{B}\}^{}_{PB}\=\delta^{A}_{B}}\~,\label{clasaction}
\eeq
where \mb{\pp^{}_{A}} denote the momenta for the \mb{\zz^{A}} variables. We
shall for simplicity not consider reparametrizations of the time variable
\mb{t}. Moreover, we assume that the Riemannian metric
\mb{g^{}_{AB}\!=\!g^{}_{AB}(\zz)}, the density \mb{\rho\!=\!\rho(\zz)}, the
potential \mb{V\!=\!V(\zz)}  and general coordinate transformations
\mb{\zz^{A}\to\zz^{\prime B}\!=\!f^{B}(\zz)} do not depend explicitly on time
\mb{t}. The naive Hamiltonian operator \mb{\hat{H}^{}_{\rho}} is
\cite{dewitt57,berezin71,mizrahi75,gj76}
\bea
\hat{H}^{}_{\rho}-V(\hat{\zz})
&=&\Hf\hat{\pp}^{r}_{A}\~g^{AB}(\hat{\zz})\~\hat{\pp}^{\ell}_{B}
\=\frac{1}{2\sqrt{\rho(\hat{\zz})}}\~\hat{\pp}^{}_{A}\~\rho(\hat{\zz})\~
g^{AB}(\hat{\zz})\~\hat{\pp}^{}_{B}\~
\frac{(-1)^{\eps_{B}}}{\sqrt{\rho(\hat{\zz})}}\label{bareham1} \\
&=&\Hf[\hat{\pp}^{}_{A}+\Hi\ln\sqrt{\rho(\hat{\zz})} \papar{\hat{\zz}^{A}}]\~ 
g^{AB}(\hat{\zz})\~[\hat{\pp}^{}_{B}(-1)^{\eps_{B}}
-\Hi\papal{\hat{\zz}^{B}}\ln\sqrt{\rho(\hat{\zz})}]\label{bareham2}  \\
&=&\Hf\hat{\pp}_{A}\~g^{AB}(\hat{\zz})\~\hat{\pp}_{B}(-1)^{\eps_{B}}
+\frac{\hbar^{2}}{2}\nurhozero(\hat{\zz}) \label{bareham3}  \\
&=&\Hf \left(\pp^{}_{A}\pp^{}_{B}g^{BA}(\zz)\right)^{\wedge}
+\frac{\hbar^{2}}{2}\left(\nurhozero(\hat{\zz})
+\frac{\nu^{(1)}(\hat{\zz})}{4}\right)\~.
\label{bareham4} 
\eea
The left, middle, and right momentum operators, denoted by
\mb{\hat{\pp}^{\ell}_{A}}, \mb{\hat{\pp}^{}_{A}}, and \mb{\hat{\pp}^{r}_{A}},
respectively, are related as
\beq
\frac{(-1)^{\eps_{A}}}{\sqrt{\rho(\hat{\zz})}}\~
\hat{\pp}^{\ell}_{A}\~\sqrt{\rho(\hat{\zz})}
\=\mb{\hat{\pp}^{}_{A}}
\=\sqrt{\rho(\hat{\zz})}\~\hat{\pp}^{r}_{A}\~
\frac{1}{\sqrt{\rho(\hat{\zz})}}\~.
\eeq
The non--zero canonical equal--time commutator relations read
\beq
-[\hat{\pp}^{\ell}_{B},\hat{\zz}^{A}]\=[\hat{\zz}^{A},\hat{\pp}^{}_{B}]
\=[\hat{\zz}^{A},\hat{\pp}^{r}_{B}]\=i\hbar\delta^{A}_{B}{\bf 1}\~.
\eeq
The hat ``\mb{\wedge}'' in \eq{bareham4} denotes the corresponding
Weyl--ordered operator. We mention for completeness a temporal point--splitting
operation ``\mb{T}'' defined as \cite{dewitt92}
\beq
T\left(\hat{F}_{1}(t)\cdots\hat{F}_{n}(t)\right)
\= \frac{1}{n!}\sum_{\pi\in{\cal S}_{n}}(-1)^{\eps_{F,\pi}}\lim_{
\begin{array}{c}t_{1}, \ldots, t_{n} \to t \cr 
t_{\pi(1)} > \ldots > t_{\pi(n)} \end{array}} 
\hat{F}_{\pi(1)}(t_{\pi(1)})\cdots\hat{F}_{\pi(n)}(t_{\pi(n)})\~,
\label{naivetimesplitting}
\eeq
where \mb{\eps_{F,\pi}} denotes the Grassmann sign factor arising from 
permuting
\beq
\hat{F}_{1}(t_{1})\cdots\hat{F}_{n}(t_{n})\~\~\longrightarrow\~\~
\hat{F}_{\pi(1)}(t_{\pi(1)})\cdots\hat{F}_{\pi(n)}(t_{\pi(n)})\~.
\eeq
{}For most practical purposes, the temporal point--splitting ``\mb{T}'' is the
same as Weyl ordering ``\mb{\wedge}''. In particular,  Weyl--ordering
``\mb{\wedge}'' and temporal point--splitting ``\mb{T}'' yield the same
two--loop quantum correction:
\beq
\left. \begin{array}{c}
\left(\pp^{}_{A}\pp^{}_{B}g^{BA}(\zz)\right)^{\wedge} \cr\cr
T\left(\hat{\pp}^{}_{A}\hat{\pp}^{}_{B}g^{BA}(\hat{\zz})\right)
\end{array} \right\}
-\hat{\pp}^{}_{A}\~g^{AB}(\hat{\zz})\~\hat{\pp}^{}_{B}(-1)^{\eps_{B}}
\=\frac{1}{4}[\hat{\pp}^{}_{A},[\hat{\pp}^{}_{B},g^{BA}(\hat{\zz})]]
\=-\frac{\hbar^{2}}{4}\nu^{(1)}(\hat{\zz})\~. \label{nuonequantumcorr}
\eeq 
Note that Weyl--ordering  ``\mb{\wedge}'' and temporal point--splitting
``\mb{T}'' are {\em not} covariant operations. 

\noi
Now what should be the quantum Hamiltonian \mb{\hat{H}} for the operator 
formalism? Obviously, one must (among other things) demand that
\begin{enumerate}{\em
\item
\mb{\hat{H}} is a scalar invariant.
\item
\mb{\hat{H}} is Hermitean.
\item
\mb{\hat{H}} has dimension of energy.
\item
\mb{\hat{H}} reduces to the classical Hamiltonian \mb{H^{}_{\cl}} in the
classical limit \mb{\hbar\to 0}.
}\end{enumerate}

\noi
The naive Hamiltonian operator \e{bareham1} satisfies all these conditions
1--4. It is a scalar invariant, since the momentum operators transform by
definition under coordinate transformations
\mb{\zz^{A}\to \zz^{\prime B}=f^{B}(\zz)} as 
\bea
\hat{\pp}^{\prime \ell}_{B}
&=&(\papal{f^{B}(\hat{\zz})}\hat{\zz}^{A})\~\hat{\pp}^{\ell}_{A} \~, \\
\hat{\pp}^{\prime r}_{B}
&=&\hat{\pp}^{r}_{A}\~(\hat{\zz}^{A}\papar{f^{B}(\hat{\zz})})\~, \\
\hat{\pp}^{\prime }_{B}
&=&(\pp_{A}\~(\zz^{A}\papar{f^{B}(\zz)}))^{\wedge}
\=\Hf\{\hat{\pp}^{}_{A},\hat{\zz}^{A}\papar{f^{B}(\hat{\zz})}\}^{}_{+}\~.
\eea
Note however that conditions 1--4 do not specify the quantum corrections to the
quantum Hamiltonian \mb{\hat{H}}. {}For instance, one could add any multiple
of \mb{\hbar^{2}\nurho(\hat{\zz})} to \mb{\mb{\hat{H}}} without affecting
conditions 1--4. Now recall that every choice of \mb{\mb{\hat{H}}} in the
operator formalism corresponds to a choice of action functional in the path
integral. One may fix the ambiguity by additionally demanding the following:
\begin{itemize}
\item[5.]{\em
The operator formalism with the Hamiltonian operator \mb{\mb{\hat{H}}} should
correspond to a Hamiltonian path integral formulation where the path integral
action is the pure classical action \mb{S^{}_{\cl}} with no quantum
corrections, \ie
\beq
\langle\zz_{f}|\exp\left[-\Ih\hat{H}\Delta t\right]|\zz_{i}\rangle
\=\langle\zz_{f},t_{f}|\zz_{i},t_{i}\rangle
\~\sim\~ \int\limits_{\zz(t_{i})=\zz_{i}}^{\zz(t_{f})=\zz_{f}} \! [d\zz][d\pp]
\~\exp\left[\Ih S^{}_{\cl}[\zz,\pp]\right]\~. \label{pathintegral}
\eeq
}\end{itemize}

\noi
The reader may wonder why we invoke the path integral formulation. The point is
that, on one hand, there is {\em no} unique way of telling what part of an
operator should be considered as quantum corrections, while on the other hand,
there is a well--defined quantum part of an action functional, namely all the
terms of order \mb{{\cal O}(\hbar)}. The phase space path integral in quantum
mechanics does not need to be renormalized (unlike configuration space path
integrals or quantum field theories), so it is consistent to demand that the
bare quantum corrections of the action functional vanish. Condition 5
determines in principle the Hamiltonian operator \mb{\hat{H}} to all orders in
\mb{\hbar}, but we shall in this paper truncate \mb{\hat{H}} at two--loop
order, \ie ignore possible higher--order quantum corrections of order
\mb{{\cal O}(\hbar^{3})} for simplicity. According to standard heuristic
arguments, it follows from condition 5 that the quantum Hamiltonian
\beq
\hat{H}\~\sim\~ T(H^{}_{\cl}) \label{qhamisthcl}
\eeq 
is equal to the time--ordered classical Hamiltonian \mb{T(H^{}_{\cl})}.
However, time--ordering ``\mb{T}'' is not a geometrically well--defined
operation, at least not if one uses the temporal point--splitting
\e{naivetimesplitting}. It should only be trusted modulo terms that contains
single--derivatives of the metric \cite{dewitt92}. In detail, the time--ordered
classical Hamiltonian \mb{T(H^{}_{\cl})} is given by the following
non--covariant expression
\beq
T(H^{}_{\cl})\=\hat{H}^{}_{\rho}-\frac{\hbar^{2}}{2}\left(
\nurhozero(\hat{\zz})+\frac{\nu^{(1)}(\hat{\zz})}{4}\right)\~,
\label{naivetimeorderedclham}
\eeq
\cf \eqs{bareham4}{nuonequantumcorr}. The combination
\beq
\nurhozero+\frac{\nu^{(1)}}{4}
\= \nurho+\frac{\nu^{(2)}}{8}+\frac{\nu^{(3)}}{16}\=\nurho
+\frac{(-1)^{\eps_{A}}}{4}g^{AB}\~\Gamma_{B}{}^{D}{}_{C}\~\Gamma^{C}{}_{DA}
\label{earlyr2}
\eeq
is the \mb{\nurho} scalar \e{evennurho} plus non--covariant terms that contain
single--derivatives of the metric, \cf \eq{earlyr}. Equations \e{qhamisthcl},
\es{naivetimeorderedclham}{earlyr2} therefore strongly suggest that the full
quantum Hamiltonian \mb{\hat{H}} is 
\beq
\hat{H}\=\hat{H}^{}_{\rho}-\frac{\hbar^{2}}{2}\nurho(\hat{\zz})\label{hop}
\=T(H^{}_{\cl})-\frac{\hbar^{2}}{16}
\left(\nu^{(2)}(\hat{\zz})+\frac{\nu^{(3)}(\hat{\zz})}{2}\right)\~,
\eeq
where we are neglecting possible quantum corrections of order
\mb{{\cal O}(\hbar^{3})}. The operator \e{hop} satisfies condition 1--5. {}For
instance, it is a scalar invariant because of Lemma~\ref{lemmaeven}. We shall
provide further details concerning condition 5 in \eq{transopform} below. The
preferred operator \e{hop} also has an extra feature:
\begin{itemize}{\em
\item[6.]
The three operators
\beq
\hat{H}^{}_{g}\~=\~
\sqrt{\rho(\hat{\zz})}\~\hat{H}\~\frac{1}{\sqrt{\rho(\hat{\zz})}}
\~,\~\~\~\~\~\~\~\~\hat{H}\~,\~\~\~\~\~\~\~\~{\it or}\~\~\~\~\~\~\~\~
\frac{1}{\sqrt{\rho(\hat{\zz})}}\~\hat{H}\~\sqrt{\rho(\hat{\zz})}
\eeq
are independent of \mb{\rho}, if one declares that the left, middle, or right
momentum operators \mb{\hat{\pp}^{\ell}_{A}}, \mb{\hat{\pp}^{}_{A}}, or
\mb{\hat{\pp}^{r}_{A}} are independent of \mb{\rho}, respectively. 
}\end{itemize}

\noi
We are now ready to relate the \mb{\Delta} operator \e{evendelta} to a particle
in a curved space. The main point is that the Hamiltonian \e{hop} becomes
\mb{\Delta\!=\!\Deltarho\!+\!\nurho} from \eq{evendelta} if we identify 
\beq
\hat{\zz}^{A}\~\leftrightarrow\~\zz^{A}\~,\~\~\~\~\~\~\~\~
\hat{\pp}^{\ell}_{A}\~\leftrightarrow\~ \Hi\papal{\zz^{A}}\~,
\eeq
\beq
\hat{H}^{}_{\rho}-\hat{V}\~
\leftrightarrow\~-\frac{\hbar^{2}}{2}\Deltarho\~,\~\~\~\~\~\~\~\~
\hat{H}-\hat{V}\~\leftrightarrow\~-\frac{\hbar^{2}}{2}\Delta\~,\~\~\~\~\~\~\~\~
\hat{H}^{}_{g}-\hat{V}\~\leftrightarrow\~-\frac{\hbar^{2}}{2}\Deltag\~.
\eeq
In detail, let \mb{\ketrho:=|\zz,t\rangle /\sqrt{\rho(\zz)}} denote the
instantaneous eigenstate \mb{\hat{\zz}^{A}(t)\ketrho=\zz^{A}\ketrho}, and let 
the eigenstate \mb{|\zz,t\rangle} be the corresponding semidensity state with
normalization \mb{\int\!d^{N}\zz\~|\zz,t\rangle\langle \zz,t|={\bf 1}} and
Grassmann--parity \mb{\eps\left(|\zz,t\rangle\right)\!=\!0}. As a check, note
that the formula \e{pathintegral} is covariant since it is implicitly
understood that the path integral contains one more momentum integration
\mb{\prod_{i=1}^{n}\int\!d\pp(t_{2i-1})} than coordinate integration 
\mb{\prod_{i=1}^{n-1}\int\!d\zz(t_{2i})} for any temporal discretization
\beq
t_{i}\equiv t_{0}<t_{1}<\ldots<t_{2n-1}< t_{2n}\equiv t_{f}\~.
\eeq 
The momentum operators \mb{\hat{\pp}^{\ell}_{A}}, \mb{\hat{\pp}^{}_{A}}, or
\mb{\hat{\pp}^{r}_{A}} act on the eigenstates as
follows:
\bea
\brarho\hat{\pp}^{\ell}_{A}(t)
&=&\Hi\papal{\zz^{A}}\~\brarho
\~,\~\~\~\~\~\~\~\~\~\~\~\~\~\~
\langle \zz,t| \hat{\pp}^{}_{A}(t)
\=\Hi\langle \zz,t| \papar{\zz^{A}}\~, \\
\hat{\pp}^{r}_{A}(t)\ketrho
&=& i\hbar\ketrho \papar{\zz^{A}}
\~,\~\~\~\~\~\~\~\~\~\~\~\~\~\~\~
\hat{\pp}^{}_{A}(t)|\zz,t\rangle
\= i\hbar|\zz,t\rangle \papar{\zz^{A}}\~.
\eea
Therefore, the Hamiltonians \mb{\hat{H}^{}_{\rho}}, \mb{\hat{H}}, and
\mb{\hat{H}^{}_{g}}  translate into the Laplace operators \mb{\Deltarho},
\mb{\Delta}, and \mb{\Deltag}:
\bea
\brarho \left( \hat{H}^{}_{\rho}(t)-\hat{V}(t)\right) 
&=& -\frac{\hbar^{2}}{2}\Deltarho\~\brarho\~, \\
\brarho \left(\hat{H}(t)-\hat{V}(t)\right) 
&=& -\frac{\hbar^{2}}{2}\Delta\~\brarho\~, \\
\langle \zz,t|\left(\hat{H}^{}_{g}(t)-\hat{V}(t)\right)
&=& -\frac{\hbar^{2}}{2}\Deltag \~\langle\zz,t|\~,
\eea
\cf eqs.\ \e{evenlapl}, \es{evendelta}{deltagdelta}, respectively. The 
time--evolution of states and operators are governed by
\bea
\brarho\hat{H}(t)&=& i\hbar \frac{d}{dt}\~\brarho
\~,\~\~\~\~\~\~\~\~\~\~\~\~\~\~
\langle \zz,t|\hat{H}(t)\= i\hbar\frac{d}{dt}\langle \zz,t| \~,  \\
\hat{H}(t)\ketrho&=& \Hi\frac{d}{dt}\ketrho 
\~,\~\~\~\~\~\~\~\~\~\~\~\~\~\~\~
\hat{H}(t)|\zz,t\rangle \= \Hi\frac{d}{dt}|\zz,t\rangle\~,
\eea
\beq
i\hbar\left(\frac{d}{dt}-\frac{\partial}{\partial t}\right)\hat{F}(t)
\=[\hat{F}(t),\hat{H}(t)]\~,
\eeq
where \mb{\hat{F}(t)\!=\!F(\hat{\zz}(t),\hat{\pp}(t),t)} is an arbitrary
operator that may depend explicitly on time \mb{t}. We should mention that
semidensity states appear in geometric quantization \cite{woodhouse92}. 

\noi
Let us now calculate the \lhs of \eq{pathintegral}, \ie the transition element
\mb{\langle\zz_{f}|e^{-\beta\hat{H}}|\zz_{i}\rangle} in the operator formalism
using expression \e{hop} as the Hamiltonian \cite{pvn93}. Here we define 
\beq
\Delta t\~:=\~t_{f}-t_{i}\~,\~\~\~\~\~\~\~\~
\beta\~:=\~\Ih\Delta t\~,\~\~\~\~\~\~\~\~
\gamma\~:=\~\frac{i}{\hbar\Delta t}\~.
\label{betagamma}
\eeq
It is better to change coordinates 
\mb{(\zz^{A}_{i};\zz^{A}_{f}) \to (\zz^{A}_{m};\Delta \zz^{A})} from the 
start--point \mb{\zz^{A}_{i}} and the endpoint \mb{\zz^{A}_{f}} to the midpoint
\mb{\zz^{A}_{m}} and the displacement \mb{\Delta \zz^{A}}, where
\beq
\zz^{A}_{m}\~:=\~\frac{\zz^{A}_{f}+\zz^{A}_{i}}{2}\~,\~\~\~\~\~\~\~\~
\Delta\zz^{A}\~:=\~\zz^{A}_{f}-\zz^{A}_{i}\~.
\label{zmdeltaz}
\eeq
In fact, we will suppress the subscript ``\mb{m}'' since it is implicitly
understood from now on that all quantities are to be evaluated at the midpoint.
We are going to rewrite all operators in terms of symbols \cite{bf88}. The Weyl
symbol \mb{H^{}_{W}} for the quantum Hamiltonian \e{hop} reads
\beq
\~H^{}_{W}\~:=\~(\hat{H})^{}_{W} \= H^{}_{\cl} + \frac{\hbar^{2}}{16}
\left(\nu^{(2)}+\frac{\nu^{(3)}}{2}\right)\~, \label{hamweyl}
\eeq
\cf \eq{hop}. Two Weyl symbols \mb{F} and \mb{G} are multiplied together via
the Groenewold/Moyal \mb{*} product. It can be graphically represented as:  
\bea
{}F*G&=& F\exp[\mapsto]G \= FG +(F\mapsto G) 
+\Hf (F\twostack{\mapsto}{\mapsto} G)
+{\cal O}(\mapsto^{3})\~, \label{moyal} \\
\frac{2}{i\hbar}(F\mapsto G)&=&(F\leftarrow G)-(F\rightarrow G) 
\= \{F,G\}^{}_{PB}\~,\\
\leftarrow&:=&\papar{\zz^{A}}\papal{\pp^{}_{A}}\~,\~\~\~\~\~\~\~\~
\rightarrow\~\~:=\~\~\papar{\pp^{}_{A}}(-1)^{\eps_{A}}\papal{\zz^{A}}\~.
\eea
We will also need the \mb{\zz\pp}--ordered and the \mb{\pp\zz}--ordered
symbols. They can be expressed in terms of the Weyl symbol \mb{(\cdot)^{}_{W}}
as 
\bea
{}F^{}_{\zz^{}_{f}\pp}
&=&\frac{\langle\zz_{f}|\hat{F}|\pp\rangle}{\langle\zz_{f}|\pp\rangle}
\=\exp\left[-\frac{i\hbar}{2}\papal{\pp^{}_{A}}\papal{\zz_{f}^{A}}\right]
{}F^{}_{W_{f}} \cr
&=&\exp\left[(\frac{\Delta z^{A}}{2}-\frac{i\hbar}{2}\papal{\pp^{}_{A}})
\papal{\zz^{A}}\right]F^{}_{W}\~,\label{zfporder}  \\
{}F^{}_{\pp\zz^{}_{i}}
&=&\frac{\langle\pp|\hat{F}|\zz_{i}\rangle}{\langle\pp|\zz_{i}\rangle}
\=\exp\left[\frac{i\hbar}{2}\papal{\pp^{}_{A}}\papal{\zz_{i}^{A}}\right]
{}F^{}_{W^{}_{i}} \cr
&=&\exp\left[(\frac{i\hbar}{2}\papal{\pp^{}_{A}}-\frac{\Delta z^{A}}{2})
\papal{\zz^{A}}\right]F^{}_{W}\label{pziorder}\~.
\eea
The transition element (or propagator) in the operator formalism now becomes
\bea
\langle\zz_{f}|e^{-\beta\hat{H}}|\zz_{i}\rangle
&=&\int \!d^{N}\pp\~\langle\zz_{f}|e^{-\Hf\beta\hat{H}}|\pp\rangle\~
\langle\pp|e^{-\Hf\beta\hat{H}}|\zz_{i}\rangle 
\=\int \!d^{N}\pp\~\langle \zz_{f}|\pp\rangle \~\langle\pp|\zz_{i}\rangle\~
(e^{-\Hf\beta\hat{H}})^{}_{\zz^{}_{f}\pp}\~
(e^{-\Hf\beta\hat{H}})^{}_{\pp\zz^{}_{i}}\cr
&=&\int \!\frac{d^{N}\pp}{(2\pi\hbar)^{N}}\~e^{\Ih\pp^{}_{A}\Delta\zz^{A}}
(e^{-\Hf\beta\hat{H}})^{}_{W}*(e^{-\Hf\beta\hat{H}})^{}_{W}
\=\int \!\frac{d^{N}\pp}{(2\pi\hbar)^{N}}\~
e^{\Ih\pp^{}_{A}\Delta\zz^{A}}(e^{-\beta\hat{H}})^{}_{W}\cr
&=&\int \!\frac{d^{N}\pp}{(2\pi\hbar)^{N}}\~
e^{\Ih\pp^{}_{A}\Delta\zz^{A}}e^{-\beta H^{}_{W}} \cr
&&\times
\left(1+\frac{\beta^{2}}{4}(H^{}_{W}\twostack{\mapsto}{\mapsto} H^{}_{W})
+\frac{\beta^{3}}{6}(H^{}_{W}\mapsto H^{}_{W}\mapsto H^{}_{W}) 
+ {\cal O}(\mapsto^{4})\right)\cr
&=&(2\pi i\hbar\Delta t)^{-\frac{N}{2}}\rhog\~
e^{\Hf\gamma\Delta\zz^{A}g^{}_{AB}\Delta\zz^{B}}e^{-\beta V}
\left(1-\frac{\hbar^{2}\beta}{16}
\left(\nu^{(2)}+\frac{\nu^{(3)}}{2}\right)\right.\cr
&&+\frac{\hbar^{2}\beta^{2}}{8}\left(
(H^{}_{\cl}\twostack{\rightarrow}{\leftarrow}H^{}_{\cl})
- (H^{}_{\cl}\twostack{\rightarrow}{\rightarrow}H^{}_{\cl})\right)  \cr
&&+\frac{\hbar^{2}\beta^{3}}{24}\left(\rule[-2.5ex]{0ex}{2.5ex}
(H^{}_{\cl}\rightarrow H^{}_{\cl}\leftarrow H^{}_{\cl})
+(H^{}_{\cl}\leftarrow H^{}_{\cl}\rightarrow H^{}_{\cl})
-2(H^{}_{\cl}\rightarrow H^{}_{\cl}\rightarrow H^{}_{\cl})\right)\cr
&&+\left.\left.\rule[-2.5ex]{0ex}{2.5ex}
{\cal O}(\rightarrow^{4},\hbar^{3})\right)\right|_{
\pp=\Hi\papar{\Delta\zz}} \cr 
&=&(2\pi i\hbar\Delta t)^{-\frac{N}{2}}\rhog
\left(e^{-\beta V}\left(1- \frac{\hbar^{2}\beta}{24}R\right)
+\frac{\hbar^{2}\beta}{6}e^{-\Hf\beta V}(\Deltarhog e^{-\Hf\beta V})
+{\cal O}((\Delta\zz)^2,\hbar^{3}) \right)\cr 
&=&\left.(2\pi i\hbar\Delta t)^{-\frac{N}{2}}\rho\~ e^{-\Hf\beta V}
\left(1+\frac{\hbar^{2}\beta}{6}\stackrel{\rightarrow}{\Delta}
+{\cal O}((\Delta\zz)^2,\hbar^{3})\right)
e^{-\Hf\beta V}\right|_{\rho=\rhog}\~.\label{transopform}
\eea
In the first equality of \eq{transopform} we summed over a complete set of 
momentum states \mb{|\pp\rangle}, so that it becomes possible to replace
operators by symbols. The \mb{\zz\pp}--ordered and the \mb{\pp\zz}--ordered
symbols \es{zfporder}{pziorder} were used in the second and third equality. 
We performed integration by part of \mb{\pp^{}_{A}} in the third equality. 
In the sixth equality, we replaced all non--Gaussian appearances of the momenta
\mb{\pp^{}_{A}} by derivatives \wrt the displacement \mb{\Delta\zz^{A}}, and
performed the \mb{\pp^{}_{A}} integration. After the integration over
\mb{\pp^{}_{A}}, the terms downstairs in the seventh expression (that are
either quadratic or cubic in \mb{H^{}_{\cl}}) read
\bea
(H^{}_{\cl}\twostack{\rightarrow}{\leftarrow}H^{}_{\cl})
&\sim&\frac{1}{\beta}\nu^{(2)}+{\cal O}((\Delta\zz)^2) \~, \label{feynman1}\\
(H^{}_{\cl}\twostack{\rightarrow}{\rightarrow}H^{}_{\cl})
&\sim&g^{AB}\papal{\zz^{B}}\papal{\zz^{A}}\left(V-\frac{1}{2\beta}\ln g\right)
-\frac{1}{2\beta}\nu^{(3)}+{\cal O}((\Delta\zz)^2) \~,\label{feynman2}\\
(H^{}_{\cl}\rightarrow H^{}_{\cl}\rightarrow H^{}_{\cl})
&\sim&\frac{1}{\beta^{2}}\nu^{(2)}
+\frac{(-1)^{\eps_{A}}}{\beta}(\papal{\zz^{A}}g^{AB})\papal{\zz^{B}}
\left(V-\frac{1}{2\beta}\ln g\right) \cr
&&+{\cal O}((\Delta\zz)^2) \~,\label{feynman3}\\
(H^{}_{\cl}\rightarrow H^{}_{\cl}\leftarrow H^{}_{\cl})
&\sim&\frac{1}{\beta^{2}}\left(\nu^{(1)}-\frac{\nu^{(3)}}{2}\right)
+\frac{1}{\beta}g^{AB}\papal{\zz^{B}}\papal{\zz^{A}}
\left(V-\frac{1}{2\beta}\ln g\right) \cr
&&+{\cal O}((\Delta\zz)^2) \~,\label{feynman4}\\
(H^{}_{\cl}\leftarrow H^{}_{\cl}\rightarrow H^{}_{\cl})
&\sim&\left(V-\frac{1}{2\beta}\ln g,\~V-\frac{1}{2\beta}\ln g\right)
-\frac{1}{2\beta^{2}}\nu^{(3)}+{\cal O}((\Delta\zz)^2)\~.\label{feynman5}
\eea
All the individual contributions of eqs.\ \e{feynman1}--\e{feynman5} have been
collected in the eighth and ninth expression of \eq{transopform}. The eighth
expression is the well-known covariant formula for the path integral, \ie the
\rhs of \eq{pathintegral}. In the phase space path integral, the \mb{R/24} term
arises from the integration over quantum fluctuations \cite{dewitt92}. In the
ninth (and last) expression of \eq{transopform}, the \mb{R/24} term conspires
with the Beltrami--Laplace operator to produce yet another appearance of the
\mb{\Delta} operator \e{evendelta}.

\subsection{First--Order \mb{\SSS^{AB}} Matrices}
\label{seccurvedriem1st}

\noi
After considering quantization of a particle on a curved space in
Subsection~\ref{secparticle}, we shall continue with the investigation of 
Riemannian manifolds. We will assume for the remainder of the Riemannian
Sections~\ref{secriem} and \ref{secriemspingeom} that the density
\mb{\rho=\rhog} is equal to the canonical density \e{riemrho}.

\noi
Because of the presence of the metric tensor \mb{g_{}^{AB}}, the symmetry of
the general linear (\mb{=gl}) Lie--algebra \e{tliealg} reduces to an
orthogonal Lie--subalgebra. Its generators \mb{\SSS^{AB}_{\mp}} read 
\beq
\SSS^{AB}_{\mp}\~:=\~\CC^{A}\~g_{}^{BC} \papal{\CC^{C}}
+ \YY^{A}\~g_{}^{BC} \papal{\YY^{C}} 
\~\mp\~ (-1)^{\eps_{A}\eps_{B}}(A\leftrightarrow B)\~,
\label{curvedriemsigma1st}
\eeq

\beq
\eps(\SSS^{AB}_{\mp})\=\eps_{A}+\eps_{B}
\~,\~\~\~\~\~\~p(\SSS^{AB}_{\mp})\=0\~,
\label{curvedriemsigma1steps}
\eeq

\beq
\SSS^{A}_{\mp C}\~:=\~ \SSS^{AB}_{\mp}\~g^{}_{BC}(-1)^{\eps_{C}}\~.
\label{curvedriemsigma1stmixed}
\eeq
The \mb{\SSS^{AB}_{\mp}} matrices are called first--order matrices, because
they are first--order differential operators in the \mb{\CC^{A}} and
\mb{\YY^{A}} variables. The \mb{\SSS^{AB}_{-}} matrices satisfy an orthogonal
Lie--algebra:
\bea
[\SSS^{AB}_{\mp},\SSS^{CD}_{\mp}]
&=& (-1)^{\eps_{A}(\eps_{B}+\eps_{C})}\left(
g_{}^{BC}\~\SSS^{AD}_{-}+\SSS^{BC}_{-}\~g_{}^{AD}\right)
\mp(-1)^{\eps_{A}\eps_{B}}(A \leftrightarrow B)\~,
\label{curvedriemsigma1stliealg} \\
{}[\SSS^{AB}_{\mp},\SSS^{CD}_{\pm}]
&=& (-1)^{\eps_{A}(\eps_{B}+\eps_{C})}\left(
g_{}^{BC}\~\SSS^{AD}_{+}-\SSS^{BC}_{+}\~g_{}^{AD}\right)
\mp(-1)^{\eps_{A}\eps_{B}}(A \leftrightarrow B)\~.
\label{curvedriemsigma1stliealg2}
\eea
Note that the \eqs{curvedriemsigma1stliealg}{curvedriemsigma1stliealg2} remain
invariant under a \mb{c}--number shift 
\beq
\SSS^{AB}_{+}\~\to\~\SSS^{\prime AB}_{+}
\~:=\~\SSS^{AB}_{+}+\alpha g^{AB}{\bf 1}\~,\label{riemshift}
\eeq
where \mb{\alpha} is a parameter.

\subsection{ \mb{\Gamma^{A}} Matrices}
\label{seccurvedriembiggamma}

\noi
The standard Dirac operator is only defined on a spin manifold, it depends
on the vielbein, and we shall describe it in
Subsections~\ref{secriemgamma}--\ref{secriem2nd}.
But first we shall introduce a poor man's version of \mb{\Gamma^{A}} matrices 
and the so--called Hodge--Dirac operator in the next
Subsections~\ref{seccurvedriembiggamma}--\ref{seccurvedriemdiracop}.
This construction will work for a general Riemannian manifold, which is not
necessarily a spin manifold.

\noi
The \mb{\Gamma^{A}} matrices can be defined via a Berezin--Fradkin operator
representation \cite{b66,f66}
\beq
\Gamma^{A}_{\lambda}\~\equiv\~\Gamma^{A}\~:=\~\CC^{A}+\lambda \PP^{A}
\~,\~\~\~\~\~\~\~\~\~\~\~\~\PP^{A}\~:=\~g_{}^{AB} \papal{\CC^{B}}
\~,\label{curvedriembiggamma}
\eeq
\beq
\eps(\Gamma^{A})\=\eps_{A}
\~,\~\~\~\~\~\~\~\~\~\~\~\~p(\Gamma^{A})\=1\~(\mod\~2)\~.
\eeq
where \mb{\lambda} is a Bosonic parameter with 
\mb{\eps(\lambda)\!=\!0\!=\!p(\lambda)}, which is introduced to bring our 
presentation of the Riemannian case in closer analogy with the antisymplectic
case, see Subsection~\ref{seccurvedantisympbiggamma}. One may interpret 
\mb{\lambda} as a Planck constant. The \mb{\Gamma^{A}} matrices satisfy a 
Clifford algebra
\beq
[\Gamma^{A},\Gamma^{B}]
\= 2 \lambda g_{}^{AB} {\bf 1}\~. \label{curvedriemclifalg}
\eeq
The \mb{\Gamma^{A}} matrices form a fundamental representation of the an 
orthogonal Lie--algebra \e{curvedriemsigma1stliealg}:
\beq
[\SSS^{AB}_{\mp},\Gamma^{C}]
\= \Gamma^{A}_{\pm\lambda}\~g_{}^{BC}
\mp(-1)^{\eps_{A}\eps_{B}}(A\leftrightarrow B)\~.
\label{curvedriemgammasigma1st}
\eeq
If one commutes a metric connection \mb{\nabla_{A}^{(T)}} in the
\mb{T^{A}{}_{B}} representation \e{nablaagammarealiz} with a \mb{\Gamma^{B}}
matrix, one gets
\beq
[\nabla_{A}^{(T)},\Gamma^{B}]
\= -\Gamma_{A}{}^{B}{}_{C}\~\Gamma^{C}\~.\label{riemdiffbiggamma}
\eeq
The minus sign on the \rhs of \eq{riemdiffbiggamma} can be explained as
follows: The contravariant flat \mb{\Gamma^{A}} matrices are passive
bookkeeping devices that ultimately should be contracted with an active
covariant tensor field \mb{\eta^{}_{A}}. It is this implicitly written
\mb{\eta^{}_{A}} that we are really differentiating. Thus there should be a
minus sign.

\noi
The \mb{\nabla_{A}^{(T)}} realization \e{nablaagammarealiz} can be
identically rewritten into the following \mb{\SSS_{\pm}} matrix realization 
\beq
\nabla_{A}^{(\SSS)}\~:=\~\papal{\zz^{A}}
-\Hf\sum_{\pm}\Gamma^{\pm}_{A,BC}\~\SSS^{CB}_{\pm}(-1)^{\eps_{B}}\~,
\label{riemnablabigsigma1st}
\eeq
\ie \mb{\nabla_{A}^{(T)}=\nabla_{A}^{(\SSS)}},
where
\beq
\Gamma^{\pm}_{A,BC}(-1)^{\eps_{C}}
\~:=\~\Hf(-1)^{\eps_{A}\eps_{B}}\Gamma^{}_{BAC}
\pm(-1)^{\eps_{B}\eps_{C}}(B \leftrightarrow C)\~.\label{bigomega1} 
\eeq
{}The Levi--Civita \mb{\Gamma^{\pm}_{A,BC}} connection reads:
\bea
\Gamma^{+}_{A,BC}&=& \Hf(\papal{\zz^{A}}g^{}_{BC})\~,\label{lcbigomegap} \cr
\Gamma^{-}_{A,BC}&=& \Hf(\tilde{g}^{}_{AB}\papar{\zz^{C}})
+(-1)^{(\eps_{B}+1)(\eps_{C}+1)}(B\leftrightarrow C)\~.\label{lcbigomegam}
\eea
Note that both the \mb{\SSS^{AB}_{-}} and the \mb{\SSS^{AB}_{+}} matrices are
needed in the matrix realization \e{riemnablabigsigma1st}.

\subsection{\mb{\CC} Versus \mb{\YY}}
\label{secriembigccyy}

\noi
The \mb{\SSS^{AB}} matrices \e{curvedriemsigma1st} treat the 
\mb{\CC^{A}} and the \mb{\YY^{A}} variables on complete equal footing, whereas
the \mb{\Gamma^{A}} matrices \e{riemflatgamma} contain only the \mb{\CC}'s.
Just from demanding that the \mb{\Gamma^{A}} matrices carry
definite Grassmann-- and form--parity, such \mb{\CC \leftrightarrow \YY} 
symmetry breaking seems unavoidable. {}Further analysis of the Riemannian
case reveals that it is only possible to write a Berezin--Fradkin operator
representation \e{riemflatgamma} of the Clifford algebra \e{riemflatclifalg}
using the \mb{\CC^{A}} variables. (The \mb{\CC^{A}} variables are also
preferred in the antisymplectic case as well, see Subsection~\ref{secattemp} 
below.) One may ponder if there are situations where the \mb{\YY}
variables are useful instead? Yes. The democracy between \mb{\CC} and \mb{\YY}
gets restored in a bigger framework that allows for both even and odd,
Riemannian and symplectic manifolds,
\cf Table~\ref{gradtable}.
{}For instance, the \mb{\YY^{A}} variables
are the only ones suitable for writing down a Berezin--Fradkin--like 
representation 
\beq
\tilde{\Gamma}^{A}\~:=\~\YY^{A}+\lambda \omega_{}^{AB}\papal{\YY^{B}}
\~,\~\~\~\~\~\~\eps(\tilde{\Gamma}^{A})\=\eps_{A}
\~,\~\~\~\~\~\~p(\tilde{\Gamma}^{A})\=0\~,
\label{symplgammaprime}
\eeq
of the Heisenberg algebra 
\beq
[\tilde{\Gamma}^{A},\tilde{\Gamma}^{B}]
\=2 \lambda \omega_{}^{AB}{\bf 1}
\= -(-1)^{\eps_{A}\eps_{B}}(A \leftrightarrow B) \label{curvedheisalg}
\eeq
in even symplectic geometry \cite{kostant74,reuter98,b08}.
(The \mb{\YY^{A}} variables are also preferred in the odd Riemannian case
\cite{b97,lavrov07,lavrov08}.)

\noi
Returning to the even Riemannian case, we will for simplicity only consider
the \mb{\CC^{A}} variables from now on, \ie we shall from now on
put the \mb{\YY^{A}} variables to zero \mb{\YY^{A}\to 0} everywhere, in
particular inside the \mb{T^{A}{}_{B}} matrices \e{tgenerator} and the 
\mb{\SSS^{AB}} matrices \e{curvedriemsigma1st}.

\subsection{Hodge \mb{*} Operation}
\label{secriemhodgestar}

\noi
One may formally define a Hodge \mb{*} operation on exterior forms
\mb{\eta=\eta(\zz;\CC)\in\Omega_{\bullet 0}(M)} as a fiberwise Fourier 
transformation
\beq
(*\eta)(\zz;\CC)\~:=\~ \int \! \frac{d^{N}\CC^{\prime}}{\rho} 
\~e^{\Ih\CC^{\prime} \wedge \CC}\eta(\zz;\CC^{\prime})
\~, \label{riemhodgestar}
\eeq
where we have introduced the shorthand notation
\beq
\CC^{\prime} \wedge \CC\~:=\~\CC^{\prime A}\~g^{}_{AB} \wedge \CC^{B}\~.
\label{riemhodgeswedge}
\eeq
The Hodge \mb{*} operation is an involution \mb{*^{2}\sim {\rm Id}}.
Note that the Hodge dual \mb{*\eta} in general is a distribution.

\noi
In detail, the Hodge \mb{*} operation is built out of two operations:
{}Firstly, a fiberwise Fourier transform 
\beq
\Gamma\left(\bigwedge{}^{\bullet}(T^{*}M)\right)\equiv\Omega_{\bullet 0}(M)
\~\ni\~\eta\~\stackrel{\cF}{\mapsto}\~ 
\pi=\cF\eta\~\in\~\Gamma\left(\bigwedge{}^{\bullet}(TM)\right)\~,
\eeq
that takes exterior forms \mb{\eta\!=\!\eta(\zz;\CC)} to multivectors 
\beq
\pi\=\pi(\zz;\BB)\= \frac{1}{m!}\pi^{A_{1}\cdots A_{m}}(\zz)
\~\BB^{\ell}_{A_{m}}\wedge\cdots\wedge \BB^{\ell}_{A_{1}}\~,
\label{picoordinatefree}
\eeq
where \mb{\BB^{\ell}_{A}\!\equiv\!(-1)^{\eps_{A}}\BB^{r}_{A}} and
\beq 
\begin{array}{rclrclrcl}
\BB^{\ell}_{A} \wedge \BB^{\ell}_{C}
&=& -(-1)^{\eps_{A}\eps_{C}} \BB^{\ell}_{C} \wedge \BB^{\ell}_{A}
\~,\~\~\~\~& \eps(\BB^{\ell}_{A}) &=& \eps_{A}\~,\~\~\~\~& 
p(\BB^{\ell}_{A}) &=& 1\~.
\end{array}\label{curvedbarcom}
\eeq
The Fourier transform \mb{\cF} itself only depends on the density \mb{\rho}:
\beq
(\cF\eta)(\zz;\BB)\~:=\~ \int \! \frac{d^{N}\CC}{\rho} 
\~e^{\Ih \CC^{A} \wedge \BB^{\ell}_{A}}\eta(\zz;\CC)\~.
\label{fouriertransf}
\eeq
Secondly, a flat map 
\beq
\Gamma(TM)\~\ni\~ X\~\stackrel{\flat}{\mapsto}\~
\eta=X^{\flat}\~\in\~\Gamma(T^{*}M)\~,
\eeq
that takes vectors \mb{X\!=\!X^{A}\BB^{\ell}_{A}} to co--vectors 
\mb{\eta\!=\!\eta^{}_{A}\CC^{A}}. The Riemannian flat map \mb{\flat} is
\mb{X^{\flat}_{A}=X_{}^{B}g^{}_{BA}}, or equivalently, in terms of basis 
elements, 
\beq
\BB^{\ell}_{A}\=g^{}_{AB}\CC^{B}\~.
\eeq
Altogether, the Hodge \mb{*} operation can be written as
\beq
(*\eta)(\zz;\CC)
\=\left.\rule[-2ex]{0ex}{2ex}(\cF\eta)(\zz;\BB) 
\right|^{}_{\BB^{\ell}_{A}=g^{}_{AB}\CC^{B}}\~.
\eeq
In contrast to the Riemannian case, there is no good way to construct an
antisymplectic Hodge \mb{*} operation. This is because the antisymplectic flat 
map \mb{\BB^{\ell}_{A}=E^{}_{AB}\CC^{B}} carries the opposite Grassmann--parity
\mb{\eps(\BB^{\ell}_{A})=\eps_{A}+1}, \cf Subsection~\ref{secantisympmetric}.

\begin{proposition}
The Hodge adjoint de Rham operator, \aka the Hodge codifferential, is:
\bea
*d*&\sim&\delta
\~:=\~(-1)^{\eps_{A}}\left(\frac{1}{\rho} \papal{\zz^{A}}\rho
-(\papal{\zz^{A}}g^{}_{BC})\CC^{C}\PP^{B}(-1)^{\eps_{B}}\right)\PP^{A} \cr
&=&~(-1)^{\eps_{A}}\left(\frac{1}{\rho} \papal{\zz^{A}}\rho
-\Hf (\papal{\zz^{A}}g^{}_{BC})\SSS_{+}^{CB}(-1)^{\eps_{B}}\right)\PP^{A}\~.
\label{hodgeadjointderham}
\eea
\label{prophodgeadjointderham0}
\end{proposition}

\noi
{\sc Proof of Proposition~\ref{prophodgeadjointderham0}:}\~
\bea
({*d*}\eta)(\zz,\CC)&=&\int \! \frac{d^{N}\CC^{\prime}}{\rho} 
\~e^{\Ih\CC^{\prime} \wedge \CC}\CC^{\prime A} \papal{\zz^{A}}
\int \! \frac{d^{N}\CC^{\prime\prime}}{\rho} 
\~e^{\Ih\CC^{\prime\prime} \wedge \CC^{\prime}}
\eta(\zz,\CC^{\prime\prime}) \cr
&=&(-1)^{\eps_{A}}\int \! \frac{d^{N}\CC^{\prime}}{\rho}
\left(\papal{\zz^{A}}+\Ih(\papal{\zz^{A}}\CC \wedge \CC^{\prime})\right)
\int \! \frac{d^{N}\CC^{\prime\prime}}{\rho}\~\CC^{\prime A}
e^{\Ih(\CC^{\prime\prime}-C)\wedge\CC^{\prime}}
\eta(\zz,\CC^{\prime\prime})\cr
&=&-(-1)^{\eps_{A}}\Ih\int \! \frac{d^{N}\CC^{\prime\prime}}{\rho}
\left(\papal{\zz^{A}}-(\papal{\zz^{A}}\CC \wedge \PP)\right)
\int \! \frac{d^{N}\CC^{\prime}}{\rho}\~\PP^{A}
e^{\Ih(\CC^{\prime\prime}-C)\wedge\CC^{\prime}}
\eta(\zz,\CC^{\prime\prime})\cr
&\sim& \frac{(-1)^{\eps_{A}}}{\rho}
\left(\papal{\zz^{A}}-(\papal{\zz^{A}}\CC \wedge\PP)\right)
\rho\PP^{A}\eta(\zz,\CC)\~.
\eea
\proofbox

\subsection{Hodge--Dirac Operator \mb{D^{(T)}=d+\lambda \delta}}
\label{seccurvedriemdiracop}

\noi
We shall for the remainder of Section~\ref{secriem} assume that the connection
is the Levi--Civita connection.

\noi
Central for our discussion are the \mb{T^{A}{}_{B}} generators \e{tgenerator}.
They act on exterior forms \mb{\eta\in\Omega_{\bullet0}(M)}, \ie functions 
\mb{\eta\!=\!\eta(\zz;\CC)} of \mb{\zz} and \mb{\CC}. (Recall that the
\mb{\YY^{A}} variables are put to zero \mb{\YY^{A}\to 0}.) 

\noi
The Dirac operator \mb{D^{(T)}} in the \mb{T^{A}{}_{B}} representation
\e{nablaagammarealiz} is a \mb{\Gamma^{A}} matrix \e{curvedriembiggamma} times
the covariant derivative \e{nablaagammarealiz} 
\beq
D^{(T)}\~:=\~\Gamma^{A}\nabla^{(T)}_{A}
\=\CC^{A}\nabla^{(T)}_{A} + \lambda \PP^{A}\nabla^{(T)}_{A}
\=d+\lambda \delta\~,\label{curvedriemdiracop}
\eeq
\beq
\eps(D^{(T)})\=0\~,\~\~\~\~\~\~p(D^{(T)})\=1\~(\mod\~2)\~.
\eeq
The component of the Dirac operator to zeroth order in \mb{\lambda},
\beq
\left. D^{(T)}\right|_{\lambda=0}\=\CC^{A}\nabla^{(T)}_{A}
\=\CC^{A}\left( \papal{\zz^{A}}
-\Gamma_{A}{}^{B}{}_{C}\~\CC^{C}\papal{\CC^{B}}\right)
\=\CC^{A} \papal{\zz^{A}}\=d\~,
\label{dreduc}
\eeq
is just the exterior de Rham derivative \mb{d}, because the connection is 
torsionfree. The component of the Dirac operator to first order in
\mb{\lambda},
\bea
[\papal{\lambda},D^{(T)}]
&=&\PP^{A}\nabla^{(T)}_{A}\=[\PP^{A},\nabla^{(T)}_{A}]
+(-1)^{\eps_{A}}\nabla^{(T)}_{A}\PP^{A} \cr
&=& \Gamma^{A}{}_{AC}\PP^{C}
+(-1)^{\eps_{A}}\left( \papal{\zz^{A}}
-(-1)^{(\eps_{A}+1)\eps_{B}+\eps_{C}}
\Gamma^{}_{BAC}\~\CC^{C}\PP^{B}\right)\PP^{A} \cr
&=&(-1)^{\eps_{A}}\left(\frac{1}{\rhog} \papal{\zz^{A}}\rhog
-(\papal{\zz^{A}}g_{BC})\CC^{C}\PP^{B}(-1)^{\eps_{B}}\right)\PP^{A}
\~\equi{\e{hodgeadjointderham}}:\~\delta\~,
\label{smallevendelta}
\eea
is the Hodge adjoint de Rham operator. Equations \es{dreduc}{smallevendelta}
prove the last equality in \eq{curvedriemdiracop}.

\noi
The Laplacian \mb{\Delta_{\rhog}^{(T)}} in the \mb{T^{A}{}_{B}} representation
\e{nablaagammarealiz} is
\bea
\Delta_{\rhog}^{(T)}&:=& (-1)^{\eps_{A}}\nabla^{}_{A}g^{AB}\nabla^{(T)}_{B}
\= (-1)^{\eps_{A}}\nabla^{(T)}_{A} g^{AB}\nabla^{(T)}_{B}
+\Gamma^{A}{}_{AC}\~ g^{CB}\~\nabla^{(T)}_{B} \cr
&=& \frac{(-1)^{\eps_{A}}}{\rhog}\nabla^{(T)}_{A}
\rhog g^{AB}\nabla^{(T)}_{B}\~.\label{curvedriemlaplact}
\eea

\begin{theorem}[Weitzenb\"ock's formula for exterior forms]
The difference between the square of the Dirac operator \mb{D^{(T)}} and the 
Laplacian \mb{\Delta_{\rhog}^{(T)}} in the \mb{T^{A}{}_{B}} representation
\e{nablaagammarealiz} is
\bea
D^{(T)}D^{(T)} \~-\~ \lambda \Delta_{\rhog}^{(T)} 
&=& -\frac{\lambda}{4}\SSS_{-}^{BA}\~R_{AB,CD}\~
\SSS_{-}^{DC}(-1)^{\eps_{C}+\eps_{D}} \label{riemslw0a}\\
&=& - \lambda\CC^{A}\~R_{AB}\~\PP^{B}
+\frac{\lambda}{2}\CC^{B}\CC^{A}\~R_{AB,CD}\~
\PP^{D}\PP^{C}(-1)^{\eps_{C}+\eps_{D}} \~.\label{riemslw0b} 
\eea
\label{theoremriem0}
\end{theorem}

\noi
Remarks: The square \mb{D^{(T)}D^{(T)}=\lambda(d \delta+\delta d)} is known as
the form Laplacian. The Laplacian \mb{\Delta_{\rhog}^{(T)}} is equal to the
Bochner Laplacian.

\noi
{\sc Proof of Theorem~\ref{theoremriem0}:}\~
The square is a sum of three terms
\beq
D^{(T)}D^{(T)} 
\= \Hf [D^{(T)},D^{(T)}] \= I + II + III\~.
\eeq
The first term is 
\beq
I\~:=\~\Hf [\Gamma^{B},\Gamma^{A}]
\nabla^{(T)}_{A}\nabla^{(T)}_{B}
\= \lambda g^{BA}\~\nabla^{(T)}_{A}\nabla^{(T)}_{B}\~.
\eeq
The second term is 
\bea
II&:=& \Gamma^{A}[\nabla^{(T)}_{A},\Gamma^{B}]\nabla^{(T)}_{B} 
\~\equi{\e{riemdiffbiggamma}}\~ -\Gamma^{A}\~\Gamma_{A}{}^{B}{}_{C}
\~\Gamma^{C}\~\nabla^{(T)}_{B}
\=-(-1)^{\eps_{C}}\Gamma^{B}{}_{CA}\~\Gamma^{A}
\~\Gamma^{C}\~\nabla^{(T)}_{B} \cr
&=&-(-1)^{\eps_{C}}\lambda \Gamma^{B}{}_{CA}\~g^{AC}\~\nabla^{(T)}_{B}
\=\lambda\frac{(-1)^{\eps_{A}}}{\rhog}
(\papal{\zz^{A}}\rhog g^{AB})\nabla^{(T)}_{B}\~.
\eea
Together, the first two terms \mb{I+II} form the Laplace operator 
\e{curvedriemlaplact}:
\beq
I+II \= \lambda\Delta_{\rhog}^{(T)}\~.
\eeq
The third term yields the curvature terms:
\bea
III&:=& -\Hf\Gamma^{B}\Gamma^{A}
[\nabla^{(T)}_{A},\nabla^{(T)}_{B}]
\= \Hf\Gamma^{B}\Gamma^{A}\~R_{AB}{}^{D}{}_{C}\~T^{C}{}_{D}
\= -\frac{1}{4}\Gamma^{B}\Gamma^{A}\~R_{AB,CD}\~
\SSS_{-}^{DC}(-1)^{\eps_{C}+\eps_{D}}\cr
&=& -\frac{1}{4}\left(\CC^{B}\CC^{A}+\lambda(\SSS_{-}^{BA}+g^{BA})
+\lambda^{2}\PP^{B}\PP^{A}\right)R_{AB,CD}\~
\SSS_{-}^{DC}(-1)^{\eps_{C}+\eps_{D}}\cr
&=&-\Hf\CC^{B}\CC^{A}\~R_{AB,CD}\~\CC^{C}\PP^{D}
(-1)^{(\eps_{C}+1)(\eps_{D}+1)} 
-\frac{\lambda}{4}\SSS_{-}^{BA}\~R_{AB,CD}\~
\SSS_{-}^{DC}(-1)^{\eps_{C}+\eps_{D}}\cr
&&-\frac{\lambda^{2}}{2}\PP^{B}\PP^{A}\~R_{AB,CD}\~\PP^{C}\CC^{D}
(-1)^{(\eps_{C}+1)(\eps_{D}+1)}\cr
&=& -\frac{\lambda}{4}\SSS_{-}^{BA}\~R_{AB,CD}\~
\SSS_{-}^{DC}(-1)^{\eps_{C}+\eps_{D}} 
\= -\lambda\CC^{B}\PP^{A}\~R_{AB,CD}\~
\CC^{D}\PP^{C}(-1)^{\eps_{C}+\eps_{D}}\cr
&=&-\lambda\CC^{B}\~R_{BA,CD}g^{DA}\~\PP^{C}
(-1)^{(\eps_{A}+1)(\eps_{C}+1)+\eps_{D}}
+\lambda\CC^{B}\~R_{BA,DC}\~\CC^{D}\PP^{C}\PP^{A}
(-1)^{\eps_{A}+(\eps_{C}+1)(\eps_{D}+1)}\cr
&=&-\lambda\CC^{B}\~R_{BAC}{}^{A}\~\PP^{C}(-1)^{(\eps_{A}+1)(\eps_{C}+1)}
+\lambda\CC^{D}\CC^{B}\~R_{BA,DC}\~\PP^{C}\PP^{A}
(-1)^{\eps_{A}(\eps_{D}+1)+\eps_{C}}\cr
&=& - \lambda\CC^{B}\~R_{BC}\~\PP^{C}
+\frac{\lambda}{2}\CC^{B}\CC^{D}\~R_{DB,AC}\~
\PP^{C}\PP^{A}(-1)^{\eps_{A}+\eps_{C}} \~.  
\eea
Here the first Bianchi identity \e{riem1bianchiid} was used to cancel terms
proportional to zeroth and second order in \mb{\lambda}.
\proofbox

\def\thesubsection{\thesection.\Alph{subsection}}
\setcounter{subsection}{0} 
\subsection{Appendix: Is There A Second--Order Formalism?}
\label{seccurvedriem2nd}

\noi
{}For the standard Dirac operator, which will be discussed in
Subsections~\ref{secriemgamma}--\ref{secriem2nd}, it is natural to replace the
first--order \mb{\sss_{-}^{ab}} matrices \e{riemsigma1st} with the 
second--order \mb{\sig_{-}^{ab}} matrices \e{riemsigma2nd}.
Therefore, it is natural to speculate if it is possible to replace the 
first--order \mb{\SSS_{\pm}^{AB}} matrices \e{curvedriemsigma1st} with the
following second--order matrices:
\beq
\Sig^{AB}_{\mp}
\~:=\~\frac{1}{4\lambda}\Gamma^{A}\Gamma^{B}
\mp (-1)^{\eps_{A}\eps_{B}}(A\leftrightarrow B)\~,
\label{riemcurvedsigma2nd}
\eeq
\beq
\eps(\Sig^{AB}_{\mp})\=\eps_{A}+\eps_{B}
\~,\~\~\~\~\~\~p(\Sig^{AB}_{\mp})\=0\~.
\label{riemcurvedsigma2steps}
\eeq
(The names first-- and second--order refer to the number of 
\mb{\CC^{A}}--derivatives.) 
On one hand, the matrices
\beq
\Sig^{AB}_{-}
\=\frac{1}{4\lambda}\{\Gamma^{A},\Gamma^{B}\}^{}_{+}
\=\frac{1}{2\lambda}\CC^{A}\CC^{B}+\Hf\SSS_{-}^{AB}
+\frac{\lambda}{2}\PP^{A}\PP^{B}\~.\label{riemcurvedsigma2ndm}
\eeq
yield precisely the same non--Abelian Lie--algebra \e{curvedriemsigma1stliealg}
and fundamental representation \e{curvedriemgammasigma1st} as the
\mb{\SSS^{AB}_{-}} matrices. Moreover, the \mb{\SSS^{AB}_{-}} matrices rotate
the \mb{\Sig^{AB}_{-}} matrices
\beq
[\Sig^{AB}_{-},\SSS^{CD}_{-}]
\= (-1)^{\eps_{A}(\eps_{B}+\eps_{C})}\left(
g_{}^{BC}\~\Sig^{AD}_{-}+\Sig^{BC}_{-}\~g_{}^{AD}\right)
-(-1)^{\eps_{A}\eps_{B}}(A \leftrightarrow B)\~.
\label{curvedriemsigsssliealg}
\eeq
However, the commutator of \mb{\Sig^{AB}_{-}} and \mb{\SSS^{CD}_{+}} does not
close,
\beq
[\Sig^{AB}_{-},\SSS^{CD}_{+}]
\=(-1)^{\eps_{A}(\eps_{B}+\eps_{C})}\left(
g_{}^{BC}\~\tilde{\Sig}_{}^{AD}
-\tilde{\Sig}_{}^{BC}\~g_{}^{AD}\right)
-(-1)^{\eps_{A}\eps_{B}}(A \leftrightarrow B)\~,
\label{curvedriemsigsssliealg2}
\eeq
where the tilde generators 
\beq
\tilde{\Sig}_{}^{AB}
\~:=\~-\frac{1}{2\lambda}\CC^{A}\CC^{B}
+\Hf\SSS_{+}^{AB}+\frac{\lambda}{2}\PP^{A}\PP^{B}
\eeq
have no \mb{(A \leftrightarrow B)} symmetry or antisymmetry.
On the other hand, the matrices
\beq
\Sig^{AB}_{+}
\~:=\~\frac{1}{4\lambda}[\Gamma^{A},\Gamma^{B}]
\~\equi{\e{curvedriemclifalg}}\~\Hf g_{}^{AB} {\bf 1}
\label{riemcurvedsigma2ndp}
\eeq
are proportional to the identity operator, and thus behave very differently
from the non--Abelian \mb{\SSS^{AB}_{+}} matrices. 

\noi
The problem with a substitution \mb{\SSS^{AB}_{\mp}\to\Sig^{AB}_{\mp}} is
that the \mb{\SSS^{AB}_{+}} matrices appear in the matrix realization
\e{riemnablabigsigma1st}. On one hand, the \mb{\Sig^{AB}_{-}} representation 
\e{riemcurvedsigma2nd} is not suitable, because it couples pathologically 
to the non--vanishing \mb{\SSS^{AB}_{+}} sector, and, on the other hand, the
\mb{\Sig^{AB}_{+}} matrices are Abelian, and therefore pathological by
themselves. Hence, it is doubtful if the substitution
\mb{\SSS^{AB}_{\mp}\to\Sig^{AB}_{\mp}} makes any sense at all. 
In any case, we shall dismiss the second--order \mb{\Sig^{AB}_{\mp}} matrices
\e{riemcurvedsigma2nd} from now on.

\def\thesubsection{\thesection.\arabic{subsection}}
\section{Antisymplectic Geometry}
\label{secantisymp}

\subsection{Metric}
\label{secantisympmetric}

\noi
Let there be given an antisymplectic metric, \ie a closed two--form
\beq
E\=\Hf \CC^{A}\~ E^{}_{AB} \wedge \CC^{B}
\=-\Hf  E^{}_{AB}\~\CC^{B} \wedge \CC^{A}\~\in\~\Omega_{20}(M)\~,
\label{defantisympe}
\eeq
of Grassmann--parity
\beq
\eps(E^{}_{AB})\=\eps_{A}+\eps_{B}+1
\~,\~\~\~\~\~\~\~\~\~\eps(E)\=1\~,\~\~\~\~\~\~\~\~\~p(E^{}_{AB})\=0\~,
\label{antisympeps}
\eeq
and with antisymmetry 
\beq
E^{}_{BA}\=-(-1)^{\eps_{A}\eps_{B}}E^{}_{AB}\~. \label{antisympsym}
\eeq
The closeness condition 
\beq
 dE\=0 \label{eclosed}
\eeq
reads in components
\beq
\sum_{{\rm cycl.}\~A,B,C}(-1)^{\eps_{A}\eps_{C}}
(\papal{\zz^{A}}E^{}_{BC} ) \= 0~. \label{eclosedabc} 
\eeq
The antisymplectic metric \mb{E_{AB}} is assumed to be non--degenerate, \ie
there exists an inverse contravariant \mb{(2,0)} tensor field 
\mb{E^{AB}} such that 
\beq
E^{}_{AB}\~E_{}^{BC} \= \delta_{A}^{C}\~.\label{antisympinv}
\eeq
The inverse \mb{E^{AB}} has Grassmann--parity
\beq
\eps(E_{}^{AB}) \= \eps_{A}+\eps_{B}+1\~, \label{antisympepsinv}
\eeq
and symmetry 
\beq
E_{}^{BA} \= -(-1)^{(\eps_{A}+1)(\eps_{B}+1)}E_{}^{AB}\~. 
\label{antisympsyminv}
\eeq
The closeness condition \e{eclosed} has no Riemannian analogue. It is the
integrability condition for the local existence of Darboux coordinates.

\subsection{Odd Laplacian \mb{\Deltarho}}
\label{secoddlapl}

The odd Laplacian \mb{\Deltarho}, which takes scalar functions in
scalar functions, is defined as
\beq
2\Deltarho\~:=\~
\frac{(-1)^{\eps_{A}}}{\rho}\papal{\zz^{A}}\rho E^{AB}\papal{\zz^{B}}
\~,\~\~\~\~\~\~\~\~\~\~\~\~\eps(\Deltarho)\=1
\~,\~\~\~\~\~\~\~\~\~\~\~\~p(\Deltarho)\=0\~.
\label{oddlapl}
\eeq
Note the factor of \mb{2} in the odd Laplacian \e{oddlapl} as compared with
the Riemannian case \e{evenlapl}. It is similar in nature to the factor of
\mb{2} in difference between \eqs{defg}{defantisympe}. Both are introduced to
avoid factors of \mb{2} in Darboux coordinates. 

\noi
The antibracket \mb{(f,g)} of two functions \mb{f\!=\!f(\zz)} and
\mb{g\!=\!g(\zz)} can be defined via a double commutator with the odd
Laplacian, acting on the constant unit function \mb{1},
\bea
(f,g)&:=& (-1)^{\eps_{f}}[[\stackrel{\rightarrow}{\Delta}^{}_{\rho},f],g]1
\~\equiv\~ (-1)^{\eps_{f}}\Deltarho(fg)
-(-1)^{\eps_{f}}(\Deltarho f)g
- f(\Deltarho g) + (-1)^{\eps_{g}} fg (\Deltarho 1) \cr
&=&(f\papar{\zz^{A}})E^{AB}(\papal{\zz^{B}}g)
\=-(-1)^{(\eps_{f}+1)(\eps_{g}+1)}(g,f)\~.
\label{antibracket}
\eea
The antibracket \e{antibracket} satisfies a Jacobi identity,
\beq
\sum_{{\rm cycl.}\~f,g,h}(-1)^{(\eps_{f}+1)(\eps_{h}+1)}
(f,(g,h))\= 0\~, \label{antijacid}
\eeq
because of the closeness condition \e{eclosed}.

\subsection{Odd Scalar \mb{\nurho}}
\label{secoddnu}

\noi
A Grassmann--odd function \mb{\nurho} can be constructed from the
antisymplectic metric \mb{E} and an arbitrary density \mb{\rho} as
\beq
\nurho\~:=\~\nurhozero+\frac{\nu^{(1)}}{8}-\frac{\nu^{(2)}}{24}\~,
\label{oddnurho}
\eeq
where 
\bea
\nu^{(0)}_{\rho}&:=&
\frac{1}{\sqrt{\rho}}(\Deltaone\sqrt{\rho})\~,\label{oddnurho0} \\
\nu^{(1)}&:=&(-1)^{\eps_{A}}(\papal{\zz^{A}}E_{}^{AB}
\papar{\zz^{B}})(-1)^{\eps_{B}}\~,\label{oddnu1} \\
\nu^{(2)}&:=&-(-1)^{\eps_{B}}(\papal{\zz^{A}}E^{}_{BC})
(\zz^{C},(\zz^{B},\zz^{A})) \cr
&=&(-1)^{\eps_{A}\eps_{D}}
(\papal{\zz^{D}}E_{}^{AB})E^{}_{BC}(E_{}^{CD}\papar{\zz^{A}})\~.\label{oddnu2}
\eea 
Here \mb{\Deltaone} is the odd Laplacian \e{oddlapl} with \mb{\rho=1}, and
\mb{(\cdot,\cdot)} is the antibracket \e{antibracket}.

\begin{lemma}
The odd quantity \mb{\nurho} is a scalar, \ie it does not depend on the 
coordinate system.
\label{lemmaodd}
\end{lemma}

\noi
The proof of Lemma~\ref{lemmaodd} is given in \Ref{b06}.
Below follows an antisymplectic version of Proposition~\ref{propositioneven}.

\begin{proposition}[Classification of \mb{2}--order differential invariants] 
If a function \mb{\nu\!=\!\nu(\zz)} has the following properties:
\begin{enumerate}
\item
The function \mb{\nu} is a scalar.
\item
\mb{\nu(\zz)} is a polynomial of the metric \mb{E^{}_{AB}(\zz)}, the density
\mb{\rho(\zz)}, their inverses, and \mb{\zz}--derivatives thereof in the point
\mb{\zz}. 
\item
\mb{\nu} is invariant under constant rescaling of the density 
\mb{\rho\to\lambda\rho}, where \mb{\lambda} is a
\mb{\zz}--independent parameter.
\item
\mb{\nu} scales as \mb{\nu\to \lambda\nu} under constant Weyl scaling
\mb{E_{}^{AB}\to\lambda E_{}^{AB}}, where \mb{\lambda} is a
\mb{\zz}--independent parameter.
\item
Each term in \mb{\nu} contains precisely two \mb{\zz}--derivatives.
\end{enumerate}
Then \mb{\nu} is proportional to the odd scalar \mb{\nurho}
\beq
\nu\=\alpha\~\nurho\~,
\label{onelittlepig}
\eeq
where \mb{\alpha} is \mb{\zz}--independent proportionality constant. 
\label{propositionodd}
\end{proposition}

\noi
The proof of Proposition~\ref{propositionodd} is similar to the
proof of Proposition~\ref{propositioneven}.

\subsection{\mb{\Delta} And  \mb{\DeltaE}}
\label{secoddkhuda}

\noi
Khudaverdian's \mb{\DeltaE} operator \cite{k99,kv02,k02,k04,bbd06,b06,b07}, 
which takes semidensities to semidensities, is defined using arbitrary
coordinates as
\beq
\DeltaE\~:=\~\Deltaone+\frac{\nu^{(1)}}{8}-\frac{\nu^{(2)}}{24}\~.
\label{deltaedef}
\eeq
It is obviously manifestly independent of \mb{\rho}. 
That it takes semidensities to semidensities will become clear because of
\eq{deltaedelta} below. The Jacobi identity \e{antijacid} precisely encodes 
the nilpotency of \mb{\DeltaE}. The Grassmann--odd nilpotent \mb{\Delta} 
operator, which takes scalar functions to scalar functions, can be defined as
defined as
\beq
\Delta\~:=\~\Deltarho+\nurho\~.\label{odddelta}
\eeq
In fact, every Grassmann--odd, nilpotent, second--order operator is of the
form \e{odddelta} up to a Grassmann--odd constant \cite{bb07}. We shall
dismiss Grassmann--odd constants since they do not satisfy all the five
assumptions of Proposition~\ref{propositionodd}. The \mb{\DeltaE} operator and
the \mb{\Delta} operator are related via \mb{\sqrt{\rho}}--conjugation
\cite{b06,bb07}
\beq
\DeltaE\=\sqrt{\rho}\Delta\frac{1}{\sqrt{\rho}}\~.
\label{deltaedelta}
\eeq
The proof is almost identical to the corresponding Riemannian calculation 
\e{deltagdeletaproof}. 

\noi
Recall how the zeroth--order term is determined in the Riemannian case, where
no nilpotency principle was available, \cf Subsections~\ref{secevenkhuda} and
\ref{secriemonshell}. There we applied a \mb{\rho} independence test. Could
one do a similar analysis in the antisymplectic case? Yes. In detail, consider
an operator
\beq
\Delta^{\prime}\~:=\~\Delta+\nu=\Deltarho+\nurho+\nu\~,\label{odddeltaprime}
\eeq 
where \mb{\nu} is a most general zeroth--order term. It is easy to see from
\eqs{deltaedelta}{odddeltaprime} that the corresponding
\mb{\sqrt{\rho}}--conjugated operator
\mb{\sqrt{\rho}\Delta^{\prime}\frac{1}{\sqrt{\rho}}} is independent of
\mb{\rho} if and only if the shift term \mb{\nu} is \mb{\rho}--independent.
{}From Proposition~\ref{propositionodd}, one then concludes that \mb{\nu=0} 
has to be zero, \ie the form of the \mb{\Delta} operator \e{odddelta} can be 
uniquely reproduced from a \mb{\rho}--independence test and knowledge about
possible scalar structures.

\subsection{Antisymplectic Connection}
\label{secantisympconn}

\noi
A connection \mb{\nabla^{(\Gamma)}} is called {\em antisymplectic}, if it
preserves the antisymplectic metric \mb{E},
\beq
0 \= (\nabla^{(\Gamma)}_{A}E)^{}_{BC}
\= (\papal{\zz^{A}}E^{}_{BC})
-\left((-1)^{\eps_{A}\eps_{B}}\Gamma^{}_{BAC}
-(-1)^{\eps_{B}\eps_{C}}(B \leftrightarrow C)\right)\~.\label{connantisymp}
\eeq
Here we have lowered the Christoffel symbol with the metric
\beq
\Gamma_{ABC}^{}\~:=\~E_{AD}\Gamma^{D}{}_{BC}(-1)^{\eps_{B}}\~.
\label{lowerantisympconn}
\eeq
We should stress that there is not a unique choice of an antisymplectic, 
torsionfree, and \mb{\rho}--compatible connection \mb{\nabla^{(\Gamma)}}, \ie a
connection that satisfies eqs.\ \e{connantisymp}, \es{torsionfree}{rhocomp}.
On the other hand, it can be demonstrated that such connections
\mb{\nabla^{(\Gamma)}} exist locally for \mb{N>2}, where \mb{N=\dim(M)} 
denotes the dimension of the manifold \mb{M}. (There are counterexamples for
\mb{N\!=\!2} where \mb{\nabla^{(\Gamma)}} need not exist.) The mere existence
of an antisymplectic and torsionfree connection \mb{\nabla^{(\Gamma)}} implies
that the two--form \mb{E} is closed \e{eclosed}, if we hadn't already assumed
it in the first place. (Curiously, while it is impossible to impose closeness
relations in Riemannian geometry, the closeness relations are almost impossible
to avoid in geometric structures defined by two--forms.) The antisymplectic
condition \e{connantisymp} reads in terms of the contravariant (inverse) metric
\beq
0\=(\nabla^{(\Gamma)}_{A}E)^{BC}
\~\equiv\~(\papal{\zz^{A}}E^{BC})+\left(\Gamma_{A}{}^{B}{}_{D}E^{DC}
-(-1)^{(\eps_{B}+1)(\eps_{C}+1)}(B\leftrightarrow C)\right)\~.
\label{upperantisympconn}
\eeq

\subsection{The Riemann Curvature}
\label{secantisympriemcurv}

\noi
{}For an antisymplectic connection \mb{\nabla^{(\Gamma)}}, we prefer to work
with a \mb{(0,4)} Riemann tensor (as opposed to a \mb{(1,3)} tensor) by
lowering the upper index with the metric \e{defantisympe}. In terms of 
Christoffel symbols it is easiest to work with expression \e{riemtensor1}:
\bea
R_{D,ABC}&:=&E^{}_{DF}R^{F}{}_{ABC}\cr
&=&(-1)^{\eps_{A}(\eps_{D}+1)}\left((-1)^{\eps_{B}}\papal{\zz^{A}}\Gamma_{DBC}
+(-1)^{\eps_{F}(\eps_{A}+\eps_{D})}\Gamma_{FAD}\Gamma^{F}{}_{BC}\right) \cr
&&-(-1)^{\eps_{A}\eps_{B}}(A\leftrightarrow B)\~.\label{antisymptensor1low}
\eea
In the second equality of \eq{antisymptensor1low} is used the antisymplectic
condition \e{connantisymp}. If the antisymplectic condition \e{connantisymp}
is used one more time on the first term in \eq{antisymptensor1low}, one
derives the following symmetry
\beq
R_{D,ABC} \= (-1)^{(\eps_{A}+\eps_{B})(\eps_{C}+\eps_{D})
+\eps_{C}\eps_{D}}(C\leftrightarrow D)\~.
\label{antisymptensor1lowsym}
\eeq
This symmetry becomes clearer if one instead starts from expression 
\e{riemtensor4} and defines
\beq
R_{AB,CD}\~:=\~ R_{ABC}{}^{F}E_{FD}\=
 -(-1)^{\eps_{A}+\eps_{B}+(\eps_{A}+\eps_{B}+\eps_{C})\eps_{D}}R_{D,ABC}\~.
\label{antisymptensor4low}
\eeq
Then the symmetry \e{antisymptensor1lowsym} simply translates into a symmetry
between the third and fourth index:
\beq
R_{AB,CD} \= (-1)^{\eps_{C}\eps_{D}}(C\leftrightarrow D)\~.
\label{antisymptensor4lowsym}
\eeq
The Ricci \mb{2}--form is then
\beq
\cR_{AB}\~=:\~R_{AB}{}^{C}{}_{C}(-1)^{\eps_{C}}
\=R_{AB,CD}E^{DC}(-1)^{\eps_{C}}\~.\label{antisympricci2form}
\eeq
We note that the torsionfree condition has not been used so far in this 
Section~\ref{secantisympriemcurv}. The first Bianchi identity \e{1bianchiid}
reads (in the torsionfree case):
\beq
0~=~\sum_{{\rm cycl.}\~A,B,C}(-1)^{\eps_{A}\eps_{C}}R_{AB,CD}~.
\label{antisymp1bianchiid}
\eeq

\subsection{Odd Scalar Curvature}
\label{secoddscalar}

\noi
The odd scalar curvature is defined as
\beq
R\~:=\~E^{BA}R^{}_{AB}\=R^{}_{AB}E^{BA}\~.\label{oddscalarcurv}
\eeq

\begin{proposition}
{}For an arbitrary, antisymplectic, torsionfree, and \mb{\rho}--compatible
connections \mb{\nabla^{\Gamma}}, the scalar curvature \mb{R} does only depend
on \mb{E} and \mb{\rho} through the odd \mb{\nurho} scalar \cite{bb07}
\beq
R\=-8\nurho\~.\label{oddrnurho}
\eeq
\label{propositionoddrnu}
\end{proposition}

\noi
The proof of Proposition~\ref{propositionoddrnu} is given in \Ref{bb07}.
It is extended to degenerate anti--Poisson structures in \Ref{b07,bb08}.
In particular, one concludes that the odd scalar curvature \mb{R} does not 
depend on the connection used, and the odd \mb{\Delta} operator \e{odddelta}
reduces to the odd \mb{\Delta} operator \e{deltaop} in the Introduction.

\noi
Altogether, we have now established a link 
between the zeroth--order terms in the even and odd \mb{\Delta} operators
\es{evenonshelldeltaop}{deltaop}:
\bea
{\rm Riemannian\~zeroth\~ order\~term}
&&{\rm Antisymplectic\~~zeroth\~order\~term}\cr 
-\frac{R}{4}\=\nurhog&\longleftrightarrow&2\nurho\=-\frac{R}{4}\~.
\eea
The left (\resp right) equality is due to Proposition~\ref{propositionevenrnu}
(\resp \ref{propositionoddrnu}). Both zeroth--order terms are characterized by
the same \mb{\rho}--independence test described in
Subsections~\ref{secriemonshell} and \ref{secoddkhuda} (up to a subtlety on
how to switch back and forth between \mb{\rho}--dependent and
\mb{\rho}--independent formalism in the Riemannian case). It is no coincidence
that the same coefficient minus--a--quarter appears on both sides of the
correspondence (after the odd \mb{\Delta} operator has been multiplied with an
appropriate factor \mb{2}). At the mathematical level, this is basically
because the zeroth--order terms are determined by the \mb{\nurhozero} building
blocks alone, where the inverse metrics \mb{g^{AB}} and \mb{E^{AB}} enter in a
similar manner, and only linearly. {}For expressions that do not depend on the
metric tensors \mb{g_{AB}} and \mb{E_{AB}}, and only have an linear dependence
of the inverse metrics \mb{g^{AB}} and \mb{E^{AB}}, respectively, one does not
see the effects that distinguish Riemannian and antisymplectic geometry, such
as \eg opposite Grassmann--parity, closeness relations and the Jacobi
identities.

\subsection{First--Order \mb{\SSS^{AB}} Matrices}
\label{seccurvedantisymp1st}

\noi
Because of the presence of the antisymplectic tensor \mb{E_{}^{AB}}, the
symmetry of the general linear (\mb{=gl}) Lie--algebra \e{tliealg} reduces to
an antisymplectic Lie--subalgebra. Its generators \mb{\SSS^{AB}_{\pm}} read
\beq
\SSS^{AB}_{\pm}\~:=\~\CC^{A} (-1)^{\eps_{B}}\PP^{B}
\~\mp\~ (-1)^{(\eps_{A}+1)(\eps_{B}+1)}(A\leftrightarrow B)
\~,\~\~\~\~\~\~\~\~\~\~\~\~
\PP^{A}\~:=\~E_{}^{AB} \papal{\CC^{B}}\~,
\label{curvedantisympsigma1st}
\eeq

\beq
\eps(\SSS^{AB}_{\pm})\=\eps_{A}+\eps_{B}+1
\~,\~\~\~\~\~\~p(\SSS^{AB}_{\pm})\=0\~,
\label{curvedantisympsigma1steps}
\eeq

\beq
\SSS^{A}_{\pm C}\~:=\~ \SSS^{AB}_{\pm}\~E^{}_{BC}(-1)^{\eps_{C}}\~. 
\label{curvedantisympsigma1stmixed}
\eeq

\noi
The \mb{\SSS^{AB}_{+}} matrices satisfy an antisymplectic Lie--algebra:
\bea
[\SSS^{AB}_{\pm},\SSS^{CD}_{\pm}]
&=& (-1)^{\eps_{A}(\eps_{B}+\eps_{C}+1)+\eps_{B}}\left(
E_{}^{BC}\~\SSS^{AD}_{+}-\SSS^{BC}_{+}\~E_{}^{AD} \right) \cr
&& \mp(-1)^{(\eps_{A}+1)(\eps_{B}+1)}(A\leftrightarrow B)\~,
\label{curvedantisympsigma1stliealg} \\
{}[\SSS^{AB}_{\pm},\SSS^{CD}_{\mp}]
&=& (-1)^{\eps_{A}(\eps_{B}+\eps_{C}+1)+\eps_{B}}\left(
E_{}^{BC}\~\SSS^{AD}_{-}+\SSS^{BC}_{-}\~E_{}^{AD} \right) \cr
&& \mp(-1)^{(\eps_{A}+1)(\eps_{B}+1)}(A\leftrightarrow B)\~.
\label{curvedantisympsigma1stliealg2} 
\eea
Note that the \eqs{curvedantisympsigma1stliealg}{curvedantisympsigma1stliealg2}
remain invariant under a \mb{c}--number shift 
\beq
\SSS^{AB}_{+}\~\to\~\SSS^{\prime AB}_{+}
\~:=\~\SSS^{AB}_{+}+\alpha E^{AB}{\bf 1}\~,\label{antisympshift}
\eeq
where \mb{\alpha} is a parameter.

\subsection{ \mb{\Gamma^{A}} Matrices}
\label{seccurvedantisympbiggamma}

\noi
Guided by the analysis of Appendix~\ref{secattemp}, we now define
antisymplectic \mb{\Gamma^{A}} matrices via the following Berezin--Fradkin
operator representation \cite{b66,f66}
\beq
\Gamma^{A}_{\theta}\~\equiv\~
\Gamma^{A}\~:=\~\CC^{A}+(-1)^{\eps_{A}}\theta \PP^{A}
\=\CC^{A}-\PP^{A}\theta
\~,\~\~\~\~\~\~\~\~\~\~\~\~\eps(\Gamma^{A})\=\eps_{A}
\~,\~\~\~\~\~\~\~\~\~\~\~\~p(\Gamma^{A})\=1\~(\mod\~2)\~,
\label{curvedantisympbiggamma}
\eeq
where \mb{\theta} is a nilpotent Fermionic parameter with
\mb{\eps(\theta)\!=\!1} and \mb{p(\theta)\!=\!0}. The \mb{\Gamma^{A}} matrices
satisfy a Clifford--like algebra
\beq
[\Gamma^{A},\Gamma^{B}]
\= 2 (-1)^{\eps_{A}}\theta E_{}^{AB} {\bf 1}\~.\label{curvedantisympclifalg}
\eeq
The \mb{\Gamma^{A}} matrices form a fundamental representation of the
antisymplectic Lie--algebra \e{curvedantisympsigma1stliealg}:
\beq
[\SSS^{AB}_{\pm},\Gamma^{C}]
\= \Gamma^{A}_{\pm\theta}(-1)^{\eps_{B}}E_{}^{BC}
\mp(-1)^{(\eps_{A}+1)(\eps_{B}+1)}(A\leftrightarrow B)\~.
\label{curvedantisympgammasigma1st}
\eeq
If one commutes an antisymplectic connection \mb{\nabla_{A}^{(T)}} in the
\mb{T^{A}{}_{B}} representation \e{nablaagammarealiz} with a \mb{\Gamma^{B}}
matrix, one gets
\beq
[\nabla_{A}^{(T)},\Gamma^{B}]
\= -\Gamma_{A}{}^{B}{}_{C}\~\Gamma^{C}\~.\label{antisympdiffbiggamma}
\eeq

\subsection{Dirac Operator \mb{D^{(T)}=d+\theta\delta}}
\label{seccurvedantisympdiracop}

\noi
We shall for the remainder of Section~\ref{secantisymp} assume that the 
connection is antisymplectic, torsionfree and \mb{\rho}--compatible.

\noi
The Dirac operator \mb{D^{(T)}} in the \mb{T^{A}{}_{B}} representation
\e{nablaagammarealiz} is a \mb{\Gamma^{A}} matrix \e{curvedantisympbiggamma}
times the covariant derivative \e{nablaagammarealiz} 
\beq
D^{(T)}\~:=\~\Gamma^{A}\nabla^{(T)}_{A}\=d+\theta\delta
\~,\~\~\~\~\~\~\eps(D^{(T)})\=0\~,\~\~\~\~\~\~p(D^{(T)})\=1\~(\mod\~2)\~.
\label{curvedantisympdiracop}
\eeq
Unlike the Riemannian case of Subsection~\ref{seccurvedriemdiracop}, the 
component \mb{\delta} of the Dirac operator to first order in \mb{\theta} 
does not have an interpretation as a Hodge codifferential, since there is no
antisymplectic Hodge \mb{*} operation. Even worse, it depends explicitly on 
the Christoffel symbols:
\bea
\delta&:=&(-1)^{\eps_{A}}\PP^{A}\nabla^{(T)}_{A}\=
(-1)^{\eps_{A}}[\PP^{A},\nabla^{(T)}_{A}]
+(-1)^{\eps_{A}}\nabla^{(T)}_{A}\PP^{A} \cr
&=& \Gamma^{A}{}_{AC}\PP^{C}
+(-1)^{\eps_{A}}\left( \papal{\zz^{A}}
+(-1)^{\eps_{A}\eps_{B}} \Gamma^{}_{BAC}\~\CC^{C}\PP^{B}\right)\PP^{A} \cr
&=&(-1)^{\eps_{A}}\left(\frac{1}{\rho} \papal{\zz^{A}}\rho
+\Gamma^{}_{ABC}\~\CC^{C}\PP^{B}\right)\PP^{A}\~.
\label{smallodddelta}
\eea
Nevertheless, there exists a close antisymplectic analogue of Weitzenb\"ock's 
formula \e{riemslw0b}, \cf \eq{antisympslw0b} below. 
The odd Laplacian \mb{\Delta_{\rho}^{(T)}} in the \mb{T^{A}{}_{B}}
representation \e{nablaagammarealiz} is
\beq
2\Delta_{\rho}^{(T)}\~:=\~ (-1)^{\eps_{A}}\nabla^{}_{A}E^{AB}\nabla^{(T)}_{B}
\=\frac{(-1)^{\eps_{A}}}{\rho}\nabla^{(T)}_{A}\rho E^{AB}\nabla^{(T)}_{B}
\~.\label{oddlaplaciant}
\eeq

\begin{theorem}[Antisymplectic Weitzenb\"ock type formula for exterior forms]
The difference between the square of the Dirac operator \mb{D^{(T)}} and twice 
the odd Laplacian \mb{\Delta_{\rho}^{(T)}} in the \mb{T^{A}{}_{B}} 
representation is
\bea
D^{(T)}D^{(T)} \~-\~ 2\theta \Delta_{\rho}^{(T)} 
&=& \frac{\theta}{4}(-1)^{\eps_{B}+\eps_{C}}
\SSS_{-}^{BA}\~R_{AB,CD}\~\SSS_{+}^{DC}\label{antisympslw0a}\\
&=& - \theta\CC^{A}\~R_{AB}\~\PP^{B}
+\frac{\theta}{2}\CC^{B}\CC^{A}\~R_{AB,CD}\~
\PP^{D}\PP^{C}(-1)^{\eps_{C}}\~.  \label{antisympslw0b}
\eea
\label{theoremantisymp0}
\end{theorem}

\noi
{\sc Proof of Theorem~\ref{theoremantisymp0}:}\~
The square is a sum of three terms
\beq
D^{(T)}D^{(T)} 
\= \Hf [D^{(T)},D^{(T)}] \= I + II + III~.
\eeq
The first term is 
\beq
I\~:=\~\Hf [\Gamma^{B},\Gamma^{A}]
\nabla^{(T)}_{A}\nabla^{(T)}_{B}
\= (-1)^{\eps_{B}}\theta E^{BA}\~\nabla^{(T)}_{A}\nabla^{(T)}_{B}\~.
\eeq
The second term is 
\bea
II&:=& \Gamma^{A}[\nabla^{(T)}_{A},
\Gamma^{B}]\nabla^{(T)}_{B} 
\~\equi{\e{antisympdiffbiggamma}}\~ 
-\Gamma^{A}\~\Gamma_{A}{}^{B}{}_{C}
\~\Gamma^{C}\~\nabla^{(T)}_{B}
\=-(-1)^{\eps_{C}}\Gamma^{B}{}_{CA}\~\Gamma^{A}
\~\Gamma^{C}\~\nabla^{(T)}_{B} \cr
&=&-(-1)^{\eps_{B}}\theta \Gamma^{B}{}_{CA}\~E^{AC}\~\nabla^{(T)}_{B}
\=\theta\frac{(-1)^{\eps_{A}}}{\rho}
(\papal{\zz^{A}}\rho E^{AB})\nabla^{(T)}_{B}\~.
\eea
Together, the first two terms \mb{I+II} form the odd Laplacian
\e{oddlaplaciant}:
\beq
I+II \= 2\theta\Delta_{\rho}^{(T)}\~.
\eeq
The third term yields the curvature terms:
\bea
III&:=& -\Hf\Gamma^{B}\Gamma^{A}
[\nabla^{(T)}_{A},\nabla^{(T)}_{B}]
\= \Hf\Gamma^{B}\Gamma^{A}\~R_{AB}{}^{D}{}_{C}\~T^{C}_{}{D}
\= \frac{1}{4}\Gamma^{B}\Gamma^{A}\~R_{AB,CD}\~
\SSS_{+}^{DC}(-1)^{\eps_{C}}\cr
&=& \frac{1}{4}\left(\CC^{B}\CC^{A}+(-1)^{\eps_{B}}\theta
(\SSS_{-}^{BA}+ E^{BA})\right)R_{AB,CD}\~
\SSS_{+}^{DC}(-1)^{\eps_{C}}\cr
&=&\Hf\CC^{B}\CC^{A}\~R_{AB,CD}\~\CC^{C}\PP^{D}(-1)^{\eps_{C}\eps_{D}} 
+\frac{\theta}{4}(-1)^{\eps_{B}+\eps_{C}}
\SSS_{-}^{BA}\~R_{AB,CD}\~\SSS_{+}^{DC}\cr
&=& \frac{\theta}{4}(-1)^{\eps_{B}+\eps_{C}} 
\SSS_{-}^{BA}\~R_{AB,CD}\~\SSS_{+}^{DC}
\= (-1)^{\eps_{A}+\eps_{B}}\theta
\CC^{B}\PP^{A}\~R_{AB,CD}\~\CC^{D}\PP^{C}\cr
&=&-(-1)^{(\eps_{A}+1)(\eps_{C}+1)}
\theta\CC^{B}\~R_{BA,CD}E^{DA}\~\PP^{C}
-\theta\CC^{B}\~R_{BA,DC}\~\CC^{D}\PP^{C}\PP^{A}
(-1)^{\eps_{A}+\eps_{C}\eps_{D}}\cr
&=&-(-1)^{(\eps_{A}+1)(\eps_{C}+1)}
\theta\CC^{B}\~R_{BAC}{}^{A}\~\PP^{C}
+\theta\CC^{D}\CC^{B}\~R_{BA,DC}\~\PP^{C}\PP^{A}
(-1)^{\eps_{A}(\eps_{D}+1)}\cr
&=& -\theta\CC^{B}\~R_{BC}\~\PP^{C}
+\frac{\theta}{2}\CC^{B}\CC^{D}\~R_{DB,AC}\~
\PP^{C}\PP^{A}(-1)^{\eps_{A}} \~.  
\eea
Here the first Bianchi identity \e{antisymp1bianchiid} was used one time in
the \mb{\theta}--independent sector.
\proofbox

\def\thesubsection{\thesection.\Alph{subsection}}
\setcounter{subsection}{0} 
\subsection{Appendix: Is There A Second--Order Formalism?}
\label{seccurvedantisymp2nd}

\noi
There are no deformations of the first--order \mb{\SSS_{-}^{AB}} matrices 
\e{curvedantisympsigma1st}. The general second--order deformation of the 
\mb{\SSS_{+}^{AB}} matrices \e{curvedantisympsigma1st} reads 
\beq
\Sig_{+}^{AB}\~:=\~\SSS_{+}^{AB}+\alpha E^{AB}{\bf 1}
+\beta \PP^{A}\PP^{B}\theta\~, 
\eeq
where \mb{\alpha} and \mb{\beta} are two parameters. 
The second--order \mb{\Sig_{+}^{AB}} matrices satisfy precisely the same
antisymplectic Lie--algebra \e{curvedantisympsigma1stliealg} as the
\mb{\SSS_{+}^{AB}} matrices. Moreover, the \mb{\SSS^{AB}_{+}} matrices rotate
the \mb{\Sig^{AB}_{+}} matrices,
\beq
[\Sig^{AB}_{+},\SSS^{CD}_{+}]
\= (-1)^{\eps_{A}(\eps_{B}+\eps_{C}+1)+\eps_{B}}\left(
E_{}^{BC}\~\Sig^{AD}_{+}-\Sig^{BC}_{+}\~E_{}^{AD} \right)
-(-1)^{(\eps_{A}+1)(\eps_{B}+1)}(A\leftrightarrow B)\~.
\label{curvedantisympsigsss1stliealg}
\eeq
The \mb{\Sig^{AB}_{+}} matrices interact with the
\mb{\Gamma^{C}} and the \mb{\SSS^{CD}_{-}} matrices as follows 
\bea
[\Sig^{AB}_{+},\Gamma^{C}]
&=& \Gamma^{A}_{(1+\beta)\theta}(-1)^{\eps_{B}}E_{}^{BC}
-(-1)^{(\eps_{A}+1)(\eps_{B}+1)}(A\leftrightarrow B)\~,
\label{curvedantisympgammasig1st} \\
{}[\Sig^{AB}_{+},\SSS^{CD}_{-}]
&=& (-1)^{\eps_{A}(\eps_{B}+\eps_{C}+1)+\eps_{B}}\left(
E_{}^{BC}\~\tilde{\Sig}^{AD}_{}+\tilde{\Sig}^{BC}_{}\~E_{}^{AD} \right) \cr
&& -(-1)^{(\eps_{A}+1)(\eps_{B}+1)}(A\leftrightarrow B)\~,
\label{curvedantisympsigsss1stliealgpm}
\eea
where the generators 
\beq
\tilde{\Sig}_{}^{AB}
\~:=\~\SSS_{-}^{AB}+\beta\PP^{A}\PP^{B}\theta
\eeq
have no \mb{(A \leftrightarrow B)} symmetry or antisymmetry. According to 
\eq{curvedantisympgammasig1st}, one must choose the parameter \mb{\beta=0} to 
be zero, in order to ensure that the \mb{\Sig_{+}^{AB}} matrices rotates the
\mb{\Gamma^{A}} matrices in the correct way. One concludes that a consistent
antisymplectic second--order formulation does not exist, regardless of 
whether the pathological \mb{\SSS_{-}^{AB}} sector decouples or not, and
we shall abandon the subject. See also comment in the Conclusions.

\subsection{Appendix: What Is An Antisymplectic Clifford Algebra?}
\label{secattemp}

\noi
In this Appendix~\ref{secattemp}, we shall motivate the definition 
\e{curvedantisympclifalg} of an antisymplectic Clifford algebra given in 
Subsection~\ref{seccurvedantisympbiggamma}. Intuitively, one would probably
assume that an antisymplectic Clifford algebra should be 
\beq
\Gamma^{A}\star\Gamma^{B}
-(-1)^{(\eps_{A}+1)(\eps_{B}+1)}(A \leftrightarrow B)
\~\equi{?}\~2 E^{AB}{\bf 1}\~,
\label{cliffordstar1}
\eeq
where the ``\mb{\star}'' denotes a Fermionic multiplication,
\mb{\eps(\star)=1}, \cf question \mb{3} in the Introduction. We will now expose
some of the weaknesses of the proposal \e{cliffordstar1}. (A question mark
``\mb{?}'' on top of an equality sign ``\mb{=}'' indicates that a formula may
be ultimately wrong.) It follows from \eq{deltastar} that the form degree of
the \mb{\star} multiplication must vanish, \mb{p(\star)=0}. Let us assume that
the \mb{\star} multiplication is invertible and commute with the
\mb{\Gamma^{A}} matrices,
\beq
\Gamma^{A}{\star}-(-1)^{\eps(\Gamma_{A})}{\star}\Gamma^{A}
\~\equiv\~[\Gamma^{A},\star]\~\equi{?}\~0\~.\label{gammastar}
\eeq
Then one can bring the Clifford algebra \e{cliffordstar1} on a Riemannian form,
\beq
\Gamma^{A}\Gamma^{B}+(-1)^{\eps_{A}\eps_{B}}(A \leftrightarrow B)
\=2 g_{}^{AB}{\bf 1} \~,\label{cliffordstar2}
\eeq
where the Riemannian metric \mb{g_{}^{AB}} is a product of \mb{\star^{-1}} and
the antisymplectic metric \mb{E_{}^{AB}},
\beq
g_{}^{AB}\~:=\~(-1)^{\eps(\Gamma_{A})} \star^{-1} E_{}^{AB}
\=(-1)^{\eps_{A}\eps_{B}}(A \leftrightarrow B)
\~,\~\~\~\~\~\~\~\~\eps(g_{}^{AB})\=\eps_{A}+\eps_{B}\~.\label{ncgab}
\eeq
The Riemannian structure \e{ncgab} is non--commutative, 
\beq
[g_{}^{AB},g_{}^{CD}]
\=-2(-1)^{\eps_{B}+\eps_{C}}\star^{-2} E_{}^{AB}E_{}^{CD}\~\neq\~0\~,
\label{metricnoncom}
\eeq
since \mb{[\star^{-1},\star^{-1}]=2\star^{-2}\neq 0}, and hence the metric
\e{ncgab} is not a classical Riemannian metric. We would like to interpret the
\lhs of \eq{cliffordstar2} as a commutator \mb{[\Gamma^{A},\Gamma^{B}]},
\cf definition \e{com}. This implies that the Grassmann-- and form--parity of 
the \mb{\Gamma^{A}} matrices are
\beq
\eps(\Gamma^{A})\=\eps_{A}\~,\~\~\~\~\~\~\~\~\~\~\~\~
p(\Gamma^{A})\=\~1\~(\mod\~2)\~.
\label{gammaparity}
\eeq
The only natural candidate for a Berezin--Fradkin operator representation
\cite{b66,f66} is
\beq
\Gamma^{A}\=\CC^{A} + g_{}^{AB} \papal{\CC^{B}}
\~\equiv\~\CC^{A}-\PP^{A}\~\star^{-1}
\~,\~\~\~\~\~\~\~\~\~\~\~\~\eps(\CC^{A})\=\eps_{A}
\~,\~\~\~\~\~\~\~\~\~\~\~\~p(\CC^{A})\=\~1\~,
\label{bfoprep}
\eeq
where the \mb{\CC^{A}} variables commute with the \mb{\star} multiplication,
\mb{[\CC^{A},\star]=0}, and they carry the same Grassmann-- and
form--parities as the \mb{\Gamma^{A}} matrices. The \mb{\PP^{A}} derivatives
are defined in \eq{curvedantisympsigma1st}. However, the Berezin--Fradkin
operator representation \e{bfoprep} does not satisfy the Clifford algebra
\e{cliffordstar2} due to the non--commutative metric \e{metricnoncom}. The
representation does also violate the commutation relation \e{gammastar}.
There appear extra terms on the respective right--hand sides, 
\bea
[\Gamma^{A},\star^{-1}]&=&-2\star^{-2}\PP^{A}\~,\label{gammastarprime} \\
{}[\Gamma^{A},\Gamma^{B}]&=&2 g_{}^{AB}{\bf 1}
-2\star^{-2} \PP^{A}\PP^{B}(-1)^{\eps_{B}}\~.\label{cliffordstar2prime}
\eea
The original antisymplectic Clifford algebra \e{cliffordstar1} looks even more
complicated:
\beq
\Hf\Gamma^{A}\star\Gamma^{B}
-(-1)^{(\eps_{A}+1)(\eps_{B}+1)}(A \leftrightarrow B)
\=\SSS_{+}^{AB}-E_{}^{AB}{\bf 1}+ \PP^{A}\PP^{B}\star^{-1}\~.
\label{cliffordstar1prime}
\eeq
One idea would be to try to correct the Clifford algebra \e{cliffordstar2prime}
by adding higher--order terms \mb{\cO(\star^{-2})} to the Berezin--Fradkin
operator representation \e{bfoprep}, but unfortunately there is no obvious way
to do that. Another idea is to take the limit \mb{\star^{-1}\to 0} in some 
appropriate way at the end of the calculations. The approach that we shall
pursuit in this paper is to take \mb{\theta\equiv\star^{-1}}
as a fundamental object, \ie forgetting that it originally was an inverse of 
\mb{\star}, and then assume that it is nilpotent \mb{\theta^{2}=\star^{-2}=0}. 
Then the \mb{\Gamma^{A}} matrices and the \mb{\theta} variable commute
\mb{[\Gamma^{A},\theta]=0}, the Riemannian metric \e{ncgab} becomes an
ordinary commutative structure, and the Clifford algebra \e{cliffordstar2} is
restored. The price is that the Fermionic \mb{\star} multiplication
\e{cliffordstar1}, which ironically was our initial clue, does not exist.

\def\thesubsection{\thesection.\arabic{subsection}}
\section{General Spin Theory}
\label{secgeneralspin}

\subsection{Spin Manifold}
\label{secspinmanifold}

\noi
Let \mb{W} be a vector space of the same dimension as the manifold \mb{M}.
Let the vectors (=points) in \mb{W} have coordinates \mb{\ww^{a}} of
Grassmann--parity \mb{\eps(\ww^{a})=\eps_{a}} (and form--degree
\mb{p(\ww^{a})=0}). It is assumed that the {\em flat} index ``\mb{a}''
(denoted with a small roman letter) of the vector space \mb{W} runs over the
same index--set as the {\em curved} index ``\mb{A}'' (denoted with a capital
roman letter) of the manifold \mb{M}. In a slight misuse of notation, let
\mb{TW:=M\times W} (\resp \mb{T^{*}W:=M\times W^{*}}) denote the trivial
vector bundle over \mb{M} with the vector space \mb{W} (\resp dual vector
space \mb{W^{*}}) as fiber. Let \mb{\partial^{r}_{a}} and
\mb{\larrow{\dww^{a}}} denote dual bases in \mb{W} and \mb{W^{*}},
respectively, of Grassmann--parity
\mb{\eps(\larrow{\dww^{a}})\!=\!\eps_{a}\!=\!\eps(\partial^{r}_{a})}.
The form--parities \mb{p(\larrow{\dww^{a}})\!=\!p(\partial^{r}_{a})} are
either all \mb{0} or all \mb{1}, depending on applications, whereas a
\mb{1}--form \mb{\dww^{a}} with no arrow ``\mb{\rightarrow}'' always carries 
odd form--parity \mb{p(\dww^{a})\!=\!1} (and Grassmann--parity
\mb{\eps(\dww^{a})\!=\!\eps_{a}}).

\noi
Let us assume that \mb{M} is a spin manifold, \ie that there exists a bijective
bundle map
\bea
e&=&\partial^{r}_{a}\~e^{a}{}_{A}\larrow{\dzz^{A}}\~:\~
\Gamma(TM) \to \Gamma(TW)\~, \\ 
e^{-1}&=&\partial^{r}_{A}\~e^{A}{}_{a}\larrow{\dww^{a}}\~:\~
\Gamma(TW) \to \Gamma(TM)\~. 
\eea
The intertwining tensor field \mb{e^{a}{}_{A}} is known as a vielbein. 
(There are topological obstructions for the existence of a global vielbein.
However, it would be out of scope to describe global notions for
supermanifolds here, such as, orientability and Stiefel--Whitney classes.
The interesting topic of index theorems for Dirac operators will for similar
reasons be omitted in this paper.)

\noi
Note that the superdeterminant \mb{\sdet(e^{a}{}_{A})\neq 0} of the vielbein 
transforms as a density under general coordinate transformations. In general,
the vielbein \mb{e^{a}{}_{A}} is called compatible with the measure density
\mb{\rho}, if
\beq
\rho\~\propto\~\sdet(e^{a}{}_{A}) \label{erhocomp}
\eeq
is proportional to the vielbein superdeterminant \mb{\sdet(e^{a}{}_{A})} with
a \mb{\zz}--independent proportionality factor. In this case, the notion of
volume is unique (up to an overall rescaling).

\subsection{Spin Connection \mb{\nabla^{(\om)}=d+\om}}
\label{secspinconn}

\noi
A connection 
\mb{\nabla^{(\om)}=d+\om\~:\~\Gamma(TM)\times\Gamma(TW)\to\Gamma(TW)} 
in the bundle \mb{TW} is known as a spin connection, where 
\beq
\nabla^{(\om)}_{A}
\= \papal{\zz^{A}}+\partial^{r}_{b}\~\om^{b}{}_{Ac} \larrow{\dww^{c}}\~.
\label{nablaaomegadef}
\eeq
The total connection \mb{\nabla=d+\Gamma+\om} contains both a Christoffel 
symbol \mb{\Gamma^{B}{}_{AC}}, which acts on curved indices, and a spin 
connection \mb{\om^{b}{}_{Ac}}, which acts on flat indices.
We will always demand that the total connection \mb{\nabla} preserves
the vielbein
\beq
0 \= (\nabla^{}_{A}e^{b}{}_{C})
\= (\papal{\zz^{A}}e^{b}{}_{C})
-(-1)^{\eps_{A}\eps_{b}}e^{b}{}_{B}\~\Gamma^{B}{}_{AC}
+ \om_{A}{}^{b}{}_{c}\~e^{c}{}_{C}\~.\label{vielbeincovpres}
\eeq
This condition \e{vielbeincovpres} fixes uniquely the spin connection as
 \bea
\om^{b}{}_{Ac}&:=&\Gamma^{b}{}_{Ac}-f^{b}{}_{Ac}\~, \label{om1}\\
\om_{A}{}^{b}{}_{c}&:=&\Gamma_{A}{}^{b}{}_{c}-f_{A}{}^{b}{}_{c}
\= (-1)^{\eps_{A}\eps_{b}}\om^{b}{}_{Ac}\~,\label{omegagammaf} \\
\om_{a}{}^{b}{}_{c}&:=&\Gamma_{a}{}^{b}{}_{c}-f_{a}{}^{b}{}_{c}
\=(e^{T})_{a}{}^{A}\~\om_{A}{}^{b}{}_{c}\~,\label{om3} \\
\om^{b}{}_{ac}&:=&\Gamma^{b}{}_{ac}-f^{b}{}_{ac}
\=(-1)^{\eps_{a}\eps_{b}}\om_{a}{}^{b}{}_{c}\~,\label{om4} 
\eea
where 
\bea
\Gamma^{b}{}_{Ac}&:=&e^{b}{}_{B}\~\Gamma^{B}{}_{AC}\~e^{C}{}_{c}\~,
\label{gam1}  \\
\Gamma_{A}{}^{b}{}_{c}&:=&(-1)^{\eps_{A}\eps_{b}}\Gamma^{b}{}_{Ac}\~,
\label{gam2} \\
\Gamma_{a}{}^{b}{}_{c}&:=&(e^{T})_{a}{}^{A}\~\Gamma_{A}{}^{b}{}_{c}\~,
\label{gam3} \\
\Gamma^{b}{}_{ac}&:=&(-1)^{\eps_{a}\eps_{b}}\Gamma_{a}{}^{b}{}_{c}\~,
\label{gam4} \\
f_{A}{}^{b}{}_{c}&:=&(\papal{\zz^{A}}e^{b}{}_{D})e^{D}{}_{c}\~,
\label{fff1} \\
f^{b}{}_{Ac}&:=&(-1)^{\eps_{A}\eps_{b}} f_{A}{}^{b}{}_{c}\~,\label{fff2} \\
f_{a}{}^{b}{}_{c}&:=&(e^{T})_{a}{}^{A}\~f_{A}{}^{b}{}_{c}\~,\label{fff3} \\
f^{b}{}_{ac}&:=&(-1)^{\eps_{a}\eps_{b}}f_{a}{}^{b}{}_{c}\~.\label{fff4}
\eea
Here the transposed vielbein is
\beq
(e^{T})_{A}{}^{a}\~:=\~(-1)^{(\eps_{a}+1)\eps_{A}}e^{a}{}_{A}\~.
\label{transposede}
\eeq
The condition \e{vielbeincovpres} implies in many cases that one can transfer
concepts/objects back and forth between \mb{TM} and \mb{TW} by simply 
multiplying with appropriate factors of the vielbein. 
{}Firstly, the spin connection 
\mb{\nabla^{(\om)}_{A}:\~\Gamma(TW)\to\Gamma(TW)} can in a certain sense be
thought of as the connection
\mb{\nabla^{(\Gamma)}_{A}:\~\Gamma(TM) \to \Gamma(TM)} conjugated with the
vielbein \mb{e:\~\Gamma(TM) \to \Gamma(TW)}, \ie roughly speaking a product
of three matrices,
\bea
e\nabla^{(\Gamma)}_{A}e^{-1}
&=&\partial^{r}_{b}\~e^{b}{}_{B}\larrow{\dzz^{B}}
(\papal{\zz^{A}}+\partial^{r}_{D}\~\Gamma^{D}{}_{AE} \larrow{\dzz^{E}})\~
\partial^{r}_{C}\~e^{C}{}_{c} \larrow{\dww^{c}} \cr
&=&\papal{\zz^{A}} 
+(-1)^{\eps_{A}\eps_{D}} \partial^{r}_{b}\~e^{b}{}_{D} 
(\papal{\zz^{A}}e^{D}{}_{c}) \larrow{\dww^{c}}
+\partial^{r}_{b}\~\Gamma^{b}{}_{Ac} \larrow{\dww^{c}}
\~\equi{\e{vielbeincovpres}}\~ \nabla^{(\om)}_{A}\~.
\label{connconj}
\eea
Secondly, the torsion tensors \mb{T^{(\om)b}{}_{AC}} for the
\mb{\nabla^{(\om)}} connection is equal to the torsion tensor
\mb{T^{(\Gamma)B}{}_{AC}} for the \mb{\nabla^{(\Gamma)}} connection up to a
vielbein factor:
\beq
T^{(\om)a}{}_{BC}\=e^{a}{}_{A}\~T^{(\Gamma)A}{}_{BC}\~.
\label{torsiongammaomega1}
\eeq
This follows from 
\bea
T^{(\om)}&\equiv&\Hf \dzz^{A} \wedge \partial^{r}_{b}\~
T^{(\om)b}{}_{AC}\~\dzz^{C}
\~:=\~ [\nabla^{(\om)} \wedgecomma e]
\= [\dzz^{A} \papal{\zz^{A}}
+\dzz^{A}\~\partial^{r}_{b}\~\om^{b}{}_{Ad}\larrow{\dww^{d}}\~
\wedgecomma \partial^{r}_{c}\~e^{c}{}_{C}\~\dzz^{C} ] \cr
&=&\dzz^{A}\wedge \partial^{r}_{b} 
\left( (-1)^{\eps_{A}\eps_{b}} \papal{\zz^{A}}e^{b}{}_{C}
+\om^{b}{}_{Ac}\~e^{c}{}_{C}\right) \dzz^{C}
\~\equi{\e{vielbeincovpres}}\~ \dzz^{A} \wedge \partial^{r}_{b}\~
e^{b}{}_{B}\~\Gamma^{B}{}_{AC}\~\dzz^{C}\cr
&=& \Hf \dzz^{A} \wedge \partial^{r}_{b}\~
e^{b}{}_{B}\~T^{(\Gamma)B}{}_{AC}\~\dzz^{C}\~.\label{torsionomega2}
\eea
In particular, the two connections \mb{\nabla^{(\Gamma)}} and
\mb{\nabla^{(\om)}} are torsionfree at the same time.

\noi
Thirdly, if the \mb{\nabla^{(\Gamma)}_{A}} connection and the vielbein 
\mb{e^{a}{}_{A}} are both compatible with the density \mb{\rho}, \cf
\eqs{rhocomp}{erhocomp}, then the spin connection \mb{\nabla^{(\om)}_{A}}
becomes traceless, 
\beq
\om_{A}{}^{b}{}_{b}(-1)^{\eps_{b}} \~\equi{\e{vielbeincovpres}}\~ 0\~.
\label{tracecond}
\eeq
{}Fourthly, the two Riemann curvature tensor \mb{R^{(\Gamma)}} and 
\mb{R^{(\om)}} are related, see next Subsection~\ref{secspinriemcurv}.
{}Fifthly, the two connections \mb{\nabla^{(\Gamma)}} and \mb{\nabla^{(\om)}}
respect an additional structure, such as a Riemannian (\resp an antisymplectic)
structure at the same time, \cf Subsection~\ref{secriemspin} (\resp 
Subsection~\ref{secantisympspin}).

\subsection{Spin Curvature}
\label{secspinriemcurv}

\noi
The spin curvature \mb{R^{(\om)}} is defined as (half) the commutator 
of the \mb{\nabla^{(\om)}} connection \e{nablaaomegadef},
\bea
 R^{(\om)}&=&\Hf [\nabla^{(\om)} \wedgecomma \nabla^{(\om)}]
\= -\Hf \dzz^{B} \wedge \dzz^{A}\~\otimes\~
[\nabla^{(\om)}_{A},\nabla^{(\om)}_{B}] \cr
&=&-\Hf \dzz^{B} \wedge \dzz^{A}\~\otimes\~\partial^{r}_{d}\~
R^{(\om)d}{}_{ABc}\~\larrow{\dww^{c}}\~,\label{spinriemtensordef1}  \\
R^{(\om)d}{}_{ABc}&=&\larrow{\dww^{d}}
\left([\nabla^{(\om)}_{A},\nabla^{(\om)}_{B}]\partial^{r}_{c}\right)\cr
&=&(-1)^{\eps_{d}\eps_{A}}(\papal{\zz^{A}}\om^{d}{}_{Bc})
+\om^{d}{}_{Ae}\~\om^{e}{}_{Bc}
-(-1)^{\eps_{A}\eps_{B}}(A\leftrightarrow B)\~.\label{spinriemtensor1} 
\eea
The two types of Riemann curvature tensors \mb{R^{(\Gamma)}} and
\mb{R^{(\om)}} are equal up to conjugation with vielbein factors
\beq
R^{(\om)d}{}_{ABc}\=\~e^{d}{}_{D}\~R^{(\Gamma)D}{}_{ABC}\~e^{C}{}_{c}
\~,\label{curvconj1}
\eeq
basically because curvature is a commutator of connections,
\bea
e\~\partial^{r}_{D}\~R^{(\Gamma)D}{}_{ABC}\larrow{\dzz^{C}}\~e^{-1}
&=&e[\nabla^{(\Gamma)}_{A},\nabla^{(\Gamma)}_{B}]e^{-1}
\~\equi{\e{connconj}}\~ [\nabla^{(\om)}_{A},\nabla^{(\om)}_{B}] \cr
&=& \partial^{r}_{d}\~R^{(\om)d}{}_{ABc}\larrow{\dww^{c}}\~.\label{curvconj2}
\eea

\subsection{Covariant Tensors with Flat Indices}
\label{secflattensor}

\noi
Let 
\beq
\Omega_{mn}(W)\~:=\~\Gamma\left(\bigwedge{}^{m}(T^{*}W)
\otimes\bigvee{}^{n}(T^{*}W)\right)
\label{flatanmalgebra}
\eeq
be the vector space of \mb{(0,m\!+\!n)}--tensors
\mb{\eta_{a_{1}\cdots a_{m}b_{1}\cdots b_{n}}(\zz)} 
that are antisymmetric \wrtt first \mb{m} indices \mb{a_{1}\ldots a_{m}}, 
and symmetric \wrtt last \mb{n} indices \mb{b_{1}\ldots b_{n}}. 
As usual, it is practical to introduce a coordinate--free notation
\beq
\eta(\zz;\cc;\yy)
\= \frac{1}{m!n!}\cc^{a_{m}}\wedge\cdots\wedge \cc^{a_{1}}\~
\eta^{}_{a_{1}\cdots a_{m}b_{1}\cdots b_{n}}(\zz)\otimes \yy^{b_{n}}
\vee\cdots\vee \yy^{b_{1}}\~.
\label{flatcoordinatefree}
\eeq
Here the variables \mb{\yy^{a}} are symmetric counterparts to the
one--form basis \mb{\cc^{a}\equiv \dww^{a}}. 
\beq
\begin{array}{rclrclrcl}
\cc^{a} \wedge \cc^{b}&=& -(-1)^{\eps_{a}\eps_{b}} \cc^{b} \wedge \cc^{a}
\~,\~\~\~\~& \eps(\cc^{a}) &=& \eps_{a}\~,\~\~\~\~& p(\cc^{a}) &=& 1\~, \\
\yy^{a} \vee \yy^{b}&=& (-1)^{\eps_{a}\eps_{b}} \yy^{b} \vee \yy^{a}
\~,& \eps(\yy^{a}) &=& \eps_{a}\~,& p(\yy^{a}) &=& 0\~.
\end{array} 
\label{flatcom}
\eeq
The covariant derivative can be realized on covariant tensors 
\mb{\eta\in\Omega_{mn}(W)} by a linear differential operator
\beq
\nabla_{A}^{(t)} \~:=\~ \papal{\zz^{A}}-\om_{A}{}^{b}{}_{c}\~t^{c}{}_{b}\~,
\label{nablaaomegarealiz}
\eeq
where
\beq
t^{a}{}_{b}\~:=\~\cc^{a}\papal{\cc^{b}}+\yy^{a}\papal{\yy^{b}}
\label{tgenerator0}
\eeq
are generators of the Lie--algebra \mb{gl(W)}, which reflects infinitesimal
change of frame/basis in \mb{W}, \cf \eq{tliealg}. The relation with the
\mb{\nabla_{A}^{(T)}} realization \e{nablaagammarealiz} is
\beq
\nabla_{A}^{(T)}\eta(\zz;e^{b}{}_{B}\CC^{B};e^{c}{}_{C}\YY^{C})
\=\left.\rule[-3ex]{0ex}{3ex}(\nabla_{A}^{(t)}\eta)(\zz;\cc;\yy)\right|_{
\mbox{\footnotesize $ 
\begin{array}{c}\cc^{b}=e^{b}{}_{B}\CC^{B} \cr \yy^{c}=e^{c}{}_{C}\YY^{C}
\end{array} $}}\~, \label{ttcorrespondence}
\eeq
because of condition \e{vielbeincovpres}, where
\mb{\eta\!=\!\eta(\zz;\cc;\yy)\!\in\!\Omega_{\bullet\bullet}(W)} is a flat
covariant tensor. The relationship \e{ttcorrespondence} between the
\mb{\nabla^{(T)}} and the \mb{\nabla^{(t)}} realizations, where one puts
\mb{\cc^{b}=e^{b}{}_{B}\CC^{B}} and \mb{\yy^{c}=e^{c}{}_{C}\YY^{C}}, is of
course just a particular case of the more general correspondence \e{connconj} 
between the \mb{\nabla^{(\Gamma)}} and the \mb{\nabla^{(\om)}} connections.

\subsection{Local Gauge Transformations}
\label{seclocalgaugetransf}

\noi
Consider for simplicity a flat one--form
\mb{\eta=\eta^{}_{a}(\zz)\cc^{a}\in\Omega_{10}(W)}. The covariant derivative 
reads 
\beq
(\nabla^{}_{A}\eta)^{}_{c}
\=(\papal{\zz^{A}}\eta^{}_{c})-\eta^{}_{b}\~\om^{b}{}_{Ac}\~.
\label{etacovder}
\eeq
Under a local gauge transformation
\beq
\eta^{}_{a}\=\eta^{\prime}_{b}\~\Lambda^{b}{}_{a}
\~,\~\~\~\~\~\~\~\~\~\cc^{\prime a}\=\cc^{a}\~,
\eeq
where the group element \mb{\Lambda^{a}{}_{b}\!=\!\Lambda^{a}{}_{b}(\zz)}
is \mb{\zz}--dependent,
the spin connection \mb{\om^{b}{}_{Ac}} obeys the well--known affine 
transformation law for gauge potentials,
\beq
\Lambda^{b}{}_{a}\~\om^{a}{}_{Ac}
\=(-1)^{\eps_{A}\eps_{b}}(\papal{\zz^{A}}\Lambda^{b}{}_{c})
+\om^{\prime b}{}_{Ad}\~\Lambda^{d}{}_{c}\~,
\label{omgaugetrans}
\eeq
so that the covariant derivative transforms covariantly,
\beq
(\nabla^{}_{A}\eta)^{}_{a}
\=(\nabla^{}_{A}\eta^{\prime})^{}_{b}\~\Lambda^{b}{}_{a}\~.
\label{covderetagaugetrans}
\eeq

\section{Riemannian Spin Geometry}
\label{secriemspingeom}

\subsection{Spin Geometry}
\label{secriemspin}

\noi
Assume that the vector space \mb{W} is endowed with a constant Riemannian
metric 
\beq
g^{(0)}\=\yy^{a}\~g^{(0)}_{ab} \vee \yy^{b}\~\in\~\Omega_{02}(W)\~,
\label{defg0}
\eeq
called the {\em flat} metric. It has Grassmann--parity
\beq
\eps(g^{(0)}_{ab}) \= \eps_{a}+\eps_{b}
\~,\~\~\~\~\~\~\~\~\~
\eps(g^{(0)}) \= 0\~,\~\~\~\~\~\~\~\~\~p(g^{(0)}_{AB})\=0\~,  
\label{epsg0}
\eeq
and symmetry 
\beq
g^{(0)}_{ba} \= -(-1)^{(\eps_{a}+1)(\eps_{b}+1)}g^{(0)}_{ab}\~. \label{symg0}
\eeq
{}Furthermore, assume that
the vielbein \mb{e^{a}{}_{A}} intertwines between the curved \mb{g^{}_{AB}} 
metric and the flat \mb{g^{(0)}_{ab}} metric:
\beq
g^{}_{AB} \= (e^{T})_{A}{}^{a}\~g^{(0)}_{ab}\~e^{b}{}_{B}\~. 
\label{gg0}
\eeq
As a consequence, the canonical Riemannian density \e{riemrho} is compatible
with the vielbein, \ie it is proportional to the vielbein superdeterminant,  
\beq
\rhog\~:=\~\sqrt{\sdet(g_{AB})}
\= \sqrt{\sdet(g^{(0)}_{ab})}\~\sdet(e^{a}{}_{A})
\~\~\propto\~\~\sdet(e^{a}{}_{A})\~,
\label{riemrhoe}
\eeq
\cf \eq{erhocomp}. A spin connection \mb{\nabla^{(\om)}} is called
{\em metric}, if it preserves the flat metric,
\beq
0\=-\nabla^{(\om)}_{A}g^{(0)}_{bc}
\=\om_{A,bc}-(-1)^{(\eps_{b}+1)(\eps_{c}+1)}\om_{A,cb}\~,\label{spinconnmetric}
\eeq
\ie the lowered \mb{\om_{A,bc}} symbol should be skewsymmetric in the flat
indices. Here we have lowered the \mb{\om_{A,bc}} symbol with the flat metric
\beq
\om_{A,bc}\~:=\~(-1)^{\eps_{A}\eps_{b}}\om_{bAc}(-1)^{\eps_{c}}
\~,\~\~\~\~\~\~\~\~
\om_{bAc}(-1)^{\eps_{c}}\~:=\~g^{(0)}_{bd}\om^{d}{}_{Ac}\~.
\label{lowerriemspinconn}
\eeq
In particular, the two connections \mb{\nabla^{(\Gamma)}} and
\mb{\nabla^{(\om)}} are metric at the same time, as a consequence of the
correspondence \es{vielbeincovpres}{gg0}.
Note that we shall from now on put the \mb{\yy^{a}} variables to zero 
\mb{\yy^{a}\to 0} everywhere, in analogy with the \mb{\YY^{a}} variables of
Subsection~\ref{secriembigccyy}.

\subsection{Levi--Civita Spin Connection}
\label{seclcspinconn}

\noi
The Levi--Civita spin connection \mb{\nabla^{(\om)}} is by definition the 
unique spin connection that corresponds to the Levi--Civita connection 
\mb{\nabla^{(\Gamma)}} via the identifications \es{vielbeincovpres}{gg0}.
It is both torsionfree \mb{T^{(\om)}\!=\!0} and preserves the metric
\e{spinconnmetric}. The Levi--Civita formula for the spin connection in terms
of the vielbein reads
\beq
-2\om_{bac}\=(-1)^{\eps_{a}\eps_{b}}f_{a[bc]}
+(-1)^{(\eps_{a}+\eps_{b})\eps_{c}}f_{c[ba]}+f_{b[ac]}\~,
\label{lcspinformula}
\eeq
where
\beq
f_{bac}\~:=\~g^{(0)}_{bd}f^{d}{}_{ac}(-1)^{\eps_{c}}\~,\~\~\~\~\~\~\~\~
\om_{bac}\~:=\~g^{(0)}_{bd}\om^{d}{}_{ac}(-1)^{\eps_{c}}\~,
\eeq
and where \mb{f_{a[bc]}:=f_{abc}-(-1)^{\eps_{b}\eps_{c}}f_{acb}}, \cf
eqs.~\e{fff1}--\e{fff4}.

\subsection{First--Order \mb{\sss^{ab}} Matrices}
\label{secriem1st}

\noi
Because of the presence of the flat metric \mb{g_{(0)}^{ab}}, the symmetry of
the general linear Lie--algebra \mb{gl(W)} reduces to an orthogonal
Lie--subalgebra \mb{o(W)}. Its generators \mb{\sss^{ab}_{\mp}} read 
\beq
\sss^{ab}_{\mp}\~:=\~\cc^{a}\pp^{b} 
\~\mp\~ (-1)^{\eps_{a}\eps_{b}}(a\leftrightarrow b)
\~,\~\~\~\~\~\~\~\~\~\~\~\~\pp^{a}\~:=\~g_{(0)}^{ab} \papal{\cc^{b}}
\~,\label{riemsigma1st}
\eeq
\beq
\eps(\sss^{ab}_{\mp})\=\eps_{a}+\eps_{b}
\~,\~\~\~\~\~\~p(\sss^{ab}_{\mp})\=0\~,
\label{riemsigma1steps}
\eeq

\beq
\sss^{a}_{\mp c}\~:=\~ \sss^{ab}_{\mp}\~g^{(0)}_{bc}(-1)^{\eps_{c}}\~.
\label{riemsigma1stmixed}
\eeq
The transposed operator of a differential operator that depend on the flat
\mb{\cc^{a}}--variables is now defined to imitate integration by part. 
(This becomes important in Lemma~\ref{lemmasstate} below.)
Explicitly, the transposed fundamental operators are 
\beq
{\bf 1}^{T}\={\bf 1}\~,\~\~\~\~\~\~\~\~
(\cc^{a})^{T}\=\cc^{a}\~,\~\~\~\~\~\~\~\~
(\pp^{a})^{T}\=-\pp^{a}\~.
\eeq
Therefore the transposed \mb{\sss^{ab}_{\mp}} matrices read
\beq
(\sss^{ab}_{-})^{T}\=-\sss^{ab}_{-}\~,\~\~\~\~\~\~\~\~
(\sss^{ab}_{+})^{T}\=2g_{(0)}^{ab} {\bf 1}-\sss^{ab}_{+}\~.
\eeq
The \mb{\nabla_{A}^{(t)}} realization \e{nablaaomegarealiz} can be
identically rewritten into the following \mb{\sss^{ab}} matrix realization 
\beq
\nabla_{A}^{(\sss)}
\~:=\~\papal{\zz^{A}}-\Hf\om_{A,bc}\~\sss^{cb}_{-}(-1)^{\eps_{b}}
\= \papal{\zz^{A}}-\Hf\om_{A}{}^{b}{}_{c}\~\sss^{c}_{-b}\~,
\label{riemnablasigma1st}
\eeq
\ie \mb{\nabla_{A}^{(t)}=\nabla_{A}^{(\sss)}} for a metric spin 
connection. One gets a projection onto the \mb{\sss_{-}^{ab}} matrices (rather 
than the \mb{\sss_{+}^{ab}} matrices), because a metric spin connection 
\mb{\om_{A,bc}} is antisymmetric, \cf \eq{spinconnmetric}. Note that in the 
\mb{\sss^{ab}} representation --- not only the connection \e{riemnablasigma1st}
--- but also the curvature --- carries a minus--a--half normalization:
\beq
[\nabla^{(\sss)}_{A},\nabla^{(\sss)}_{B}]
\= -\Hf R_{AB}{}^{d}{}_{c}\~\sss^{c}_{-d}~.\label{riemcurvsigma}
\eeq
This can be explained as follows: The minus sign is caused by that the 
\mb{\sss^{ab}} representation acts on covariant tensors (as opposed to
contravariant tensors), and the factor \mb{\Hf} because the \mb{t^{a}{}_{b}}
generator \e{tgenerator0} becomes \mb{\Hf\sss^{a}_{-b}} after the metric
symmetrization.

\noi
The \mb{\sss^{ab}_{-}} matrices satisfy an \mb{o(W)} Lie--algebra:
\beq
[\sss^{ab}_{\mp},\sss^{cd}_{\mp}]
\= (-1)^{\eps_{a}(\eps_{b}+\eps_{c})}\left(
g_{(0)}^{bc}\~\sss^{ad}_{-}+\sss^{bc}_{-}\~g_{(0)}^{ad}\right)
\mp(-1)^{\eps_{a}\eps_{b}}(a\leftrightarrow b)\~.\label{riemsigma1stliealg}
\eeq

\subsection{ \mb{\gamma^{a}} Matrices And Clifford Algebras}
\label{secriemgamma}

\noi
The flat \mb{\gamma^{a}} matrices can be defined via a Berezin--Fradkin
operator representation \cite{b66,f66}
\beq
\gamma^{a}_{\lambda}\~\equiv\~\gamma^{a}\~:=\~\cc^{a}+\lambda \pp^{a}
\~,\~\~\~\~\~\~\~\~\~\~\~\~\eps(\gamma^{a})\=\eps_{a}
\~,\~\~\~\~\~\~\~\~\~\~\~\~p(\gamma^{a})\=1\~(\mod\~2)\~.\label{riemflatgamma}
\eeq
The transposed \mb{\gamma^{a}} matrices correspond to a change in the 
parameter \mb{\lambda \leftrightarrow -\lambda}:
\beq
(\gamma^{a})^{T}\~:=\~\cc^{a}-\lambda \pp^{a}\=\gamma^{a}_{-\lambda}\~.
\eeq
The \mb{\gamma^{a}} matrices satisfy a Clifford algebra
\beq
[\gamma^{a},\gamma^{b}]
\= 2 \lambda g_{(0)}^{ab} {\bf 1}\~. \label{riemflatclifalg}
\eeq
The \mb{\gamma^{a}} matrices commute with the transposed \mb{(\gamma^{b})^{T}} 
matrices
\beq
[\gamma^{a},(\gamma^{b})^{T}] \=0\~.\label{gammagammatcommute}
\eeq
Let \mb{V} be the vector space
\beq
V\~:=\~{\rm span}\~\cc^{a} \~\oplus\~ {\rm span}\~\pp^{a}
\={\rm span}\~\gamma^{a} \~\oplus\~ {\rm span}\~(\gamma^{a})^{T}\~,
\label{defv}
\eeq
and let
\beq 
T(V)\~:=\~\bigoplus_{m=0}^{\infty} V^{\otimes m}
\=({\rm span}\~{\bf 1})\~\oplus\~ V\~ \oplus\~ V\!\otimes\! V 
\~\oplus\~ V\!\otimes\! V\!\otimes\! V\~\oplus \ldots \label{deftv}
\eeq
be the corresponding tensor algebra. Let \mb{I(V)} be the two--sided ideal
generated by 
\beq
[\cc^{a}\otimescomma \cc^{b}]\~,\~\~\~\~
[\pp^{a}\otimescomma \cc^{b}]-g^{ab}{\bf 1}\~,\~\~\~\~
[\pp^{a}\otimescomma \pp^{b}]\~,
\eeq
or equivalently, the two--sided ideal generated by 
\beq
[\gamma^{a}\otimescomma \gamma^{b}]-2g^{ab}{\bf 1}\~,\~\~\~\~
[\gamma^{a}\otimescomma (\gamma^{b})^{T}]\~,\~\~\~\~
[(\gamma^{a})^{T}\otimescomma (\gamma^{b})^{T}]+2g^{ab}{\bf 1}\~.
\eeq
Then the Heisenberg algebra, or equivalently, the Clifford algebra
\mb{{\rm Cl}(V)} is isomorphic to the quotient
\beq
{\rm Cl}(V)\~\cong\~T(V)/I(V)\~.\label{clvquotient}
\eeq
Each element of \mb{{\rm Cl}(V)} is a differential operator in the
\mb{\cc^{a}}--variables, and may be Wick/normal--ordered in a unique way, so
that all the \mb{\cc}--derivatives (the \mb{\pp}'s) stands to the right of all
the \mb{\cc}'s. This is \aka \mb{\cc\pp}--ordering.

\noi
There is another important description of the Clifford algebra
\mb{{\rm Cl}(V)} as a tensor product of two (mutually commutative) Clifford 
algebras
\beq
{\rm Cl}(V)\~\cong\~{\rm Cl}(\gamma) \otimes {\rm Cl}(\gamma^{T})\~,
\label{clvtensor} 
\eeq
where
\bea
{\rm Cl}(\gamma)&=&\bigoplus_{m=0}^{\infty}{\rm span}\~
\gamma^{a_{1}}\gamma^{a_{2}}\cdots\gamma^{a_{m}}
\~\cong\~T(\gamma)/I(\gamma)~,\label{clgamma}  \\
{\rm Cl}(\gamma^{T})&=&\bigoplus_{m=0}^{\infty}{\rm span}\~
(\gamma^{a_{1}})^{T}(\gamma^{a_{2}})^{T}\cdots(\gamma^{a_{m}})^{T}
\~\cong\~T(\gamma^{T})/I(\gamma^{T})\~.
\label{clgammat}
\eea
Since the \mb{\gamma} matrices commute with the transposed \mb{\gamma^{T}}
matrices, it is possible to unshuffle an arbitrary element in \mb{{\rm Cl}(V)}
into a \mb{\gamma\gamma^{T}}--ordered form, \ie so that all the \mb{\gamma} 
matrices stand to the left of all the \mb{\gamma^{T}} matrices. {}For
instance, the \mb{\gamma\gamma^{T}}--ordered form of the \mb{\gamma^{a}} and
the \mb{(\gamma^{a})^{T}} matrices are
\bea
\gamma^{a} &=& \gamma^{a}\otimes{\bf 1}\~, \cr
(\gamma^{a})^{T} &=& {\bf 1} \otimes (\gamma^{a})^{T}\~,
\eea
respectively. {}For more complicated expressions, the
\mb{\gamma\gamma^{T}}--ordered form will in general not be unique, since \eg
the \mb{\gamma} matrices do not commute among themselves. Nevertheless, the
\mb{\gamma\gamma^{T}}--ordering bears some resemblance with, \eg the method
of holomorphic and antiholomorphic blocks in conformal field theory.

\noi
The \mb{\gamma^{a}} matrices form a fundamental representation of the
\mb{o(W)} Lie--algebra \e{riemsigma1stliealg}:
\beq
[\sss^{ab}_{\mp},\gamma^{c}]
\= \gamma^{a}_{\pm\lambda}\~g_{(0)}^{bc}
\mp(-1)^{\eps_{a}\eps_{b}}(a\leftrightarrow b)\~.
\label{riemgammasigma1st}
\eeq
As a consequence, if one commutes a metric spin connection
\e{riemnablasigma1st} with a flat \mb{\gamma^{a}} matrix, one gets
\beq
[\nabla_{A}^{(\sss)},\gamma^{b}]
\= -\om_{A}{}^{b}{}_{c}\~\gamma^{c}\~.\label{diffflatgamma}
\eeq
A curved \mb{\gamma^{A}} matrix is now defined as a flat \mb{\gamma^{a}}
matrix dressed with the inverse vielbein in the obvious way: 
\beq
\gamma^{A}\~:=\~e^{A}{}_{a}\~\gamma^{a}
\= \gamma^{a}\~(e^{T})_{a}{}^{A}
\~,\~\~\~\~\~\~\eps(\gamma^{A})\=\eps_{A}
\~,\~\~\~\~\~\~p(\gamma^{A})\=1\~(\mod\~2)\~. \label{riemcurvedgamma}
\eeq
(Similar straightforward rules applies to other objects when switching
between flat and curved indices.)

\noi
If one commutes a metric spin connection \e{riemnablasigma1st} with a curved
\mb{\gamma^{A}} matrix, one gets
\beq
[\nabla_{A}^{(\sss)},\gamma^{B}]
\= -\Gamma_{A}{}^{B}{}_{C}\~\gamma^{C}\~,\label{diffcurvedgamma}
\eeq
\cf \eqs{omegagammaf}{diffflatgamma}. The result \e{diffcurvedgamma} can be
summarized as saying that the total connection \mb{\nabla=d+\Gamma+\om}
commutes with the \mb{\gamma^{A}} matrices: \mb{[\nabla^{}_{A},\gamma^{B}]=0}.

\subsection{Dirac Operator \mb{D^{(\sss)}}}
\label{secriemdiracop}

\noi
{}For a general discussion of Dirac operators, see \eg \Ref{bgv92}.
We shall for the remainder of the Section~\ref{secriemspingeom} assume that
the connection is the Levi--Civita connection.

\noi
Central for our discussion are the \mb{\sss^{ab}} matrices \e{riemsigma1st}.
They act on flat exterior forms \mb{\eta\in\Omega_{\bullet0}(W)}, \ie
functions \mb{\eta\!=\!\eta(\zz;\cc)} of the \mb{\zz^{A}} and \mb{\cc^{a}}
variables. 

\noi
The Dirac operator \mb{D^{(\sss)}} in the \mb{\sss^{ab}} representation 
\e{riemnablasigma1st} is a \mb{\gamma^{A}} matrix \e{riemcurvedgamma}
times a covariant derivative \e{riemnablasigma1st}
\beq
D^{(\sss)}\~:=\~\gamma^{A}\nabla^{(\sss)}_{A}
\~,\~\~\~\~\~\~\eps(D^{(\sss)})\=0
\~,\~\~\~\~\~\~p(D^{(\sss)})\=1\~(\mod\~2)\~.
\label{riemdiracop}
\eeq
The Laplace operator \mb{\Delta_{\rhog}^{(\sss)}} in the \mb{\sss^{ab}}
representation \e{riemnablasigma1st} is
\bea
\Delta_{\rhog}^{(\sss)}
&:=&(-1)^{\eps_{A}}\nabla^{}_{A}g^{AB}\nabla^{(\sss)}_{B}
\= (-1)^{\eps_{A}}\nabla^{(\sss)}_{A} g^{AB}\nabla^{(\sss)}_{B}
+\Gamma^{A}{}_{AC}\~ g^{CB}\~\nabla^{(\sss)}_{B} \cr
&=& \frac{(-1)^{\eps_{A}}}{\rhog}\nabla^{(\sss)}_{A}
\rhog g^{AB}\nabla^{(\sss)}_{B}\~.\label{riemlaplaceopsigma}
\eea

\begin{theorem}[\mb{\cc\pp}--ordered Weitzenb\"ock formula for flat exterior 
forms]
The difference between the square of the Dirac operator \mb{D^{(\sss)}}
and the Laplace operator \mb{\Delta_{\rhog}^{(\sss)}} in the \mb{\sss^{ab}} 
representation \e{riemnablasigma1st} is
\bea
D^{(\sss)}D^{(\sss)} \~-\~ \lambda \Delta_{\rhog}^{(\sss)} 
&=& -\frac{\lambda}{4}\sss_{-}^{BA}\~R_{AB,CD}\~
\sss_{-}^{DC}(-1)^{\eps_{C}+\eps_{D}} \label{riemslw1a} \\ 
&=& - \lambda\cc^{A}\~R_{AB}\~\pp^{B}
+\frac{\lambda}{2}\cc^{B}\cc^{A}\~R_{AB,CD}\~
\pp^{D}\pp^{C}(-1)^{\eps_{C}+\eps_{D}} \~.  \label{riemslw1b}
\eea
\label{theoremriem1}
\end{theorem}

\noi
{\sc Proof of Theorem~\ref{theoremriem1}:}\~
Almost identical to the proof of Theorem~\ref{theoremriem0} because of 
\eq{curvconj1}.
\proofbox

\subsection{Second--Order \mb{\sig^{ab}} Matrices}
\label{secriem2nd}

\noi
We now replace the first--order \mb{\sss_{\mp}^{ab}} matrices \e{riemsigma1st}
with second--order matrices:
\beq
\sig^{ab}_{\mp}(\lambda)\~\equiv\~\sig^{ab}_{\mp}
\~:=\~\frac{1}{4\lambda}\gamma^{a}\gamma^{b}
\mp (-1)^{\eps_{a}\eps_{b}}(a\leftrightarrow b)
\=\sig^{ab}_{\mp} \otimes {\bf 1}\~,
\label{riemsigma2nd}
\eeq
\beq
\eps(\sig^{ab}_{\mp})\=\eps_{a}+\eps_{b}
\~,\~\~\~\~\~\~p(\sig^{ab}_{\mp})\=0\~,
\label{riemsigma2steps}
\eeq
\beq
\sig^{a}_{\mp c}\~:=\~ \sig^{ab}_{\mp}\~g^{(0)}_{bc}(-1)^{\eps_{c}}\~. 
\label{riemsigma2nmdixed}
\eeq
(The names first-- and second--order refer to the number of 
\mb{\cc^{a}}--derivatives.) The transposed \mb{\sig^{ab}_{\mp}} matrices read
\beq
(\sig^{ab}_{\mp})^{T}\=\pm\frac{1}{4\lambda}(\gamma^{a})^{T}(\gamma^{b})^{T}
\mp (-1)^{\eps_{a}\eps_{b}}(a\leftrightarrow b)
\=\mp\sig^{ab}_{\mp}(-\lambda)\={\bf 1} \otimes (\sig^{ab}_{\mp})^{T}\~.
\label{riemsigma2ndt}
\eeq
In the last expression of \eqs{riemsigma2nd}{riemsigma2ndt} we wrote the
\mb{\sig^{ab}_{\mp}} and the \mb{(\sig^{ab}_{\mp})^{T}} matrices on a
\mb{\gamma\gamma^{T}}--ordered form. In particular, the \mb{\sig^{ab}_{\mp}} 
matrices decouple completely from the \mb{(\sig^{ab}_{\mp})^{T}} matrices, 
\beq
[\sig^{ab}_{\mp},(\sig^{cd}_{\mp})^{T}]\=0\~,\~\~\~\~\~\~\~\~
[\sig^{ab}_{\mp},(\sig^{cd}_{\pm})^{T}]\=0\~.
\eeq
On one hand, the matrices
\beq
\sig^{ab}_{-}
\=\frac{1}{4\lambda}\{\gamma^{a},\gamma^{b}\}^{}_{+}
\=\frac{1}{2\lambda}\cc^{a}\cc^{b}+\Hf\sss_{-}^{ab}
+\frac{\lambda}{2}\pp^{a}\pp^{b} \label{riemflatsigma2ndm}
\eeq
satisfy precisely the same non--Abelian \mb{o(W)} Lie--algebra
\e{riemsigma1stliealg} and fundamental representation \e{riemgammasigma1st}
as the \mb{\sss^{ab}_{-}} matrices. On the other hand, the matrices
\beq
\sig^{ab}_{+}
\~:=\~\frac{1}{4\lambda}[\gamma^{a},\gamma^{b}]
\~\equi{\e{riemflatclifalg}}\~\Hf g_{(0)}^{ab} {\bf 1}
\eeq
are proportional to the identity operator, and thus Abelian.

\noi
The \mb{\sss_{-}^{ab}} matrices can be expressed in terms of the
\mb{\sig_{-}^{ab}} matrices and their transposed, 
\beq
\sss_{-}^{ab}
\= \sig_{-}^{ab}+\sig_{-}^{ab}(-\lambda)
\= \sig_{-}^{ab}\otimes{\bf 1} - {\bf 1}\otimes(\sig_{-}^{ab})^{T}\~,
\eeq
as a consequence of \eq{riemflatsigma2ndm}. In contrast, the \mb{\sss_{+}^{ab}}
matrices can {\em not} be expressed in terms of the \mb{\sig_{\mp}^{ab}}
matrices and their transposed.

\noi
The first--order \mb{\nabla_{A}^{(\sss)}} realization \e{riemnablasigma1st}
can be identically rewritten into the following second--order
\mb{\sig\sig^{T}} matrix realization 
\beq
\nabla_{A}^{(\sig\sig^{T})}
\~:=\~\papal{\zz^{A}}
-\Hf\om_{A,bc}\left(\sig^{cb}_{-}\otimes{\bf 1}
-{\bf 1}\otimes(\sig^{cb}_{-})^{T}\right)(-1)^{\eps_{b}}
\=\papal{\zz^{A}}-\Hf\om_{A}{}^{b}{}_{c}\left(\sig^{c}_{-b}\otimes{\bf 1}
-{\bf 1}\otimes(\sig^{c}_{-b})^{T}\right)\~,
\label{riemnablasigma2nd}
\eeq
\ie \mb{\nabla_{A}^{(t)}=\nabla_{A}^{(\sss)}=\nabla_{A}^{(\sig\sig^{T})}} for
a metric spin connection. In contrast, the first--order
\mb{\nabla_{A}^{(\SSS)}} realization \e{riemnablabigsigma1st} does in general
not have a second--order formulation for a metric connection, even if the
manifold is a spin manifold, \cf Appendix~\ref{seccurvedriem2nd}. This is
despite the fact that the first--order realizations \mb{\nabla_{A}^{(\SSS)}}
and \mb{\nabla_{A}^{(\sss)}} are closely related via condition
\e{vielbeincovpres},
\beq
\nabla_{A}^{(\SSS)}\eta(\zz;e^{b}{}_{B}\CC^{B})
\=\left.\rule[-2ex]{0ex}{2ex}(\nabla_{A}^{(\sss)}\eta)(\zz;\cc)\right|_{
\cc^{b}=e^{b}{}_{B}\CC^{B}}\~, \label{sscorrespondence}
\eeq
where \mb{\eta\!=\!\eta(\zz;\cc;\yy)\!\in\!\Omega_{\bullet 0}(W)} is a flat
exterior form. Here the \mb{\SSS_{\mp}^{AB}} and \mb{\sss_{\mp}^{ab}} matrices
act by adjoint action on the \mb{\CC^{C}} and \mb{\cc^{c}} variables as
\beq
[\SSS_{\mp}^{AB},\CC^{C}]
\= \CC^{A}g_{}^{BC}\mp(-1)^{\eps_{A}\eps_{B}}(A \leftrightarrow B)
\~,\~\~\~\~\~\~\~\~
[\sss_{\mp}^{ab},\cc^{c}]
\= \cc^{a}g_{(0)}^{bc}\mp(-1)^{\eps_{a}\eps_{b}}(a \leftrightarrow b)\~,
\eeq
\cf \eqs{curvedriemsigma1st}{riemsigma1st}, respectively. The crucial
difference is that the \mb{\nabla_{A}^{(\SSS)}} realization
\e{riemnablabigsigma1st} contains a non--trivial \mb{\SSS_{+}} sector, while
the \mb{\nabla_{A}^{(\sss)}} realization \e{riemnablasigma1st} has {\em no}
\mb{\sss_{+}} sector. This has its root in the fact that the flat metric
condition \e{spinconnmetric} is an algebraic condition, while the curved metric
condition \e{connmetric} is a differential condition. (Curiously, it is just
opposite for the torsionfree conditions: the curved torsionfree condition is
an algebraic condition, while the flat torsionfree condition is a differential
condition, \cf \eqs{torsiongamma}{torsionomega2}.)

\subsection{Lichnerowicz' Formula}
\label{secriemlformula}

\noi
It is convenient to define a totally symmetrized combination of three
\mb{\gamma^{a}} matrices as
\beq
\gamma^{a^{}_{1}a^{}_{2}a^{}_{3}}
\~:=\~\frac{1}{3!}\sum_{\pi\in S_{3}}(-1)^{\eps_{\pi,a}}
\gamma^{a^{}_{\pi(1)}}\~\gamma^{a^{}_{\pi(2)}}\~\gamma^{a^{}_{\pi(3)}}\~,
\label{threegammadef}
\eeq
where \mb{(-1)^{\eps_{\pi,a}}} is the sign factor that arises when one does a
\mb{\pi}--permutation of three supercommuting objects with the same Grassmann--
and form--parity as the \mb{\gamma^{a}} matrices, say, the \mb{\cc}'s 
\beq
\cc^{a^{}_{1}} \wedge \cc^{a^{}_{2}} \wedge \cc^{a^{}_{3}} 
\= (-1)^{\eps_{\pi,a}} 
\cc^{a^{}_{\pi(1)}} \wedge \cc^{a^{}_{\pi(2)}} \wedge \cc^{a^{}_{\pi(3)}}\~,
\eeq
\cf \e{flatcom}.
The symmetrized \mb{\gamma^{abc}} matrix can be reduced \wthot Clifford
relation \e{riemflatclifalg} as 
\beq
\gamma^{abc}
\= \gamma^{a}\gamma^{b}\gamma^{c}-\lambda g_{(0)}^{ab}\~\gamma^{c}
+(-1)^{\eps_{b}\eps_{c}}\lambda g_{(0)}^{ac}\~\gamma^{b}
-\gamma^{a}\~\lambda g_{(0)}^{bc}\~.
\label{threegammareduc}
\eeq

\begin{theorem}[\mb{\gamma\gamma^{T}}--ordered Lichnerowicz' formula
\cite{lich63}] The square of the Dirac operator \mb{D^{(\sig\sig^{T})}} in the
\mb{\sig\sig^{T}} representation \e{riemsigma2nd} is
\beq
D^{(\sig\sig^{T})}D^{(\sig\sig^{T})}
\= \lambda \Delta_{\rhog}^{(\sig\sig^{T})} - \frac{\lambda}{4}R
+\frac{\lambda}{2}\sig^{BA}_{-}\~R_{AB,CD}\otimes
(\sig^{DC}_{-})^{T}(-1)^{\eps_{C}+\eps_{D}}\~.  
\label{rieml2}
\eeq
\label{theoremriem2}
\end{theorem}

\noi
{\sc Proof of Theorem~\ref{theoremriem2}:}\~
One derives that the square of the Dirac operator \mb{D^{(\sig\sig^{T})}} is
the Laplacian \mb{\Delta_{\rhog}^{(\sig\sig^{T})}} plus a curvature term, by
proceeding along the lines of the proof of Theorem~\ref{theoremriem0}:
\beq
D^{(\sig\sig^{T})}D^{(\sig\sig^{T})} 
\= \Hf [D^{(\sig\sig^{T})},D^{(\sig\sig^{T})}]
\=\lambda \Delta_{\rhog}^{(\sig\sig^{T})}
-\Hf\gamma^{B}\gamma^{A}
[\nabla^{(\sig\sig^{T})}_{A},\nabla^{(\sig\sig^{T})}_{B}]\~.
\eeq
When one \mb{\gamma\gamma^{T}}--decomposes the curvature term, it splits in two
parts:
\beq
-\Hf\gamma^{B}\gamma^{A}
[\nabla^{(\sig\sig^{T})}_{A},\nabla^{(\sig\sig^{T})}_{B}]
\=\frac{1}{4}\gamma^{B}\gamma^{A}\~R_{AB}{}^{d}{}_{c}
\left(\sig^{c}_{-d}\otimes{\bf 1}-{\bf 1}\otimes(\sig^{c}_{-d})^{T}\right)
\=III + III^{T}\~,
\eeq
where 
\beq
III^{T}\~:=\~\frac{\lambda}{2}\sig^{BA}_{-}\~R_{AB,CD}\otimes
(\sig^{DC}_{-})^{T}(-1)^{\eps_{C}+\eps_{D}}\~,  
\eeq
and
\bea
III&:=& -\frac{1}{4}\gamma^{B}\gamma^{A}\~R_{AB,CD}\~
\sig^{DC}_{-}(-1)^{\eps_{C}+\eps_{D}}
\= -\frac{1}{8\lambda}\gamma^{B}\gamma^{A}\~R_{AB,CD}
\~\gamma^{D}\gamma^{C} (-1)^{\eps_{C}+\eps_{D}} \cr
&=& \frac{1}{8\lambda}(-1)^{(\eps_{A}+\eps_{B})\eps_{C}}
\gamma^{B}\gamma^{A}\gamma^{C}
\~R_{AB,CD}\~\gamma^{D}(-1)^{\eps_{D}} \cr
&\equi{\e{threegammareduc}}& \frac{1}{8\lambda}\left(
\gamma^{CBA}
+\gamma^{C}\~\lambda g^{BA}
-\lambda g^{CB}\~\gamma^{A}
+(-1)^{\eps_{A}\eps_{B}}
\lambda g^{CA}\~\gamma^{B}
\right)R_{AB,CD}\~\gamma^{D}(-1)^{\eps_{D}}\cr
&=& -\frac{1}{4}g^{CB}\~\gamma^{A}\~
R_{AB,CD}\~\gamma^{D}(-1)^{\eps_{D}}
\= \frac{1}{4}(-1)^{(\eps_{A}+\eps_{B})(\eps_{D}+1)}R_{ABD}{}^{B}\~
\gamma^{A}\gamma^{D}(-1)^{\eps_{D}} \cr
&=& -\frac{1}{4}R_{DA}\~\gamma^{A}\gamma^{D}(-1)^{\eps_{D}}
\= -\frac{\lambda}{4}R_{DA}\~g^{AD}(-1)^{\eps_{D}}
\= -\frac{\lambda}{4}R\~.
\eea
Here the first Bianchi identity \e{riem1bianchiid} was used one time.
\proofbox

\subsection{Clifford Representations}
\label{seccliffordrepr}

\noi
The spinor representations \mb{\cS} and \mb{\cS^{T}} can be defined as Fock
spaces
\bea
\cS&:=&{\rm Cl}(\gamma)|0\rangle\~\~\~
\=\bigoplus_{m=0}^{\infty}{\rm span}\~
\cc^{a_{1}}\cc^{a_{2}}\cdots\cc^{a_{m}}|0\rangle
\~,\~\~\~\~\~\~\~\~\~\~\~\~\pp^{a}|0\rangle\=0\~,\label{srepr} \\
\cS^{T}&:=&{\rm Cl}(\gamma^{T})|0^{T}\rangle
\=\bigoplus_{m=0}^{\infty}{\rm span}\~
\pp^{a_{1}}\pp^{a_{2}}\cdots\pp^{a_{m}}|0^{T}\rangle
\~,\~\~\~\~\~\~\~\~\cc^{a}|0^{T}\rangle \=0\~.\label{strepr}
\eea
The constraints \mb{\pp^{a}|0\rangle\!=\!0} (\resp
\mb{\cc^{a}|0^{T}\rangle\!=\!0}) are consistent, because the \mb{\pp^{a}}'s
(\resp the \mb{\cc^{a}}'s) commute. 
The representation \es{srepr}{strepr} are of course just two possibilities
out of infinitely many equivalent choices of Fock space representations. A 
different class of vacua \mb{|1\rangle} and \mb{|1^{T}\rangle} are defined via
\beq
\sig_{-}^{ab}|1\rangle\=0\~,\~\~\~\~\~\~\~\~\~\~\~
(\sig_{-}^{ab})^{T}|1^{T}\rangle\=0\~.\label{vacuumone}
\eeq
They both represent the singlet/trivial representation of the orthogonal 
Lie--group \mb{O(W)}. Again, the constraints \e{vacuumone} for the vacua are
consistent, since the \mb{\sig_{-}^{ab}} (\resp the \mb{(\sig_{-}^{ab})^{T}}) 
matrices form Lie--algebras. 
All the above constraints are examples of first--class constraints. More 
generally, assume that \mb{|\Omega\rangle} and \mb{|\Omega^{T}\rangle} are two
arbitrary consistent vacua (that are not necessarily related).
Let \mb{\cV} and \mb{\cV^{T}} denote the corresponding vector spaces
\beq
\cV\~:=\~{\rm Cl}(\gamma)|\Omega\rangle~,\~\~\~\~\~\~\~\~\~\~\~ 
\cV^{T}\~:=\~{\rm Cl}(\gamma^{T})|\Omega^{T}\rangle~.\label{vrepr}
\eeq
The Clifford algebra
\mb{{\rm Cl}(V)\cong{\rm Cl}(\gamma)\otimes{\rm Cl}(\gamma^{T})} is defined
to act on the tensor product \mb{\cV\otimes\cV^{T}} via a
\mb{\gamma\gamma^{T}}--ordered form, \ie the \mb{\gamma^{a}} matrices act on 
the first factor \mb{\cV} and the transposed \mb{(\gamma^{a})^{T}} matrices
act on the second factor \mb{\cV^{T}}. In detail, if \mb{|v\rangle\in \cV} and
\mb{|v^{T}\rangle\in \cV^{T}} are two (not necessarily related) states, then 
\bea
\gamma^{a}.(|v\rangle \otimes |v^{T}\rangle)
&:=&(\gamma^{a}|v\rangle) \otimes |v^{T}\rangle\~, 
\label{cliffordaction} \\
(\gamma^{a})^{T}.(|v\rangle \otimes |v^{T}\rangle)&:=&
(-1)^{\vec{\eps}(\gamma^{a})\cdot\vec{\eps}(v)}|v\rangle \otimes 
(\gamma^{a})^{T}|v^{T}\rangle\~.\label{cliffordactiont}
\eea
By definition, \mb{\cV} is a Clifford bundle, while \mb{\cV^{T}} is a
dual/contragredient Clifford bundle.

\noi
A Lie--algebra element \mb{x\in so(W)} is of the form
\beq
x\=\Hf (-1)^{\eps_{a}}x^{}_{ab}\sss_{-}^{ba}
\=\Hf x^{a}{}_{b}\sss^{b}_{-a}
\=\Hf x^{a}{}_{b}\left(\sig^{b}_{-a}\otimes{\bf 1}
-{\bf 1}\otimes(\sig^{b}_{-a})^{T}\right)\~,
\eeq
where
\beq
x^{}_{ab}\=(-1)^{(\eps_{a}+1)(\eps_{b}+1)}(a \leftrightarrow b)
\~,\~\~\~\~\~\~\~\~
x^{a}{}_{c}\~:= g_{(0)}^{ab}x^{}_{bc}\~.
\eeq
A \mb{\gamma\gamma^{T}}--ordered form of a generic special orthogonal
Lie--group element \mb{g\!=\!e^{x}\in SO(W)} is
\beq
\exp\left[\Hf x^{a}{}_{b}\sss^{b}_{-a}\right]\=
\exp\left[\Hf x^{a}{}_{b}\sig^{b}_{-a}\right]\~\otimes\~
\exp\left[-\Hf x^{c}{}_{d}(\sig^{d}_{-c})^{T}\right]\~.
\eeq
In this way the vector space \mb{\cV^{T}} becomes a dual/contragredient 
representation of the special orthogonal Lie--group \mb{SO(W)}, hence the name.

\subsection{Intertwining Operator}
\label{secintertwining}

\noi
Consider the intertwining operator
\beq
s\~:=\~ \int \! d^{N}\theta\~ e^{\theta^{}_{a}\gamma^{a}}
\otimes e^{\theta^{}_{b}(\gamma^{b})^{T}}\~, \label{operatorsinglet}
\eeq
where \mb{\theta^{}_{a}} are integration variables with Grassmann--parity
\mb{\eps(\theta^{}_{a})\!=\!\eps_{a}} and form--parity
\mb{p(\theta^{}_{a})=1\~(\mod\~2)}.

\begin{lemma}
The intertwining operator \mb{s} is invariant under the adjoint action 
\mb{e^{x}s e^{-x}\!=\!s} of the special orthogonal Lie--group \mb{SO(W)}.
Equivalently, the intertwining operator \mb{s} commute with the 
\mb{so(W)} Lie--algebra generators \mb{[\sss^{ab}_{-},s]\!=\!0}.
\label{lemmasop}
\end{lemma}

\noi
{\sc Proof of Lemma~\ref{lemmasop}:}\~The adjoint action rotates the
\mb{\gamma^{a}} matrices,
\bea
\exp\left[\Hf x^{c}{}_{d}\sig^{d}_{-c}\right]
\gamma^{a}\exp\left[-\Hf x^{e}{}_{f}\sig^{f}_{-e}\right]
&=&(e^x)^{a}{}_{b}\gamma^{b}\~,\cr
\exp\left[-\Hf x^{c}{}_{d}(\sig^{d}_{-c})^{T}\right]
(\gamma^{a})^{T}\exp\left[\Hf x^{e}{}_{f}(\sig^{f}_{-e})^{T}\right]
&=&(e^x)^{a}{}_{b}(\gamma^{b})^{T}\~,
\eea
where
\beq
(e^x)^{a}{}_{b}\~:=\~\delta^{a}_{b}+ x^{a}{}_{b}
+\frac{1}{2!}x^{a}{}_{c} x^{c}{}_{b}
+\frac{1}{3!}x^{a}{}_{c} x^{c}{}_{d}x^{d}{}_{b}
+\frac{1}{4!}x^{a}{}_{c} x^{c}{}_{d}x^{d}{}_{e}x^{e}{}_{b}+\ldots\~.
\eeq
Hence one may change integration variables 
\mb{\theta^{}_{a}\to\theta^{\prime}_{b}=\theta^{}_{a}(e^x)^{a}{}_{b}}
in the integral \e{operatorsinglet}. The Jacobian vanishes for special
orthogonal transformations 
\beq 
\ln\sdet(e^x)^{a}{}_{b}\=(-1)^{\eps_{a}} x^{a}{}_{a}
\=(-1)^{\eps_{a}}g_{(0)}^{ab}x^{}_{ba}\=0\~.
\eeq
\proofbox

\begin{lemma}
The corresponding intertwining state
\beq
\sstate\~:=\~s.(|\Omega\rangle\otimes|\Omega^{T}\rangle)
\=\int \! d^{N}\theta\~ e^{\theta^{}_{a}\gamma^{a}}|\Omega\rangle
\otimes e^{\theta^{}_{b}(\gamma^{b})^{T}}|\Omega^{T}\rangle
\label{statesinglet}
\eeq
is invariant under the action of the special orthogonal Lie--group \mb{SO(W)}.
Equivalently, the \mb{so(W)} Lie--algebra generators \mb{\sss^{ab}_{-}} 
annihilate the intertwining state \mb{\sss^{ab}_{-}\sstate\!=\!0}.
\label{lemmasstate}
\end{lemma}

\noi
{\sc Proof of Lemma~\ref{lemmasstate}:}\~
\bea
e^{x}\sstate&=&
\int \! d^{N}\theta\~e^{\theta^{}_{a}\gamma^{a}}
\exp\left[\frac{1}{4\lambda}(-1)^{\eps_{c}}
x^{}_{cd}\gamma^{d}\gamma^{c}\right]|\Omega\rangle
\otimes e^{\theta^{}_{b}(\gamma^{b})^{T}}
\exp\left[-\frac{1}{4\lambda}(-1)^{\eps_{e}}
x^{}_{ef}(\gamma^{f})^{T}(\gamma^{e})^{T}\right]|\Omega^{T}\rangle \cr
&=&\int \! d^{N}\theta\~
\exp\left[\frac{1}{4\lambda}(-1)^{\eps_{c}}
x^{}_{cd}\tilde{\gamma}^{d}\tilde{\gamma}^{c}\right] 
e^{\theta^{}_{a}\gamma^{a}}|\Omega\rangle
\otimes \exp\left[-\frac{1}{4\lambda}(-1)^{\eps_{e}}
x^{}_{ef}(\tilde{\gamma}^{f})^{T}(\tilde{\gamma}^{e})^{T}\right]
e^{\theta^{}_{b}(\gamma^{b})^{T}}|\Omega^{T}\rangle \cr
&=&\sstate\~,\label{proofoneinv}
\eea
where we have introduced (a kind of) Fourier transformed \mb{\gamma} matrices
\beq
\tilde{\gamma}^{a}
\~:=\~\papal{\theta^{}_{a}}+g_{(0)}^{ab}\theta^{}_{b}
\~,\~\~\~\~\~\~\~\~\~\~\~\~
(\tilde{\gamma}^{a})^{T}
\~:=\~-\papal{\theta^{}_{a}}+g_{(0)}^{ab}\theta^{}_{b}\~,
\eeq
which satisfy
\beq
\tilde{\gamma}^{a}\exp\left[\theta^{}_{b}\gamma^{b}\right]
\=\exp\left[\theta^{}_{b}\gamma^{b}\right]\gamma^{a}\~,\~\~\~\~\~\~\~\~\~\~\~\~
-(\tilde{\gamma}^{a})^{T}\exp\left[\theta^{}_{b}(\gamma^{b})^{T}\right]
\=\exp\left[\theta^{}_{b}(\gamma^{b})^{T}\right](\gamma^{a})^{T}\~.
\eeq
In the last equality of \eq{proofoneinv}, we performed integration by part,
which turns \mb{\tilde{\gamma}^{a}} into \mb{(\tilde{\gamma}^{a})^{T}}, and
vice--versa.
\proofbox

\noi
The algebra bundle \e{clvquotient} of differential operators in the 
\mb{\cc^{a}}--variables, or equivalently polynomials in \mb{\gamma} and 
\mb{\gamma^{T}}, is isomorphic to the bispinor bundle \mb{\cS\otimes \cS^{T}}. 
The bundle isomorphism is
\beq
{\rm Cl}(V)\~\cong\~{\rm Cl}(\gamma)\otimes{\rm Cl}(\gamma^{T})
\~\ni\~ F\~ \stackrel{\cong}{\mapsto}\~F\sstate
\~\in\~\cS\otimes \cS^{T}\~\cong\~{\rm End}(\cS)\~.
\eeq
The bispinor bundle \mb{\cS\otimes \cS^{T}\cong{\rm End}(\cS)} is, in turn,
isomorphic (as vector bundles) to the bundle of endomorphisms: \mb{\cS\to\cS}. 
Let us also mention that the Weyl symbol \mb{\stackrel{\cong}{\mapsto}}
operator isomorphism
\mb{\bigwedge{}^{\bullet}(V)\stackrel{\cong}{\to}{\rm Cl}(V)} from the exterior
algebra \mb{(\bigwedge{}^{\bullet}(V);*)}, equipped with the Groenewold/Moyal
\mb{*} product, to the Heisenberg algebra \mb{({\rm Cl}(V);\circ)}, is known as
the Chevalley isomorphism in the context of Clifford algebras.

\subsection{Schr\"odinger--Lichnerowicz' Formula}
\label{secriemslformula}

\noi
We will be interested in how the Dirac operator acts on a Clifford bundle
\mb{\cV\otimes|1^{T}\rangle\cong\cV} and a tensor Clifford bundle 
\mb{\cV\otimes \cV^{T}}.

\begin{theorem}[Schr\"odinger--Lichnerowicz' formula \cite{schroed32,lich63}]
On a Clifford bundle \mb{\cV\otimes|1^{T}\rangle\cong\cV}, the square of the
Dirac operator \mb{D^{(\sig)}} is equal to the Laplacian
\mb{\Delta_{\rhog}^{(\sig)}} minus a quarter of the scalar curvature \mb{R},
\beq
D^{(\sig)}D^{(\sig)}
\= \lambda \Delta_{\rhog}^{(\sig)} - \frac{\lambda}{4}R\~.  
\label{riemsl2}
\eeq
\label{theoremriem3}
\end{theorem}

\noi
{\sc Proof of Theorem~\ref{theoremriem3}:}\~This is a Corollary to
Lichnerowicz' formula \e{rieml2}.
\proofbox

\noi
The Schr\"odinger--Lichnerowicz' formula \e{riemsl2} corresponds to naively
substituting the first--order matrices \mb{\sss_{-}^{ab}\to\sig_{-}^{ab}} in
the \mb{\nabla^{(\sss)}} realization \e{riemnablasigma1st} with the
second--order matrices \mb{\sig_{-}^{ab}}. The analysis in
Subsections~\ref{secriem2nd} and \ref{seccliffordrepr} shows in detail why
this replacement is geometrically sound and in fact very natural.

\begin{theorem}
The square of the Dirac operator \mb{D^{(\sig\sig^{T})}} on a tensor Clifford
bundle \mb{\cV\otimes\cV^{T}} becomes equal to the Laplace--Beltrami operator
\mb{\Deltarhog} when it is projected on the singlet representation
\mb{\sstate},
\beq
D^{(\sig\sig^{T})}D^{(\sig\sig^{T})}f\sstate
\=\lambda(\Deltarhog f)\sstate~,  
\label{riemspinorscalar}
\eeq
where \mb{f\!=\!f(\zz)} is an arbitrary scalar function.
\label{theoremriem4}
\end{theorem}

\noi
{\sc Proof of Theorem~\ref{theoremriem4}:}\~This is a Corollary to the
Weitzenb\"ock formula \e{riemslw1a}.
\proofbox

\section{Antisymplectic Spin Geometry}
\label{secantisympspingeom}

\subsection{Spin Geometry}
\label{secantisympspin}

\noi
Assume that the vector space \mb{W} is endowed with a constant antisymplectic
metric 
\beq
E^{(0)}\=\Hf \cc^{a}\~E^{(0)}_{ab} \wedge \cc^{b}
\=-\Hf  E^{(0)}_{ab}\~\cc^{b} \wedge \cc^{a}\~\in\~\Omega_{20}(W)\~,
\~\~\~ \label{defe0}
\eeq
called the {\em flat} metric. It has Grassmann--parity
\beq
\eps(E^{(0)}_{ab}) \= \eps_{a}+\eps_{b}+1
\~,\~\~\~\~\~\~\~\~\~
\eps(E^{(0)}) \= 1\~,\~\~\~\~\~\~\~\~\~p(E^{(0)}_{AB})\=0\~, 
\label{epse0}
\eeq
and symmetry 
\beq
E^{(0)}_{ba} \= -(-1)^{\eps_{a}\eps_{b}}E^{(0)}_{ab}\~. \label{syme0}
\eeq
{}Furthermore, assume that the vielbein \mb{e^{a}{}_{A}} intertwines between
the curved \mb{E^{}_{AB}} metric and the flat \mb{E^{(0)}_{ab}} metric:
\beq
E^{}_{AB} \= (e^{T})_{A}{}^{a}\~E^{(0)}_{ab}\~e^{b}{}_{B}\~. 
\label{ee0}
\eeq
A spin connection \mb{\nabla^{(\om)}} is called {\em antisymplectic}, if
it preserves the flat metric,
\beq
0\=-\nabla^{(\om)}_{A}E^{(0)}_{bc}
\=\om_{A,bc}-(-1)^{\eps_{b}\eps_{c}}\om_{A,cb}\~,\label{spinconnantisymp}
\eeq
\ie the lowered \mb{\om_{A,bc}} symbol should be symmetric in the flat indices.
Here we have lowered the \mb{\om_{A,bc}} symbol with the flat metric
\beq
\om_{A,bc}\~:=\~(-1)^{\eps_{A}\eps_{b}}\om_{bAc}\~,\~\~\~\~\~\~\~\~
\om_{bAc}\~:=\~E^{(0)}_{bd}\om^{d}{}_{Ac}(-1)^{\eps_{A}}\~.
\label{lowerantisympspinconn}
\eeq
In particular, the two connections \mb{\nabla^{(\Gamma)}} and
\mb{\nabla^{(\om)}} are antisymplectic at the same time, as a consequence of 
the correspondence \es{vielbeincovpres}{ee0}.

\subsection{First--Order \mb{\sss^{ab}} Matrices}
\label{secantisymp1st}

\noi
Because of the presence of the flat metric \mb{E_{(0)}^{ab}}, the symmetry of
the general linear Lie--algebra \mb{gl(W)} reduces to an antisymplectic
Lie--subalgebra. Its generators \mb{\sss^{ab}_{\pm}} read 
\beq
\sss^{ab}_{\pm}\~:=\~\cc^{a} (-1)^{\eps_{b}}\pp^{b}
\~\mp\~ (-1)^{(\eps_{a}+1)(\eps_{b}+1)}(a\leftrightarrow b)
\~,\~\~\~\~\~\~\~\~\~\~\~\~
\pp^{a}\~:=\~E_{(0)}^{ab} \papal{\cc^{b}}\~,
\label{antisympsigma1st}
\eeq

\beq
\eps(\sss^{ab}_{\pm})\=\eps_{a}+\eps_{b}+1
\~,\~\~\~\~\~\~p(\sss^{ab}_{\pm})\=0\~,
\label{antisympsigma1steps}
\eeq

\beq
\sss^{a}_{\pm c}\~:=\~ \sss^{ab}_{\pm}\~E^{(0)}_{bc}(-1)^{\eps_{c}}\~. 
\label{antisympsigma1stmixed}
\eeq

\noi
The \mb{\nabla_{A}^{(t)}} realization \e{nablaaomegarealiz} can be
identically rewritten into the following \mb{\sss^{ab}} matrix realization 
\beq
\nabla_{A}^{(\sss)}
\~:=\~\papal{\zz^{A}}+\Hf\om_{A,bc}\~\sss^{cb}_{+}(-1)^{\eps_{b}}
\= \papal{\zz^{A}}-\Hf\om_{A}{}^{b}{}_{c}\~\sss^{c}_{+b}\~,
\label{antisympnablasigma1st}
\eeq
\ie \mb{\nabla_{A}^{(t)}=\nabla_{A}^{(\sss)}} for an antisymplectic spin
connection. One gets a projection onto the \mb{\sss_{+}^{ab}} matrices (rather 
than the \mb{\sss_{-}^{ab}} matrices), because an antisymplectic spin
connection \mb{\om_{A,bc}} is symmetric, \cf \eq{spinconnantisymp}.

\noi
The \mb{\sss^{ab}_{+}} matrices satisfy an antisymplectic Lie--algebra:
\beq
[\sss^{ab}_{\pm},\sss^{cd}_{\pm}]
\= (-1)^{\eps_{a}(\eps_{b}+\eps_{c}+1)+\eps_{b}}\left(
E_{(0)}^{bc}\~\sss^{ad}_{+}-\sss^{bc}_{+}\~E_{(0)}^{ad} \right)
\mp(-1)^{(\eps_{a}+1)(\eps_{b}+1)}(a\leftrightarrow b)\~.
\label{antisympsigma1stliealg}
\eeq

\subsection{ \mb{\gamma^{a}} Matrices}
\label{secantisympgamma}

\noi
The flat \mb{\gamma^{a}} matrices can be defined via a
Berezin--Fradkin operator representation \cite{b66,f66}
\beq
\gamma^{a}_{\theta}\~\equiv\~
\gamma^{a}\~:=\~\cc^{a}+(-1)^{\eps_{a}}\theta \pp^{a}
\=\cc^{a}-\pp^{a}\theta
\~,\~\~\~\~\~\~\~\~\~\~\~\~\eps(\gamma^{a})\=\eps_{a}
\~,\~\~\~\~\~\~\~\~\~\~\~\~p(\gamma^{a})\=1\~(\mod\~2)\~.
\label{antisympflatgamma}
\eeq
The \mb{\gamma^{a}} matrices satisfy a Clifford--like algebra
\beq
[\gamma^{a},\gamma^{b}]
\= 2 (-1)^{\eps_{a}}\theta E_{(0)}^{ab} {\bf 1}\~. \label{antisympflatclifalg}
\eeq
The \mb{\gamma^{a}} matrices form a fundamental representation of the 
antisymplectic Lie--algebra \e{antisympsigma1stliealg}:
\beq
[\sss^{ab}_{\pm},\gamma^{c}]
\= \gamma^{a}_{\pm\theta}(-1)^{\eps_{b}}E_{(0)}^{bc}
\mp(-1)^{(\eps_{a}+1)(\eps_{b}+1)}(a\leftrightarrow b)\~.
\label{antisympgammasigma1st}
\eeq
As a consequence, if one commutes an antisymplectic spin connection
\e{antisympnablasigma1st} with a flat \mb{\gamma^{a}} matrix, one gets
\beq
[\nabla_{A}^{(\sss)},\gamma^{b}]
\= -\om_{A}{}^{b}{}_{c}\~\gamma^{c}\~.\label{antisympdiffflatgamma}
\eeq
Similarly, if one commutes an antisymplectic spin connection
\e{antisympnablasigma1st} with a curved \mb{\gamma^{A}} matrices, one gets
\beq
[\nabla_{A}^{(\sss)},\gamma^{B}]
\= -\Gamma_{A}{}^{B}{}_{C}\~\gamma^{C}\~,
\label{antisympdiffcurvedgamma}
\eeq
\cf \eqs{omegagammaf}{antisympdiffflatgamma}.

\subsection{Dirac Operator \mb{D^{(\sss)}}}
\label{secantisympdiracop}

\noi
We shall for the remainder of Section~\ref{secantisympspingeom} assume that
the connection is antisymplectic, torsionfree and \mb{\rho}--compatible.

\noi
The Dirac operator \mb{D^{(\sss)}} in the \mb{\sss^{ab}} representation 
\e{antisympnablasigma1st} is a \mb{\gamma^{A}} matrix \e{antisympflatgamma}
times a covariant derivative \e{antisympnablasigma1st}
\beq
D^{(\sss)}\~:=\~\gamma^{A}\nabla^{(\sss)}_{A}
\~,\~\~\~\~\~\~\eps(D^{(\sss)})\=0
\~,\~\~\~\~\~\~p(D^{(\sss)})\=1\~(\mod\~2)\~.
\label{antisympdiracop}
\eeq
The odd Laplacian \mb{\Delta_{\rho}^{(\sss)}} in the \mb{\sss} representation
\e{antisympnablasigma1st} is
\beq
2\Delta_{\rho}^{(\sss)}\~:=\~ (-1)^{\eps_{A}}
\nabla^{}_{A}E^{AB}\nabla^{(\sss)}_{B}
\=  \frac{(-1)^{\eps_{A}}}{\rho}\nabla^{(\sss)}_{A}
\rho E^{AB}\nabla^{(\sss)}_{B}\~.\label{oddlaplaciansigma}
\eeq

\begin{theorem}[Antisymplectic Weitzenb\"ock type formula for flat exterior
forms]
The difference between the square of the Dirac operator \mb{D^{(\sss)}}
and twice the odd Laplacian \mb{\Delta_{\rho}^{(\sss)}} in the \mb{\sss^{ab}} 
representation \e{antisympnablasigma1st} is
\bea
D^{(\sss)}D^{(\sss)} \~-\~ 2\theta \Delta_{\rho}^{(\sss)} 
&=& \frac{\theta}{4}(-1)^{\eps_{B}+\eps_{C}}
\sss_{-}^{BA}\~R_{AB,CD}\~\sss_{+}^{DC}\label{antisympslw1a} \\
&=& - \theta\cc^{A}\~R_{AB}\~\pp^{B}
+\frac{\theta}{2}\cc^{B}\cc^{A}\~R_{AB,CD}\~
\pp^{D}\pp^{C}(-1)^{\eps_{C}}\~.  \label{antisympslw1b}
\eea
\label{theoremantisymp1}
\end{theorem}

\noi
{\sc Proof of Theorem~\ref{theoremantisymp1}:}\~
Almost identical to the proof of Theorem~\ref{theoremantisymp0} because of 
\eq{curvconj1}.
\proofbox

\def\thesubsection{\thesection.\Alph{subsection}}
\setcounter{subsection}{0}
\subsection{Appendix: Shifted \mb{\sss^{\prime ab}_{+}} Matrices}
\label{secantisympshiftsigma}

\noi
We have already seen in Appendix~\ref{seccurvedantisymp2nd} that there
are no consistent antisymplectic second--order deformations of the
\mb{\sss^{ab}_{+}} matrices. The only remaining deformation is a \mb{c}--number
shift, 
\bea
\sss^{\prime ab}_{+}&:=&\sss^{ab}_{+}+\alpha E_{(0)}^{ab} {\bf 1} \~, \\
\sss^{\prime a}_{+b}&:=&\sss^{a}_{+b}
+\alpha (-1)^{\eps_{a}}\delta^{a}_{b} {\bf 1} \~,
\eea
with a parameter \mb{\alpha}, \cf \eq{antisympshift}. These shifted 
\mb{\sss^{\prime ab}_{+}} matrices satisfy the same Lie--algebra 
\e{antisympsigma1stliealg} and fundamental representation
\e{antisympgammasigma1st} as the \mb{\sss^{ab}_{+}} matrices.
Moreover, the shift does not affect the \mb{\sss^{ab}} representation 
\e{antisympnablasigma1st} of the spin connection, because of tracefree
condition \e{tracecond}. Similarly, the curvature
\beq
[\nabla^{(\sss)}_{A},\nabla^{(\sss)}_{B}]
\= -\Hf R_{AB}{}^{d}{}_{c}\~\sss^{c}_{+d}\~.\label{antisymplcurvsigma}
\eeq 
is unaffected, since the shift--term is proportional to the Ricci two--form 
\mb{\cR_{AB}\!=\!0}, which is zero. Thus we conclude that the \mb{c}--number
shift \mb{\sss^{ab}_{+}\to\sss^{\prime ab}_{+}} has no effects at all on the
construction.

\def\thesubsection{\thesection.\arabic{subsection}}
\setcounter{equation}{0}
\section{Conclusions}
\label{secconcl}

\noi
The main objective of the paper is to gain knowledge about the deepest and most
profound geometric levels of the field--antifield formalism
\cite{bv81,bv83,bv84}. It is imperative to better understand the geometric
meaning of the odd scalar curvature \mb{R}, which sits as a zeroth--order term
in the odd \mb{\Delta} operator \e{deltaop}, and which descends to the quantum
master equation \mb{\Delta\exp[\Ih W]=0} as a two--loop contribution:
\beq
(W,W) \= 2i\hbar\Deltarho W - \hbar^{2}\frac{R}{4}\~. \label{mqme}
\eeq
We have in this paper investigated the hypothesis that the zeroth--order term
\mb{-R/4} of (twice) the odd \mb{\Delta} operator \e{deltaop} is related to
the zeroth--order term \mb{-R/4} in the Schr\"odinger--Lichnerowicz formula
\e{riemsl2}. We have so far been unable to give a closed argument that such
relationship exists. In fact, Theorem~\ref{theoremriem4} indicates that there
is no relation, as explained in the Introduction. Some of the main results of
the paper are the following.

\begin{itemize}
\item
We have classified scalar invariants of suitable dimensions that depend on
the density \mb{\rho} and the metric, \cf Proposition~\ref{propositioneven}
and Proposition~\ref{propositionodd}.
\item
We have identified (via a \mb{\rho}--independence argument) a Riemannian 
counterpart \e{evenanaloguedelta} of the antisymplectic \mb{\Delta} operator
\e{deltaop}, that takes scalars to scalars, and we have, in terms of formulas,
traced the minus--a--quarter coefficient in front of \mb{R} from the Riemannian
to the antisymplectic side, \cf Subsection~\ref{secoddscalar}.
\item
We have tied the Riemannian \mb{\Delta} operator \e{evendelta} to the quantum
Hamiltonian \mb{\hat{H}} for a particle moving in a curved Riemannian space,
\cf Subsection~\ref{secparticle}. 
\item
We have derived the Laplace-Beltrami operator \mb{\Deltarhog} by projecting
the square of the bispinor Dirac operator \mb{D^{(\sig\sig^{T})}} to a
singlet state \mb{\sstate}, \cf Theorem~\ref{theoremriem4}.
\item
We have found a first--order formalism for antisymplectic spinors and proved
two Weitzenb\"ock--type identities (Theorem~\ref{theoremantisymp0} and
Theorem~\ref{theoremantisymp1}) that are in exact one--to--one correspondence
with their Riemannian siblings (Theorem~\ref{theoremriem0} and
Theorem~\ref{theoremriem1}).
\end{itemize}

\noi
However, there is in our approach {\em no} antisymplectic analogue of the
Riemannian second--order formalism and the Schr\"odinger--Lichnerowicz formula
\e{riemsl2}. A bit oversimplified, this is because the canonical choice for
antisymplectic second--order \mb{\Sigma_{\pm}^{AB}} matrices is
\beq
\Sigma_{\pm}^{AB}\~\equi{?}\~ \frac{1}{4}\Gamma^{A}\star\Gamma^{B}
\mp(-1)^{(\eps_{A}+1)(\eps_{B}+1)}(A \leftrightarrow B)\~,\~\~\~\~\~\~\~\~
\eps(\Sigma_{\pm}^{AB})\=\eps_{A}+\eps_{B}+1\~,\~\~\~\~\~\~\~\~
p(\Sigma_{\pm}^{AB})\=0\~,
\label{naiveantisympsigma2nd}
\eeq
where ``\mb{\star}'' is a Fermionic multiplication, \mb{\eps(\star)=1}. This
choice \e{naiveantisympsigma2nd} meet all the requirements of
Grassmann--parity and symmetry, and is a direct analogue of the Riemannian
second--order \mb{\Sigma_{\pm}^{AB}} matrices \e{riemcurvedsigma2nd}.
Unfortunately, such \mb{\star} multiplication does not admit a Berezin--Fradkin
representation of the \mb{\Gamma^{A}} matrices, \cf Appendix~\ref{secattemp}.
We instead introduced a Fermionic nilpotent parameter \mb{\theta}, which
may formally be identified with the inverse \mb{\star^{-1}}, and which serves
as a Fermionic analogue of the ``Planck constant'' \mb{\lambda} from the
Riemannian case. Then the \mb{\star} multiplication itself should be identified
with the inverse \mb{\theta^{-1}}, which is an ill--defined quantity, and hence
the above formula \e{naiveantisympsigma2nd} for the \mb{\Sigma_{\pm}^{AB}}
matrices does not make sense. Note however that the nilpotent \mb{\theta}
parameter breaks the non--degeneracy of the Clifford algebra
\e{curvedantisympclifalg}.

\vspace{0.8cm}

\noi
{\sc Acknowledgement:}~We would like to thank Poul Henrik Damgaard, the Niels
Bohr Institute and the Niels Bohr International Academy for warm hospitality.
I.A.B.\ would also like to thank Michal Lenc and the Masaryk University for
warm hospitality. The work of I.A.B.\ and K.B.\ is supported by the Ministry
of Education of the Czech Republic under the project MSM 0021622409. The work
of I.A.B.\ is supported by grants RFBR 08--01--00737, RFBR 08--02--01118 and
LSS--1615.2008.2.

\end{document}